\documentclass[acmsmall]{acmart}
\AtBeginDocument{%
  }

\setcopyright{acmlicensed}
\copyrightyear{2024}
\acmYear{2024}
\acmDOI{XXXXXXX.XXXXXXX}

\acmJournal{Tosem}

\usepackage{graphicx}
\usepackage{subcaption}
\usepackage{pifont} 
\usepackage{tikz}
\usepackage{threeparttable}
\usepackage{url}
\usepackage{colortbl}
\usepackage{multirow}
\usepackage{algorithm}
\usepackage[noend]{algpseudocode}

\begin{document}

\title{MOS: Towards Effective Smart Contract Vulnerability Detection through Mixture-of-Experts Tuning of Large Language Models}

\author{Hang Yuan}
\authornote{Both authors contributed equally to this research.}

\author{Lei Yu}
\authornotemark[1]

\affiliation{
  \institution{Institute of Software, Chinese Academy of Sciences; University of Chinese Academy of Sciences}
  \city{Beijing}
  \country{China}
}

\author{Zhirong Huang}
\affiliation{%
  \institution{Institute of Software, Chinese Academy of Sciences}
  \city{Beijing}
  \country{China}}

\author{Jingyuan Zhang}
\affiliation{%
  \institution{Institute of Software, Chinese Academy of Sciences}
  \city{Beijing}
  \country{China}}

\author{Junyi Lu}
\affiliation{%
  \institution{Institute of Software, Chinese Academy of Sciences}
  \city{Beijing}
  \country{China}}

\author{Shiqi Cheng}
\affiliation{%
  \institution{Institute of Software, Chinese Academy of Sciences}
  \city{Beijing}
  \country{China}}

\author{Li Yang}
\authornote{Corresponding author.}
\affiliation{%
  \institution{Institute of Software, Chinese Academy of Sciences}
  \city{Beijing}
  \country{China}}

\author{Fengjun Zhang}
\affiliation{%
  \institution{Institute of Software, Chinese Academy of Sciences}
  \city{Beijing}
  \country{China}}

\author{Jiajia Ma}
\affiliation{%
  \institution{Institute of Software, Chinese Academy of Sciences}
  \city{Beijing}
  \country{China}}

\author{Chun Zuo}
\affiliation{%
  \institution{Sinosoft Company Limited}
  \city{Beijing}
  \country{China}}
\email{zuochun@sinosoft.com.cn}

\renewcommand{\shortauthors}{Hang Yuan et al.}

\begin{abstract}
Smart contract vulnerabilities pose significant security risks to blockchain systems, potentially leading to severe financial losses. Existing methods face several limitations: (1) Program analysis-based approaches heavily rely on predefined patterns, lacking flexibility and adaptability to new vulnerability types; (2) Deep learning-based methods lack explanations for their detection results; (3) Large language model-based approaches suffer from high false positive rates. To address these challenges, we propose MOS, a smart contract vulnerability detection framework based on mixture-of-experts tuning (MOE-Tuning) of large language models. First, we conduct continual pre-training on a large-scale smart contract dataset to provide a domain-enhanced initialization for the subsequent MOE-Tuning process. Second, to provide reliable explanations for detected vulnerabilities, we construct a high-quality MOE-Tuning dataset through a multi-stage pipeline combining large language model generation and expert verification. Third, we design a vulnerability-aware routing mechanism that activates the most relevant expert networks by analyzing code features and their matching degree with experts, addressing the limitation of predefined patterns in program analysis approaches. Finally, we extend the feed-forward layers of the LLM after continual pre-training into multiple parallel expert networks, each specializing in specific vulnerability patterns. 
We also employ a dual-objective loss function: one for optimizing vulnerability detection and explanation performance, and another for ensuring reasonable distribution of different vulnerability types to corresponding experts through entropy calculation of expert load distribution, thereby improving model performance in complex vulnerability scenarios. 
Extensive experiments show that MOS significantly outperforms existing state-of-the-art methods with average improvements of 6.32\% in F1 score and 4.80\% in accuracy. The vulnerability explanations generated by MOS also demonstrate high quality, achieving positive ratings (scores of 3 and 4 on a 4-point scale) of 82.96\%, 85.21\% and 94.58\% for correctness, completeness, and conciseness, respectively, through a combined approach of human evaluation and LLM evaluation.
\end{abstract}



\received{20 February 2007}
\received[revised]{12 March 2009}
\received[accepted]{5 June 2009}

\maketitle

\section{Introduction}
Blockchain technology has been rapidly adopted across various domains due to its decentralized architecture \cite{swan2015blockchain}. This innovative technology enables the creation of secure, distributed digital ledgers for recording transactions \cite{hewa2021survey}. By utilizing advanced cryptographic methods, blockchain ensures the integrity and verification of each transaction, establishing a highly reliable technological framework \cite{wood2014ethereum, yu2023money}. Within this ecosystem, smart contracts function as self-executing programs on the blockchain, automating the management of digital assets such as cryptocurrencies. These contracts are activated when specific conditions are met and, once deployed, become permanent components of the blockchain \cite{zou2019smart}. However, the immutability and inherent complexity of smart contracts pose significant security challenges \cite{zou2019smart}. The well-known DAO incident \cite{dhillon2017dao,mehar2019understanding} serves as a cautionary example, demonstrating the potential severity of such vulnerabilities. This security breach resulted in the illegal transfer of \$60 million worth of Ethereum, causing widespread impact on the blockchain community \cite{alharby2017blockchain, hegedHus2018towards}. This incident highlights the importance of enhancing smart contract security to prevent similar catastrophic consequences in the future.

\textbf{Program analysis-based} smart contract vulnerability detection tools, despite making progress in recent years, still face significant limitations. First, these methods \cite{feist2019slither,torres2019art, mueller2017mythril, tsankov2018securify, luu2016making, bose2022sailfish, choi2021smartian} heavily rely on expert-defined detection rules and patterns. Manually defined patterns are not only error-prone but also struggle to cover complex vulnerability patterns, especially given the rapid growth in smart contract numbers where a few experts cannot design precise detection rules for all contracts \cite{liu2021combining}. Second, these rule-based methods often show unsatisfactory performance in practical applications. Rigid rules lead to high false-positive and false-negative rates, and attackers can easily bypass pattern checking through specific techniques. Particularly for complex vulnerabilities such as DoS and Front Running, due to their intricate nature, traditional program analysis methods struggle to define rules that cover all possible vulnerability patterns \cite{chen2023chatgpt}. Additionally, some program analysis-based tools are limited in their vulnerability type coverage. As shown in \cite{chen2023chatgpt}, tools like Maian and AChecker can only detect one type of vulnerability (Access Control), with detection rates of 51.9\% and 50.0\% respectively.


Furthermore, researchers have begun exploring \textbf{deep learning-based} approaches \cite{gao2019smartembed, qian2020towards, zhuang2020smart, liu2021combining, cai2023combine, luo2024scvhunter} for smart contract vulnerability detection to address the limitations of traditional methods. These approaches learn by treating source code as text sequences or converting it into graphs containing control and data flow semantics, achieving promising results in vulnerability detection. However, deep learning-based methods still have their shortcomings. While these methods can effectively determine whether contracts contain vulnerabilities, they perform poorly in explaining model decisions and locating specific vulnerability positions. For smart contract developers and auditors, understanding why and where vulnerabilities exist is crucial for efficient debugging and security enhancement. Considering that smart contract code size ranges from dozens to thousands of lines, manually locating erroneous statements remains time-consuming and error-prone even when developers are aware of the presence of vulnerabilities \cite{zhang2022reentrancy}.


Recent research\cite{david2023you,hu2023large,ma2024combining,sun2024gptscan} has shown that large language models (LLMs) demonstrate potential in smart contract auditing. Compared to existing analysis tools, LLMs can emulate human linguistic understanding and reasoning, describing any type of vulnerability using natural language, thus potentially detecting a wider range of vulnerabilities, including those unknown or uncategorized a priori\cite{hu2023large}. However, LLMs face several challenges: they can produce numerous false positives leading to low precision; often fail to identify all vulnerabilities when used naively; show varying performance across different vulnerability types; and struggle with complex scenarios involving multiple vulnerabilities. A recent systematic evaluation\cite{sun2024llm4vuln} shows that even when equipped with state-of-the-art approaches like enhancing GPT-4 with RAG\cite{lewis2020retrieval}, the precision only reaches $\sim$30\% when both decision and justification are correct. Researchers have proposed various \textbf{LLM-based} techniques: GPTScan\cite{sun2024gptscan} combines GPT with static analysis by breaking down each type of logical vulnerability into scenarios and attributes for GPT matching and employs static confirmation to reduce false positives; Hu et al.\cite{hu2023large} proposed a two-stage adversarial collaborative framework; while Ma et al.\cite{ma2024combining} developed a general framework integrating fine-tuning with proxy models, introducing Ranker and Critic proxies for vulnerability analysis. However, these methods still have limitations: GPTScan heavily relies on manually defined scenarios and attributes for each vulnerability type, making it difficult to adapt to undefined vulnerabilities, while both \cite{sun2024gptscan} and \cite{ma2024combining}'s approaches primarily focus on logical vulnerabilities with limited adaptability to other vulnerability types. Additionally, GPTScan and GPTLens heavily rely on the inherent capabilities of GPT-3.5-Turbo and GPT-4.

Based on the above analysis, existing smart contract vulnerability detection approaches face three key challenges: (1) Program analysis-based methods rely heavily on predefined patterns, making them inflexible and unable to adapt to new vulnerability types, with each tool limited to detecting specific vulnerability types; (2) Deep learning-based methods lack explanation for detected vulnerabilities; (3) LLM-based approaches suffer from high false positive rates and inconsistent performance across different vulnerability types, or are specifically designed for limited type of vulnerability.

To address these challenges, we propose MOS, a framework that leverages Mixture-of-Experts tuning of large language models for effective smart contract vulnerability detection. To address the lack of explanations for vulnerabilities detected by deep learning-based methods, we constructed specialized datasets for MOE-tuning, enabling the model to provide corresponding explanations and analyses while detecting vulnerabilities. In the data construction process, we used Qwen2.5-72B-Instruct and Mistral-Large-Instruct-2407-123B to generate initial vulnerability explanations. To ensure content accuracy and relevance, we designed label-guided prompts for each vulnerability type, where prompts are specifically crafted based on vulnerability labels and their unique characteristics. For instance, prompts for reentrancy vulnerabilities focus on analyzing call.value() and call() usage, operation order, external calls and access control, and internal function implementation. These explanations are then evaluated using Llama-3.1-70B-Instruct based on correctness, completeness, and conciseness. The highest-scoring explanations undergo review and refinement by security experts grouped by vulnerability type, ensuring high-quality training data. Our MOS consists of two key components: \textbf{Vulnerability-aware Routing} and \textbf{Specialized Mixture of Experts Network}. First, the Vulnerability-aware Routing mechanism converts smart contract code and prompt information into vectorized features as input for the routing network. The routing network calculates the matching degree between code features and each expert through trainable parameter matrices and activates the top-k experts with the highest scores. This design avoids the rigidity of traditional program analysis methods and enables adaptive handling of emerging vulnerability types. Second, the Specialized Mixture of Experts Network extends the feed-forward network layers of pre-trained language models by replacing each layer with multiple parallel expert networks. These expert networks share the same network structure but maintain independent parameters, enabling each expert to develop specialized understanding of specific vulnerability patterns while maintaining general comprehension capabilities. The expert networks are specialized in different vulnerability types, including reentrancy, integer overflow/underflow, timestamp dependency, and delegatecall. Third, for contracts containing multiple vulnerabilities, we design a dynamic combination mechanism that integrates detection results from multiple activated experts and optimizes their combination based on routing network weights, thereby improving model's performance in complex vulnerability scenarios.

The training process consists of two stages: continual pre-training and MoE-tuning. We first conduct continual pre-training on a large-scale smart contract code dataset to enhance the LLM's understanding of smart contract domain. In the MoE-tuning stage, we extend the feed-forward layers of the LLM after continual pre-training into expert networks and initialize routing parameters using corresponding parameters from the LLM after continual pre-training. The fine-tuning employs a dual-part loss function: one for vulnerability detection and explanation tasks, and another for ensuring reasonable allocation of different vulnerability types to different experts through entropy calculation of expert load distribution. This design improves the model's consistency across various vulnerability types and its explanation capability. For training strategy, we freeze the parameters of the LLM after continual pre-training and only train the newly added expert networks and routing layers, thus maintaining the model's fundamental capabilities while enhancing the expert networks' vulnerability detection abilities.

To validate the effectiveness of MOS, we conducted a comprehensive experimental evaluation. Our evaluation dataset integrates samples from two significant sources: the smartbugs-curated dataset \cite{durieux2020empirical} and Qian et al.'s dataset \cite{qian2023cross}, covering four major vulnerability types: reentrancy, timestamp dependence, integer overflow/underflow, and delegatecall. The results show that MOS significantly outperforms state-of-the-art methods across all vulnerability types. Notably, MOS achieves F1 score 10.02\%, 1.92\%, 3.60\%, and 9.75\% higher than the previous best performance for reentrancy, timestamp dependence, integer overflow/underflow, and delegatecall vulnerabilities, respectively. In terms of accuracy, MOS surpasses the previous SOTA methods by 6.70\%, 0.70\%, 6.88\%, and 4.91\% for these four vulnerability types.
In addition to detection performance, we also evaluated the quality of vulnerability explanations generated by MOS. Inspired by \cite{sun2023real}, our evaluation framework incorporates assessments from both four security experts and an advanced LLM (Llama-3.1-70B-Instruct). The Kappa coefficient was used to ensure consistency among evaluators. The evaluation results indicate that the explanations generated by MOS are more correct, complete, and concise compared to LLaMA-3.1-8B-Instruct. MOS received positive scores (4 points and 3 points) in 82.96\%, 85.21\%, and 94.58\% of cases for correctness, completeness, and conciseness, respectively, significantly outperforming the baseline method.


The main contributions of this paper are as follows:

\begin{itemize}
    \item  We propose MOS, the first framework that leverages Mixture-of-Experts tuning of large language models for smart contract vulnerability detection. Our approach integrates vulnerability-aware expert routing with specialized expert networks, achieving state-of-the-art performance on four major vulnerability types while providing reliable explanations. 

    \item We conduct comprehensive experiments demonstrating MOS's effectiveness, achieving average improvements of 6.41\% in F1 score and 4.80\% in accuracy compared to existing methods. Through a combined approach of human evaluation and LLM evaluation, we validate the model's capability to generate accurate, comprehensive, and concise vulnerability explanations, with strong inter-rater reliability (Kappa > 0.4) across all evaluation dimensions.
    

    \item To promote further research in related fields, we make available our dataset and the source code of MOS.
\end{itemize}

The organization of the rest of this paper is as follows. In Section \ref{background_motivation}, we provide background information about smart contracts and present a motivational study that illustrates the limitations of current approaches. In Section \ref{method}, we detail our proposed MOS methodology, including the data construction process, model architecture, continual pre-training stage, MOE-tuning stage, and evaluation approaches for vulnerability explanations. In Section \ref{evaluation}, we present comprehensive experimental evaluations addressing six research questions, demonstrating MOS's effectiveness compared to state-of-the-art baselines and analyzing its internal mechanisms. In Section \ref{limitations}, we discuss the limitations of our approach. In Section \ref{related_work}, we review related work in smart contract vulnerability detection and mixture-of-experts models. In Section \ref{conclusion}, we conclude the paper and outline future research directions. 

\section{Background and Motivation}
\label{background_motivation}
\subsection{Smart Contract}

Smart contracts, introduced by Nick Szabo in 1994 \cite{szabo1996smart}, are self-executing digital agreements that automatically enforce contractual terms. With blockchain technology, particularly Ethereum, these contracts have found practical implementation. On Ethereum, smart contracts are programs that automatically execute when predetermined conditions are met. For example, if Bob and Alice have a contract where Bob must pay a penalty for breach, the system will automatically deduct the penalty from Bob's deposit if he violates the terms.

\begin{figure}[htp]
\centerline{\includegraphics[width=1.0\textwidth]{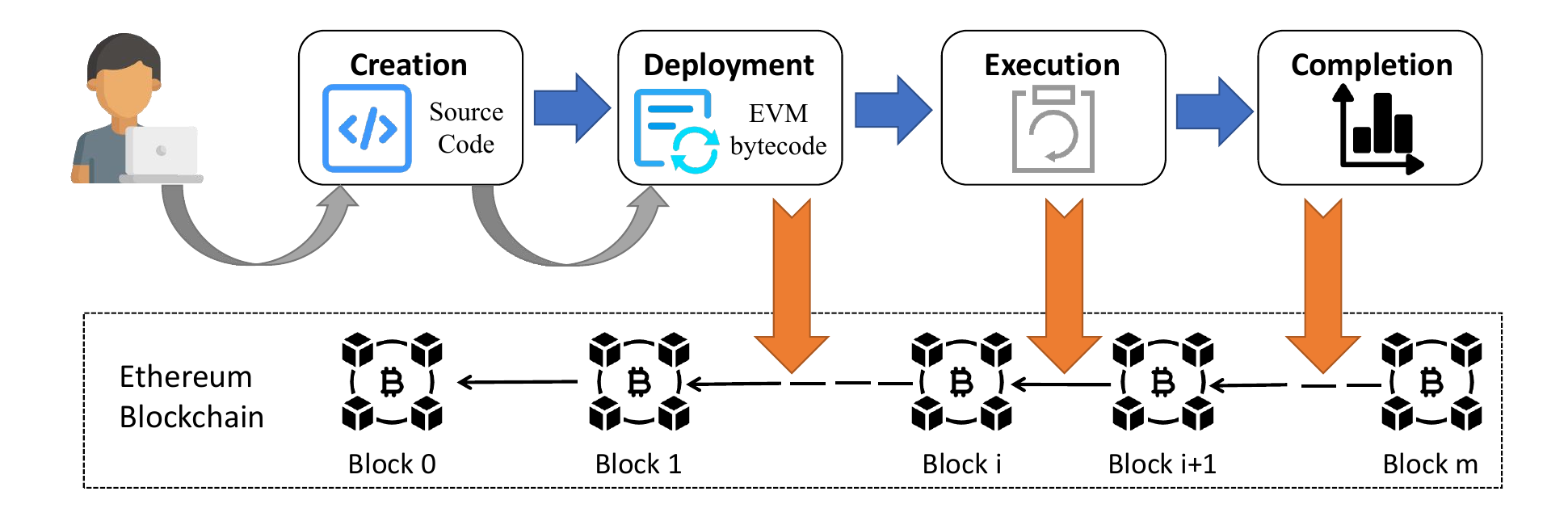}}
\caption{The Life Cycle of Smart Contract}
\label{sc_life_cycle}
\end{figure}

As shown in Fig. \ref{sc_life_cycle}, the life cycle of smart contracts begins with contract creation, where multiple parties (such as stakeholders and lawyers) reach a contractual agreement after discussions. Software developers then convert this agreement from natural language into smart contracts written in computer languages, predominantly using Solidity. Next, in the deployment phase, the validated smart contracts are deployed on the Ethereum blockchain. During the execution phase, smart contracts are triggered by events, which can be external (such as receiving a payment) or internal (such as reaching a specific date or time). Once an event satisfies the predefined conditions, the corresponding functions automatically execute. Finally, in the completion phase, all transactions during execution and updated states are permanently stored in the blockchain, and digital assets are transferred between parties as specified in the contract.

Due to the immutable nature of smart contracts once deployed on the blockchain, detecting all vulnerabilities before deployment is absolutely critical. If only some vulnerabilities are identified during the pre-deployment phase, the remaining undetected vulnerabilities could have catastrophic consequences. These vulnerabilities could lead to irreversible financial losses, as demonstrated by infamous incidents like the DAO hack \cite{dhillon2017dao,mehar2019understanding} where \$50 million worth of Ether was stolen, or the Parity wallet bug \cite{destefanis2018smart} that permanently froze \$280 million in assets. Unlike traditional software where patches can be applied post-deployment, smart contract vulnerabilities cannot be fixed after deployment, making them permanent security risks that malicious actors can exploit repeatedly.

\subsection{Problem Definition}

Our research focuses on developing an automated vulnerability detection system for smart contracts. The system performs binary classification on each smart contract, outputting a prediction $\hat{y}$, where $\hat{y}=1$ signifies a vulnerable contract and $\hat{y}=0$ indicates a secure one. Beyond mere detection, our system provides comprehensive vulnerability analysis including specific vulnerability types, their locations within the code, and potential security implications. We concentrate on four critical vulnerability categories:

\textbf{Reentrancy:} This vulnerability emerges when a contract makes external calls before finalizing its internal state updates. Malicious actors can exploit this by recursively calling the vulnerable function, potentially causing multiple unauthorized operations like duplicate withdrawals.

\textbf{Timestamp Dependency:} This weakness manifests when contracts rely on block timestamps for crucial decisions. Since miners have some flexibility in setting block timestamps, contracts using these timestamps for randomization or decision-making become susceptible to manipulation.

\textbf{Integer Overflow/Underflow:} These vulnerabilities occur when arithmetic operations produce results outside the variable's valid range. Overflow happens when a number exceeds its maximum value and cycles to its minimum, while underflow occurs when a number drops below its minimum and cycles to its maximum, leading to incorrect calculations and potential security breaches.

\textbf{Delegatecall Vulnerability:} This involves risks associated with dynamic code execution where one contract executes code from another. The unique characteristic of delegatecall is that it preserves the context of the calling contract, meaning the called code can directly manipulate the caller's state variables, potentially compromising contract security if not properly implemented.

\textbf{These four vulnerabilities are our primary focus due to the following considerations:} (i) Research indicates that these vulnerabilities are responsible for roughly 70\% of financial damages in Ethereum smart contract security incidents \cite{chen2020survey}. (ii) Studies conducted by \cite{chen2020survey,gao2019easyflow,praitheeshan2019security} reveal that these vulnerabilities appear more frequently in Ethereum smart contracts than other types of vulnerabilities. (iii) The OWASP Smart Contract Top 10~\cite{owasp2023} lists reentrancy, integer overflow/underflow, and timestamp dependence as the top three security concerns, while also recognizing delegatecall vulnerability under the category of unchecked external calls. (iv) The exploitation of these vulnerabilities can severely compromise contract security and operations, potentially resulting in substantial financial losses. (v) Despite widespread awareness, the inherent complexity and immutable nature of smart contracts often lead to these vulnerabilities being either overlooked or inadequately addressed.

\subsection{Motivational Study}

\begin{figure}[htbp]
\centerline{\includegraphics[width=0.8\textwidth]{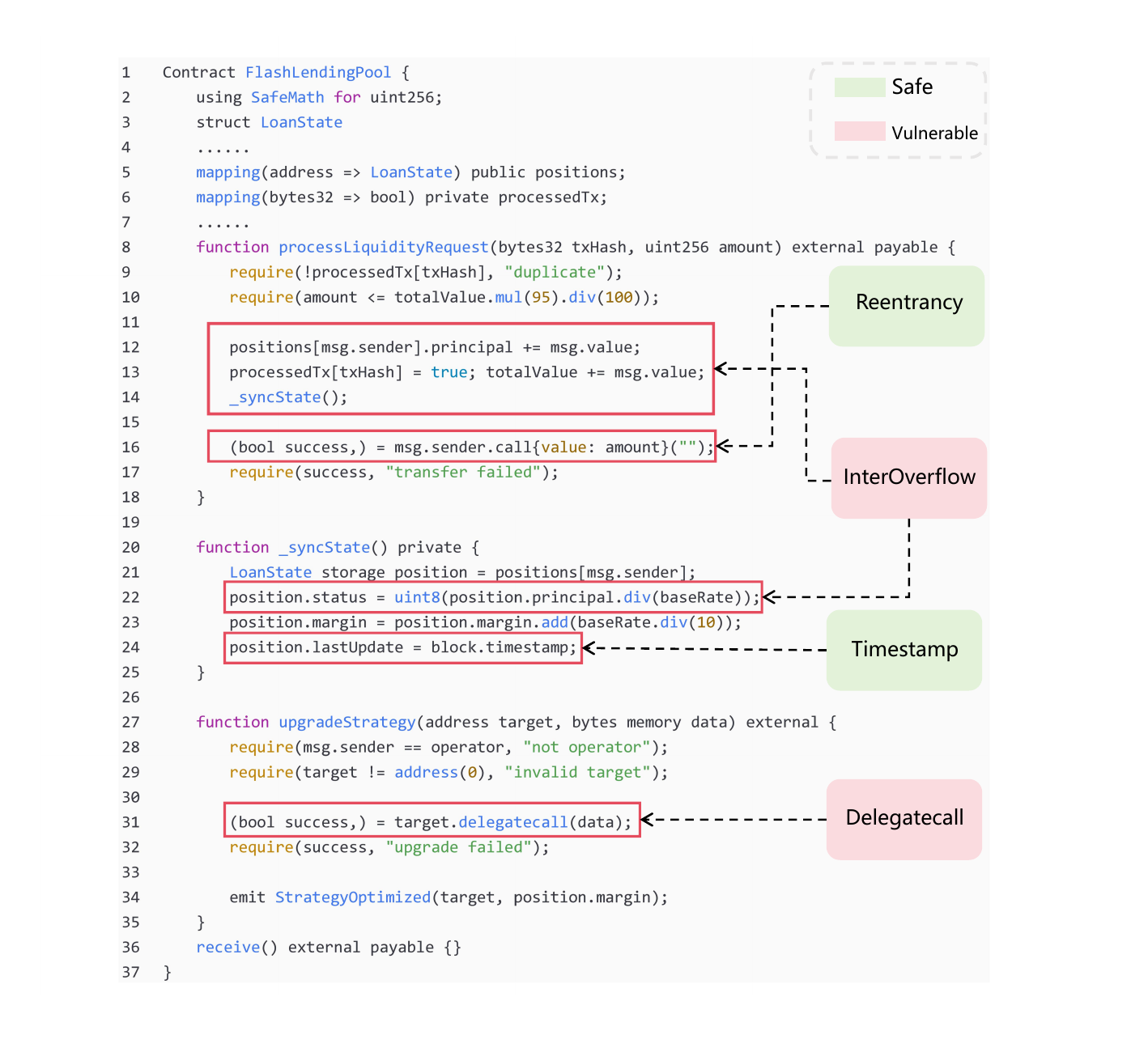}}
\caption{Motivation: A complex smart contract with four vulnerabilities }
\label{motivation}
\end{figure}

\subsubsection{Case Study with Flash Loan Contract}

A simplified Flash Loan contract (Fig. \ref{motivation}) illustrates the complex security analysis scenarios that large language models struggle to handle. While this contract implements basic lending functionality, it contains security features that require deep contextual understanding. Detailed analysis reveals that the contract is secure against reentrancy attacks, as the \texttt{processLiquidityRequest} function strictly follows the Checks-Effects-Interactions pattern, ensuring state updates (\texttt{positions} modification and \texttt{processedTx} marking) occur before external calls. Similarly, \texttt{block.timestamp} usage is limited to record updates, avoiding timestamp dependency vulnerabilities. The contract contains two critical vulnerabilities: integer overflow risks in multiple operations (the multiplication of \texttt{position.principal} with \texttt{baseRate}, addition of accumulated interest to principal, and unsafe conversion to \texttt{uint8}), and an unsafe \texttt{delegatecall} operation that allows arbitrary code execution without sufficient validation.

\begin{table}[htbp]
\caption{A Case Study of Popular LLMs in Smart Contract Vulnerability Detection. (1: vulnerability exists/detected, 0: vulnerability does not exist/not detected)}
\centering
\begin{tabular}{l|c|c|c|c}
\hline
\textbf{Model} & \textbf{Reentrancy} & \textbf{Timestamp} & \textbf{Integer Overflow} & \textbf{Delegatecall} \\
\hline
LLaMA3.1-8B-Instruct & 1 & 1 & 1 & 0 \\
LLaMA3.2-3B-Instruct &1 & 1 & 0 & 1 \\
Qwen2.5-7B-Instruct &1 & 1 & 0 & 1 \\
Qwen2-7B-Instruct &1 & 1 & 0 & 1 \\
GPT-4o &1 &0 &0 &1 \\
Claude-3.5-Sonnet &1 &1 &1 &1 \\
\hline
\textbf{Ground Truth} & 0 & 0 & 1 & 1 \\
\hline
\end{tabular}
\label{tab:moe_motivation}
\end{table}

The empirical results from various large language models, as shown in Table \ref{tab:moe_motivation}, reveal significant limitations in handling these complex scenarios. Comparing with the ground truth (where only integer overflow and delegatecall vulnerabilities exist), all models produced false positives by incorrectly identifying reentrancy vulnerabilities. Most models also showed false positives for timestamp dependency issues, with only GPT-4o correctly identifying the absence of timestamp vulnerability. While most models successfully detected the delegatecall vulnerability, only LLaMA3.1-8B-Instruct and Claude-3.5-Sonnet correctly identified the integer overflow vulnerability. These results demonstrate that current LLMs face two major challenges in smart contract security analysis: (1) They struggle to differentiate between secure coding patterns and actual vulnerabilities, as evidenced by all models incorrectly flagging the secure Checks-Effects-Interactions pattern as a reentrancy vulnerability; (2) Their detection performance varies significantly across different vulnerability types - while they perform well in identifying delegatecall vulnerabilities, they show poor accuracy in detecting integer overflow issues and generate false positives for reentrancy and timestamp vulnerabilities.

\subsubsection{Empirical Analysis}
A large-scale empirical study conducted by \cite{chen2023chatgpt} examined 14 program analysis-based detection tools and GPT models (including GPT-3.5-Turbo, GPT-4, and optimized GPT-4) using the SmartBugs dataset containing 142 annotated vulnerabilities. Based on their comprehensive evaluation results, our analysis presents the most representative and widely-used program analysis-based tools from their study, along with their complete findings on large language models, to illustrate the key limitations in current smart contract vulnerability detection methods.

The experimental results shown in Table \ref{rule_compare} reveal two notable issues with program analysis-based detection tools. First, their detection scope is limited, with most tools capable of detecting only specific types of vulnerabilities. The data shows that tools like Maian only support detection of Access Control vulnerabilities (F1-score 51.9\%), while Honeybadger only covers Reentrancy and Arithmetic Issues. Second, they perform poorly when handling complex vulnerabilities: mainstream tools like Oyente completely fail to identify Access Control vulnerabilities (F1-score 0\%), and multiple tools including Oyente and Securify are entirely ineffective ("/") in detecting Time Manipulation vulnerabilities that require understanding temporal logic. These data clearly demonstrate that program analysis-based tools not only suffer from limited detection scope but also face significant technical bottlenecks in processing complex vulnerabilities requiring deep semantic understanding.

\begin{table}[h]
\small
\centering
\caption{Performance (F1 Score) Comparison of Program Analysis-based Smart Contract Vulnerability Detection Tools}
\label{rule_compare}
\begin{tabular}{|l|c|c|c|c|c|}
\hline
\textbf{Tool} & \textbf{Reentrancy} & \textbf{Access} & \textbf{Arithmetic} & \textbf{Unchecked} & \textbf{Time} \\
 & & \textbf{Con-} & \textbf{Issues} & \textbf{Return} & \textbf{Manipu-} \\
 & & \textbf{trol} & & \textbf{Values} & \textbf{lation} \\
\hline
Oyente & 87.5\% & 0.0\% & 23.0\% & / & 0.0\% \\
Maian & / & 51.9\% & / & / & / \\
Honeybadger & 76.0\% & / & 0.0\% & / & / \\
Securify & 72.0\% & 8.7\% & / & 91.7\% & / \\
Slither & 87.0\% & 23.1\% & / & 72.7\% & 60.0\% \\
\hline
\end{tabular}
\end{table}

\begin{table}[h]
\small
\centering
\caption{Performance (F1 Score) Comparison of GPT Models on Smart Contract Vulnerability Detection}
\label{llm_compare}
\begin{tabular}{|l|c|c|c|c|c|}
\hline
\textbf{Model} & \textbf{Reentrancy} & \textbf{Access} & \textbf{Arithmetic} & \textbf{Unchecked} & \textbf{Time} \\
 & & \textbf{Con-} & \textbf{Issues} & \textbf{Return} & \textbf{Manipu-} \\
 & & \textbf{trol} & & \textbf{Values} & \textbf{lation} \\
\hline
GPT-3.5-turbo & 33.8\% & 23.9\% & 34.8\% & 59.0\% & 19.0\% \\
GPT-4 & 49.8\% & 28.4\% & 35.7\% & 67.9\% & 28.4\% \\
GPT-4 (optimized) & 59.8\% & 32.2\% & 44.0\% & 68.2\% & 29.6\% \\
\hline
\end{tabular}
\end{table}

Although emerging large-language model approaches theoretically overcome the limited detection scope of traditional tools by covering all vulnerability types, their detection performance is concerning and the detection performance varies across different types of vulnerabilities as shown in Table \ref{llm_compare}. Specifically, GPT-3.5-turbo shows poor performance across several critical vulnerability types. Even the more capable GPT-4 demonstrates notable deficiencies in detection capabilities: achieving only 28.4\% detection rate for Time Manipulation vulnerabilities, 28.4\% for Access Control vulnerabilities, and 35.7\% for Arithmetic Issues. Researchers attempted to enhance model performance through temperature adjustment (set to 0) and the introduction of MITB \cite{sun2024gptscan} (Mimic In-The-Background prompting). Experimental data shows that while these methods can improve performance for specific vulnerability types (e.g., GPT-4's performance on Reentrancy improved from 49.8\% to 59.8\%, and on Arithmetic Issues from 35.7\% to 44.0\%), the overall detection capability improvement remains limited. Even for Unchecked Return Values, where detection rates are relatively higher, the optimized performance only improved marginally from 67.9\% to 68.2\%. These results indicate that even state-of-the-art large language models face significant challenges in precisely identifying and analyzing smart contract vulnerabilities.

It is important to note that the four vulnerability types in the case study closely correspond and intersect with the five vulnerability types in the empirical analysis. Reentrancy vulnerabilities directly correspond as independent types in both analyses; Timestamp Dependency is a specific manifestation of Time Manipulation vulnerabilities; Integer Overflow is a specific subtype of Arithmetic Issues. Notably, the delegatecall vulnerability in the case study actually involves multiple security dimensions: when lacking proper checks on delegatecall return values, it falls under the Unchecked Return Values category; while the risk of arbitrary code execution due to insufficient validation of delegatecall target contracts primarily belongs to the Access Control category.

The experimental data clearly demonstrates the limitations of both approaches: Program analysis-based tools are constrained by their predefined detection rules and patterns, making comprehensive vulnerability coverage difficult; while large language models, despite their potential to detect all types of vulnerabilities, generally show insufficient detection performance. Furthermore, detection performance varies across different types of vulnerabilities\cite{chen2023chatgpt}. This situation highlights the importance and urgency of developing new detection methods. Consequently, we propose the MOE-Tuning method, training specialized expert models for each vulnerability type to precisely identify and handle specific categories of vulnerabilities.

\begin{figure}[htbp]
\centerline{\includegraphics[width=1.0\textwidth]{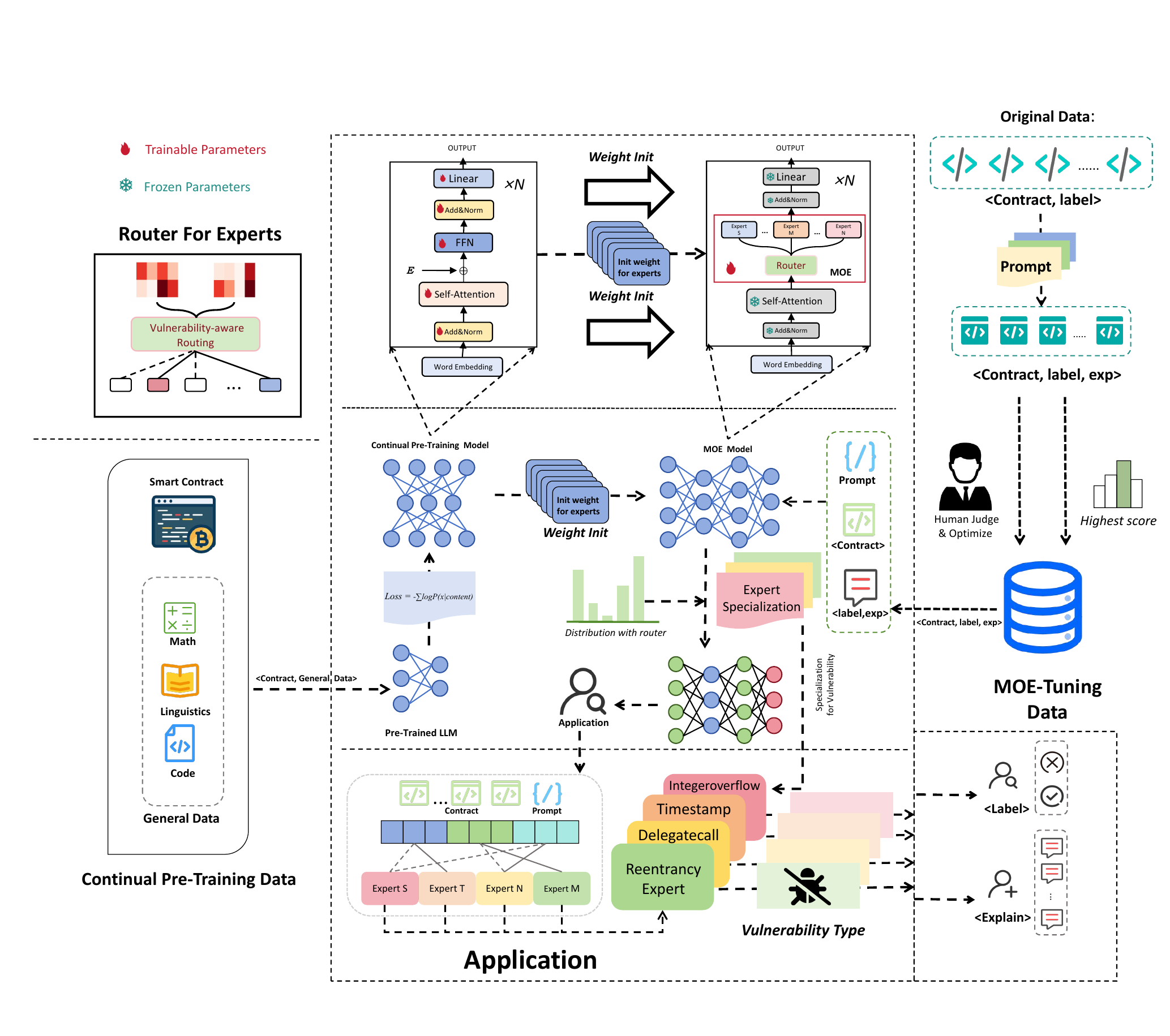}}
\caption{The Overview of  MOS.}
\label{overview}
\end{figure}

\section{Method}
\label{method}
In this section, we introduce our method MOS. 
First, based on the existing large language model LLaMA, we start with continual pre-training of the model, with the aim of providing a better initialization for the subsequent transformation of the model into a MOE model.
Subsequently, we transform the model into a MOE architecture by expanding the pre-trained feed-forward network (FFN) layers and incorporating the dynamic routing mechanism.
In the next stage, we specialize the mixture of experts network into the expert network for detecting different vulnerabilities by freezing parameters outside the MOE layers and employing a dual loss function strategy.

\subsection{Data Construct}

We adopt a dataset construction method combining multiple LLM generation and expert review (MOE-Tuning), as shown in Algorithm \ref{alg:annotation}. We initially employ two open-source large language models, Qwen2.5-72B-Instruct and Mistral-Large-Instruct, to generate preliminary explanations. These models demonstrate capabilities comparable to GPT-4-Turbo in natural language processing and code comprehension while being more cost-effective for deployment. To ensure the quality of generated content, we designed targeted prompt templates that guide models to explain vulnerabilities such as reentrancy, timestamp dependence, delegatecall, and integer overflow, while requiring precise localization of problematic code segments.

To further enhance the quality of explanations, we introduce Llama-3.1-70B-Instruct as an evaluation model. This model scores the generated explanations across three dimensions: correctness, completeness, and conciseness (on a scale of 1-10), providing detailed scoring rationales. As implemented in the EvaluateExplanations function, only explanations scoring 6 or above proceed to the next phase.

In the expert review phase, we enlisted 12 PhD candidates specializing in smart contract vulnerability detection. These experts were organized into four groups, each comprising two review experts and one refinement expert, responsible for quality control of a specific vulnerability type. As illustrated in the ExpertVerification and expertReview functions, the two review experts first independently assess the quality of explanations. When consensus is reached, the refinement expert optimizes the content based on their feedback. When disagreements arise or serious errors are identified (such as incorrect state update sequences in reentrancy attacks, inaccurate timestamp impact assessments, or misjudgments in integer overflow calculation ranges), the group members collectively discuss and revise the content.

After 20 days of intensive development, we completed a dataset containing 6,524 high-quality vulnerability explanations, comprising 3,567 reentrancy vulnerabilities, 1,246 timestamp dependency vulnerabilities, 1,089 integer overflow vulnerabilities, and 722 delegatecall vulnerability explanations, totaling approximately 8.15M tokens. The entire process strictly follows the generation-evaluation-verification workflow defined in Algorithm \ref{alg:annotation}, ensuring the dataset's high quality and reliability.

\begin{algorithm}[t]
\caption{Smart Contract Vulnerability Explanation Annotation}
\label{alg:annotation}
\begin{algorithmic}[1]
\Require Smart contract code $C$, vulnerability type $V$, Experts $F$
\Ensure High-quality explanation $E$
\Function{GenerateExplanations}{$C, V$}
    \State $E_{q} \gets \text{Qwen2.5-72B}(C, V)$ \Comment{Generate with Qwen2.5}
    \State $E_{m} \gets \text{Mistral-Large}(C, V)$ \Comment{Generate with Mistral}
    \State \Return $\{E_{q}, E_{m}\}$
\EndFunction
\Function{EvaluateExplanations}{$\{E_q, E_m\}$}
    \State $s_q,s_m \gets \text{Llama-70B}.\text{evaluate}(\{E_q,E_m\})$ \Comment{Score for $E_q,E_m$ (1-10)}
    
    \If{$s_q \geq 6$ \textbf{or} $s_m \geq 6$}
         \State \Return $E_q$ \textbf{if} $s_q \geq s_m$ \textbf{else} $E_m$
    \Else
        \State \Return \textbf{null} \Comment{Neither explanation is satisfactory}
    \EndIf
\EndFunction
\Function{ExpertVerification}{$E,F$}
    \State $\{f_i\} \gets \text{assignExpertGroup}(F)$ \Comment{Divide Experts  for specific vulnerability }
    \State $E' \gets \text{ExpertReview}(E, \{f_i\})$
    \State \Return $E'$
\EndFunction
\Function{expertReview}{$(E, \{f_i\})$}
   \State Assign $E_i$ to corresponding group $f_i$ 
   \State $R \gets f_i.\text{review}(E_i)$
   \If{$E_i$ \textbf{contains errors, omissions, or ambiguities} base on $R$} 
        \State $E_i' \gets f_i.\text{refine}(E_i)$
    \EndIf
    \If{$\text{requireConsensus}(E',R)$}
        \State $E' \gets \text{resolveDiscrepancies}(E',R)$
    \EndIf
    \State \Return $E'$
\EndFunction
\end{algorithmic}
\end{algorithm}

\subsection{Model Selection}
We employ LLaMA-3.2-3B \cite{dubey2024llama} as the base model for Continual Pre-training and MOE-Tuning, based on the following considerations: 1) Its open-source nature ensures transparency and accessibility; 2) The smaller parameter scale (3B) significantly reduces training and inference costs, making it particularly suitable for validating the effectiveness of MOE-Tuning; 3) It demonstrates strong fine-tuning potential for better task adaptation; 4) Previous research \cite{alrashedy2023language, cifarelli2023safurai, roumeliotis2023llama} has validated its reliable performance across various scenarios.

To verify the model selection, the research team conducted a comparative evaluation of the base versions of LLaMA-3.2-1B, LLaMA-3.2-3B, Qwen2.5-7B, and CodeLLaMA-7B using 20 randomly selected samples from four types of smart contract vulnerabilities (5 each from reentrancy, timestamp dependency, integer overflow, and delegate call). The evaluation revealed that while LLaMA-3.2-1B offers advantages in resource consumption, its performance was significantly lower than other versions and insufficient for the complex requirements of smart contract vulnerability detection. Qwen2.5-7B demonstrated the best detection performance but exhibited inconsistent responses. CodeLLaMA-7B showed relatively weak detection performance, potentially due to older training data and insufficient Solidity smart contract data. In contrast, LLaMA-3.2-3B demonstrated performance levels comparable to larger models while maintaining a relatively small parameter scale. Particularly in balancing computational efficiency and performance, the 3B version provides the optimal choice for validating novel fine-tuning methods. Based on these evaluation results and research objectives, LLaMA-3.2-3B was ultimately selected as the base model.

\subsection{Architecture of Model}
The overall architecture of our approach for Smart Contract Vulnerability Detection is illustrated in Fig. \ref{overview},  where the framework consists of multi-head self-attention (MSA), Vulnerability-aware Routing and Specialized Mixture of Experts Network.
Given a code snippet 
\begin{math}
  \mathbf{x}_0 = \left[ t_1, t_2, \cdots, t_L \right]\in \mathbb{R}^{L \times d}
\end{math},
$L$ is the length of input $x$, $d$ is the dimension of the model, which is encoded as input in the text embedding layer. 
For an N-layer MoE model, the forward process can be formulated as follows:
\begin{align}
   \mathbf{x}'_\ell &= \text{MSA}\left( \text{LN}\left( \mathbf{x}_{\ell-1} \right) \right) + \mathbf{x}_{\ell-1}, \quad \ell = 1, \dots, N \tag{1} \\
   \mathbf{x}_\ell &= \text{MoE}\left( \text{LN}\left( \mathbf{x}'_\ell \right) \right) + \mathbf{x}'_\ell, \quad \ell = 1, \dots, N \tag{2}
\end{align}
where MSA represents the multi-head self-attention module and LN represents layer normalization.
The final input to the model is 
\begin{math}
  \text{LN}({x}_\ell)
\end{math}.
MoE is the mixture-of-experts layer, and the MoE layer consists of multiple experts and a router.
Details regarding routing, MoE layers, and MOE-Tuing in our approach for Smart Contract Vulnerability are elaborated in the subsequent subsections.

\subsubsection{Multi-head Self-Attention}
The multi-head self-attention mechanism further refines and explores the deep semantics of smart contract vulnerability by capturing complex semantic relationships and contextual dependencies within contract code. By processing attention heads in parallel, multi-head self-attention acquires information from different perspectives.
First, a set of linear transformations is used to map the input vector into collections of query, key, and value vectors, with each vector collection corresponding to a specific attention head. These vectors correspond to different attention subspaces, enabling the model to focus on specific aspects like logical dependencies or anomalous patterns in code behavior. Then, by calculating the dot product between the queries and keys and applying scaling and the softmax function, we generate a set of attention weights for each head, which reflect the relationships between different parts of the input vector and their importance within the current context.
Next, based on these weights, we perform a weighted sum of the value vectors to produce the output for each head. The outputs from all heads are concatenated and passed through a linear transformation to restore the original dimensionality.

\subsubsection{Specialized Vulnerability Detection MoE Layers }
The Mixture-of-Experts (MoE) layer consists of the Specialized Mixture of Experts Network
\begin{math}
    [ E_1, E_2, \cdots, E_n ]
\end{math}
, and Vulnerability-aware Routing. 
Similarly to real-world smart contract auditing teams that consist of security experts with diverse specializations, each expert $E_i$ focuses on detecting a specific type of vulnerability. For example, the reentrancy expert specializes in analyzing features such as the order of state updates and the placement of external calls within the contract. Integer Overflow expert focuses on inspecting the safety of numerical operations, including mathematical calculations and type conversions. Timestamp Dependency expert primarily examines how block timestamps are utilized and their potential impact on the contract's logic. Delegatecall expert verifies the access control and context inheritance associated with the use of delegatecall.
Each expert will handle its own specialization, and the selection process is implemented by routing. Inputs are routed and assigned to select the appropriate expert for processing.

The output of the MoE layers can be formulated as
\begin{equation}
    \text{MoE}(x) = \sum_{i=1}^{E} G(x)_i E_i(x), \quad i = 1, \dots, E\tag{3}
\end{equation}
To realize sparse selection for experts, we only compute on the selected experts based on the output of routing. Specifically we select the first k experts, and the selected experts will participate in the computation with different weights, while the experts not selected for are not involved in the computation. For example, when the contract is detected to contain external calls, the Reentrancy experts or Delegatecall experts  are prioritized for activation.

{\bfseries Vulnerability-aware  Routing:}
A certain type of vulnerability is related to some specific tokens and the context, however, the dense model is unbalanced for critical information handling during processing. 
Based on the unique code characteristics and risk patterns associated with different types of vulnerabilities, we designed a specialized {\bfseries Vulnerability-aware  Routing} mechanism to allocate detection tasks. This router analyzes key features of the code, such as external calls, mathematical operations, and timestamp usage, to determine which experts should be activated.
It will be able to handle the relevant vulnerability characteristics well in order to better match and prepare the inputs for later expert processing. For the routing, we take:

\begin{equation}
    \mathbf{G(x)_i} = \frac{exp({f(\mathbf{x,k})_i})}{\sum_{j}^{E} exp({f(\mathbf{x,k})_j})}, \quad i = 1, \dots, E\tag{4}
\end{equation}
\begin{equation}
     \mathbf{f(x,k)_i=}
     \begin{cases}
      (x \cdot W_v)_i & \text{if } i \text{ is in the top } k \text{ elements of } E. \\
      \quad {0} & \text{otherwise.}
      \end{cases}\tag{5}
\end{equation}

where
\begin{math}
    \mathbf{W_v}
\end{math}
is a trainable parameter matrix used to evaluate the alignment between code features and the expertise of each specialist and k is the selected experts and
\begin{math}
    \mathbf{f(x,k)_i}
\end{math}
represents the routing output of the 1st expert for input x. The purpose is to realize the sparse activation of the expert. In addition,
\begin{math}
    {G(x)_i}
\end{math}
is the corresponding output weight of each expert. Through the routing, only the first k experts are activated during each inference, and then the sparsely weighted combination of the k experts is used as output.

{\bfseries Specialized Mixture of Experts Network:}
In our expert system, the total number of experts is denoted as $N$, which serves as a hyperparameter.The system selectively activates only the top $k$ experts, with each expert's computational procedure defined as $E(x)$.
In order to facilitate the computation for experts, we will perform a summarization after assigning the token to different expert. 
For each expert, the input they receive is the subset of the input $x$, 
\begin{math}
  {e}_n = \left[ t_{n1}, t_{n2}, \cdots, t_{nl} \right]\in \mathbb{R}^{l \times d}
\end{math}.
Each expert processes their subset. 
As a consequence of implementing the top k mechanism, which activates the best-performing k experts for processing, different experts may receive the same token.
Then the respective outputs are weighted after their calculation, and the weights are the output $G(x)$ on the gate route, which are weighted and summed with the expert $E(e_n)$to produce the final output.

\subsection{Continual Pre-training  }

To specialize our model for smart contract tasks, we implement continual pre-training on the dense model architecture. This phase enables the LLM to develop a deep understanding of contract-specific patterns and concepts. The weights obtained from this stage serve as the foundation for the subsequent MOE implementation, facilitating more efficient training of the sparse architecture.

The continual pre-training process equips the model with comprehensive knowledge in several critical areas: (1) Solidity language fundamentals, encompassing contract structures and specialized modifiers such as 'payable' and 'view'; (2) architectural design patterns unique to smart contracts, including defensive programming practices like 'Checks-Effects-Interactions'; (3) EVM-specific considerations, particularly regarding gas efficiency and memory management; (4) security-critical operations, focusing on value transfer mechanisms and their associated risks; (5) semantic understanding capabilities, enabling thorough security auditing, contract interaction analysis, and verification of security controls.

The adaptation objective is formulated as:
\begin{equation}
L_{adapt} = -\sum \log P(x_i | context)\tag{6}
\end{equation}

Here, $x_i$ denotes individual tokens in the sequence, $context$ represents the contextual information surrounding $x_i$, and $P(x_i | context)$ reflects the model's token prediction probability.

To maintain the model's versatility while preventing knowledge degradation, we augment the training data with diverse content spanning mathematics, programming, and linguistics. This balanced approach ensures the model maintains its broad knowledge foundation while developing specialized expertise in smart contract analysis, leading to improved generalization and robust performance across varied contract scenarios.

\subsection{MOE-Tuning}
The MOE-Tuning phase aims to specialize the mixture of experts network through targeted parameter optimization and dual loss function implementation. To optimize vulnerability detection capabilities, the MOE layer experts are initialized with FFN parameters from the continual pre-training phase, leveraging prior knowledge. Subsequently, we implement vulnerability-aware routing mechanisms to direct smart contract vulnerability patterns to their corresponding expert networks. The fine-tuning process is constrained to specific parameters, with non-MOE layer parameters frozen to concentrate on routing mechanism development and expert network specialization.

First, to achieve specialization of the mixture of experts network and develop the model's dual capabilities in vulnerability detection and explanation, we define the following formula:
\begin{equation}
L_{MOE-Tuning} = L_{MOE} + \alpha\cdot L_{balance}  \tag{7}
\end{equation}

$L_{MOE-Tuning}$ consists of $L_{MOE}$ and $L_{balance}$, where $L_{MOE}$  is used as a loss function for the model on the downstream tasks: explanations and vulnerability detection; $L_{balance}$ is an auxiliary loss used to balance the expert load, which is scaled by the balancing coefficient $\alpha$.

\begin{equation}
L_{MOE} = - \sum[\log P(y_g|x;\theta_{moe}) + \log P(y_d|x;\theta_{moe})]/2\tag{8}
\end{equation}

In this formula, $x$ represents the input smart contract code, while $y_g$ and $y_d$ denote the target outputs for the generation task (explanations) and the detection task (vulnerability labels), respectively. The parameter $\theta_{moe}$ encompasses MOE layers parameters of the model and all other parameters are frozen.

Due to the presence of multiple experts, it is necessary to impose load balancing constraints on the MoE layer. Therefore, we add an auxiliary loss. 
For each MOE layer, this auxiliary loss is added to the total model loss during training.
Given $E$ experts and a batch of input with $L$ tokens, the auxiliary loss is computed as the scaled dot-product between vectors $F$ and $P$,

\begin{equation}
    L_{balance} =  E \cdot \sum_{i=1}^{E} F_i \cdot P_i \tag{9}
\end{equation}

where $F_i$ is the fraction of tokens dispatched to expert $i$, and $P_i$ represents the average routing probability of  expert $i$, which can be expressed by the following formulas:

\begin{equation}
    F_i = \frac{1}{L} \sum_{i=1}^{E} \mathbf{1}\{\text{argmax} \ G(x) = i\} \tag{10}
\end{equation}

\begin{equation}
    P_i = \frac{1}{L} \sum_{i=1}^{L} G_i(x) \tag{11}
\end{equation}

This loss function ensures that the model is adequately trained on both tasks—vulnerability detection and explanation generation. The domain-specific knowledge gained during the Smart Contract-Specific Continual Pre-training stage serves as the initialization for MOE layers, enabling the model to grasp the context and nuances of smart contracts while effectively learning to detect vulnerabilities.

\subsection{Prompt Design}

\begin{figure}[htbp]
    \centering
    \begin{subfigure}{0.45\textwidth}
        \includegraphics[width=\linewidth]{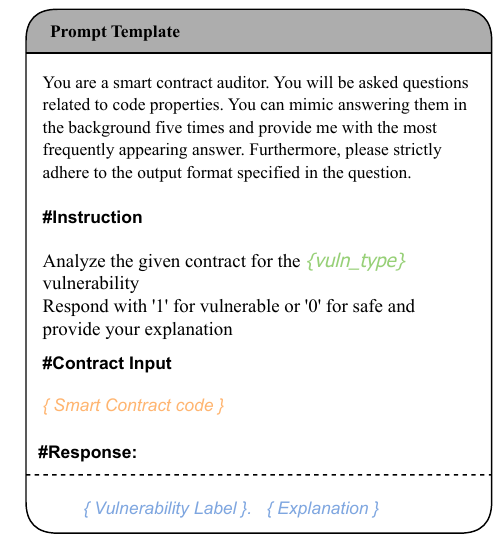}
        \caption{Prompt for inference }
        \label{prompt_a}
    \end{subfigure}
    \begin{subfigure}{0.44\textwidth}
        \includegraphics[width=\linewidth]{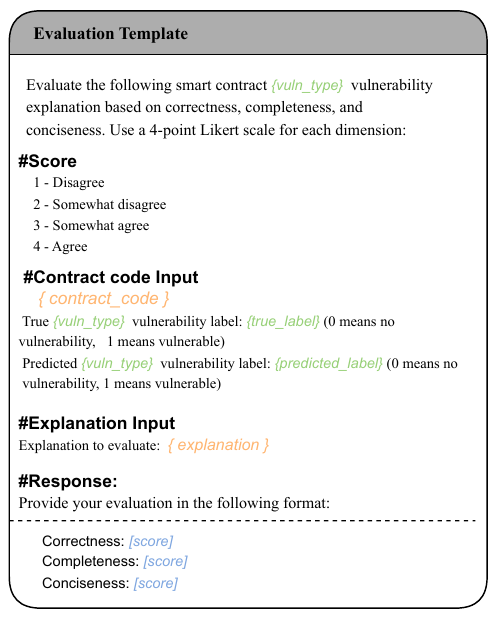}
        \caption{Prompt for evaluation of explanations}
        \label{prompt_b}
    \end{subfigure}

    \caption{Prompt Design for Vulnerability Detection}
    \label{prompt}
    
\end{figure}

\subsubsection{Prompt Design for inference}



During the inference stage, our system extends beyond binary classification (Safe/Vulnerable) to provide detailed analytical explanations. Given the inherent stochastic nature of large language models, we implemented specific measures to ensure output stability.

To minimize the impact of stochastic outputs from large models and ensure output stability, we adopted multiple stabilization techniques drawing from GPTScan's\cite{sun2024gptscan}. The temperature parameter was set to zero to reduce output variability. We implemented their "mimic-in-the-background" strategy, instructing the model to internally simulate five iterations and select the most consistent response, as illustrated in Fig. \ref{prompt_a}. While preserving essential elements from MOE-Tuning prompts (including instructions and contract code), we streamlined the inference prompts to ensure strict adherence to output specifications.

\subsubsection{Prompt Design for explanation evaluation}


To effectively integrate LLMs as evaluators, we developed a structured evaluation framework. Our approach builds upon Wang et al.'s\cite{wang2023generating} research, incorporating three key assessment dimensions: correctness, completeness, and conciseness. Each dimension is evaluated using a 4-point Likert\cite{joshi2015likert} scale with detailed scoring criteria, as shown in Fig. \ref{prompt_b}.
To enhance evaluation accuracy, particularly for correctness and completeness assessments, we incorporated the relevant contract code within the prompts. The evaluation framework maintains a standardized output format to ensure consistent and comparable results across assessments.

\subsection{Evaluation of Explanations}
\label{subsubsec:evaluation_criteria}

We established a systematic evaluation framework to assess MOS-generated smart contract vulnerability explanations, building upon research by \cite{wang2023generating}. Our framework incorporates three fundamental dimensions rated on a 4-point Likert scale:

\textbf{Correctness}. This criterion examines how well the explanation aligns with fundamental security principles and accuracy in vulnerability analysis.

1 - Strongly disagree: Contains critical flaws in security reasoning and vulnerability assessment.
2 - Disagree: Shows inconsistent analysis with several security oversights.
3 - Agree: Generally sound analysis with occasional minor inaccuracies.
4 - Strongly agree: Demonstrates precise security analysis with accurate vulnerability mapping.

\textbf{Completeness}. This criterion evaluates the thoroughness of vulnerability coverage and depth of security analysis.

1 - Strongly disagree: Fails to address fundamental security concerns, analysis lacks substance.
2 - Disagree: Coverage is fragmentary, missing critical security aspects.
3 - Agree: Addresses primary security concerns with some gaps in secondary issues.
4 - Strongly agree: Provides thorough security analysis covering all vulnerability aspects.

\textbf{Conciseness}. This criterion assesses the explanation's effectiveness in communicating security insights.

1 - Strongly disagree: Explanation is convoluted and impractical for implementation.
2 - Disagree: Contains relevant points but lacks clear organization.
3 - Agree: Maintains clarity with occasional redundancy.
4 - Strongly agree: Delivers essential information efficiently without excess detail.







\subsubsection{Consistency Evaluation and Discrepancy Handling}
We adopt a joint evaluation mechanism involving both large models and experts. For both, we implement macro-level consistency validation and micro-level discrepancy resolution.

\textbf{Consistency Validation:} We employ a combined approach of human evaluation and large model assessment. To ensure minimal divergence between evaluators (both human evaluators and large models), we utilize the Kappa coefficient\cite{hubert1977kappa} as a measure of agreement. A sample of cases is selected, and the correlation between the evaluations of correctness, completeness, and conciseness for the generated explanations is analyzed. 
The value of the Kappa coefficient ranges from -1 to 1, where a value of 0.75 or higher indicates strong consistency, and a value of 0.4 or higher indicates medium consistency.This metric helps determine the macro-level consistency of the evaluations. Evaluators exhibiting significant discrepancies are excluded from the evaluation group.
Meanwhile, strong consistency ensures that disagreements occur infrequently, thereby enhancing the confidence and reliability of the evaluation results.

The following is the formula for Kappa coefficient:

\begin{equation}
    \kappa = \frac{P_o - P_e}{1 - P_e}
\end{equation}

where $P_o$ is the observed agreement, which refers to the proportion of actual agreement between two evaluators. It represents the percentage of instances in which both evaluators provide the same evaluation.
$P_e$ is the expected agreement, which refers to the expected level of consistency between two evaluators if their ratings were completely random. It represents the probability that both evaluators would agree by chance.

\begin{equation}
    P_o = \frac{\sum_{i=1}^{k} a_i}{N}
\end{equation}
\begin{equation}
     P_e = \sum_{i=1}^{k} p_i q_i
\end{equation}

In this formula, $k$ represents the total number of categories, and $N$ is the total number of ratings. $p_i$ is the probability that evaluator assigns a rating to the $i$-th category, while $q_i$ is the probability that anothor evaluator assigns a rating to the $i$-th category.

\textbf{Discrepancy Handling:} We categorize evaluation scores for the explanations into positive and negative categories, with scores of 1 and 2 considered negative, and scores of 3 and 4 considered positive. If a discrepancy arises between negative and positive evaluations across all samples, it is considered a divergence. For individual sample discrepancies, a third expert is introduced to resolve the issue. The third expert must meet the Kappa coefficient criterion, meaning their evaluation must show macro-level consistency with the original evaluators.

\subsubsection{LLM Evaluation}
We employed the advanced LLM Llama-3.1-70B-Instruct for automated assessment. We carefully crafted a set of prompts containing detailed scoring guidelines, score examples, and reasoning to guide the LLM in evaluating smart contract vulnerability explanations. The LLM rated each explanation generated by MOS on correctness, completeness, and conciseness using a 4-point Likert scale \cite{joshi2015likert}. Beyond providing specific scores, the LLM also offered detailed justifications for each aspect's rating. We then developed automated scripts to compute the score distribution for each vulnerability category across all aspects as the final evaluation outcome.

\subsubsection{Human Evaluation}
The evaluation of generated vulnerability explanations is a labor-intensive task that requires significant time and effort from evaluators. Inspired by\cite{sun2023real} , we enlisted four experienced smart contract security experts for human evaluation, each with at least three years of experience in smart contract development and an average of 4.2 years of development experience. Each expert was responsible for detecting one category of vulnerabilities. To ensure the validity of this approach, we employed the Kappa coefficient to assess the agreement  among these four experts.
The experts utilized the same 4-point Likert scale \cite{joshi2015likert} as the LLM, scoring each explanation on correctness, completeness, and conciseness. They also provided detailed scoring rationales, suggestions for improvement, and overall quality assessments. Finally, similar to the LLM evaluation, we calculated the score distribution for each category across all aspects as the final evaluation result.


\section{Evaluation}
\label{evaluation}
\subsection{Research Questions}
To evaluate our proposed MOS approach, we conduct experiments to answer the following research questions:

\textbf{RQ1}: How does MOS perform compared to baselines in Smart Contract Vulnerability Detection ?

\textbf{RQ2}: What is the specialization of different experts in Smart Contract Vulnerability Detection?

\textbf{RQ3}: How do the hyperparameters of total experts and activations affect the effectiveness of MOS?

\textbf{RQ4}: How do different training stages contribute to MOS?

\textbf{RQ5}: How effective is MOS for generating smart contract explanations?

\textbf{RQ6}: What insights can be gained from analyzing failed cases using MOS?

\subsection{Dataset}
\textbf{Continual Pre-training:} we employ a dataset derived from the work of Storhaug et al. \cite{storhaug2023efficient}. This dataset comprises 186,397 unique smart contract instances from the Ethereum blockchain, totalling 501.62M tokens. We also augment this dataset with an additional 100,000 instances from various domains, including general code, mathematics, English, and Chinese text, totalling 118.94M tokens. This results in a comprehensive dataset of 286,397 instances, totaling 620.56M tokens.

\textbf{MOE-Tuning:} Our MOE-Tuning dataset is curated from multiple sources, primarily drawing from Liu et al. \cite{liu2023rethinking} and Yu et al. \cite{yu2023pscvfinder}. We incorporate manually verified vulnerabilities from \cite{liu2023rethinking} and the SmartBugs dataset \cite{ferreira2020smartbugs} with reentrancy vulnerabilities annotated by Wu et al. \cite{wu2021peculiar}. To enrich the dataset, we extracted and manually annotated additional contracts from \cite{yu2023pscvfinder} and Etherscan. The final dataset comprises 3,567 reentrancy, 1,246 timestamp dependency, 1,089 integer overflow/underflow, and 722 delegatecall vulnerability instances, totalling 8.15M tokens.

\textbf{Evaluation: }In this study, our evaluation dataset integrates samples from two significant sources. The smartbugs-curated dataset, originally established by Durieux et al.\cite{durieux2020empirical}, was utilized by Chen et al.\cite{chen2023chatgpt} in their research evaluation. Building upon Chen et al.'s work, we expanded the evaluation scope by incorporating the dataset from Qian et al. Our comprehensive dataset contains 58 vulnerable and 166 non-vulnerable samples for Reentrancy vulnerability, 172 vulnerable and 60 non-vulnerable samples for Timestamp Dependency, 64 vulnerable and 184 non-vulnerable samples for Integer Overflow/Underflow, and 50 vulnerable and 132 non-vulnerable samples for Delegatecall vulnerability. The selection of these two datasets as evaluation benchmarks was primarily motivated by the fact that both contain all four key vulnerability types and all samples have undergone detailed manual verification. Notably, in a single dataset (such as Qian et al.'s dataset), there are often numerous duplicate samples and similar code patterns, which could lead to biased evaluation results. Therefore, relying on a single dataset for evaluation might lack representativeness.


\subsection{Baselines}

Our evaluation includes a range of baselines for Smart Contract Vulnerability Detection, representing state-of-the-art approaches in four categories: traditional tools, neural network-based, pre-trained model-based, and LLM-based techniques. These baselines were selected based on several criteria: (1) their recognition in top conferences such as ICSE, ASE, S\&P, USENIX Security and WWW, with numerous citations demonstrating their impact in the field (excluding preprints and non-peer-reviewed manuscripts); (2) the public availability of their source code or detailed implementation descriptions on GitHub, allowing for reproducible research; and (3) their representation of diverse technical approaches to ensure comprehensive comparison across different approaches.


The first category includes traditional vulnerability detection tools such as Securify \cite{tsankov2018securify}, Smartian \cite{choi2021smartian}, Osiris \cite{torres2018osiris}, Oyente \cite{luu2016making}, Slither \cite{feist2019slither}, which employ various techniques including rule-based analysis, symbolic execution, and fuzzing.

Neural network-based techniques harness the power of deep learning algorithms to detect vulnerabilities in smart contracts. We consider several cutting-edge methods in this category:  GCN \cite{kipf2016semi}, TMP \cite{zhuang2020smart}, AME \cite{liu2021smart}, SMS \cite{qian2023cross} and DMT \cite{qian2023cross}. It should be noted that we faced challenges in reproducing some baselines, and in other cases, our reproduced results differed significantly from the originally reported values. For fairness, we used the higher of our reproduced results and the originally reported figures for each baseline in our comparisons.

Pre-trained model-based techniques rely on pre-trained models like CodeT5 \cite{wang2021codet5}, CodeBERT \cite{feng2020codebert}, GraphCodeBERT \cite{guo2020graphcodebert} and fine-tuning approaches to identify smart contract vulnerabilities, including Peculiar \cite{wu2021peculiar} and PSCVFinder \cite{yu2023pscvfinder}.

LLM-based techniques leverage large language models for vulnerability detection, including instruction-tuned versions of LLaMA3.1-8B \cite{dubey2024llama} and LLaMA3.1-70B \cite{dubey2024llama}, GPT-3.5-Turbo \cite{openai2023gpt35}, GPT-4o \cite{openai2024gpt4o}, GPTScan \cite{sun2024gptscan}, GPTLens \cite{hu2023large}, and iAudit \cite{ma2024combining}.

We evaluated our model performance from two dimensions: vulnerability detection capability and explanation quality. For vulnerability detection, we adopted four standard metrics: Precision measures the ratio of actual vulnerabilities among predicted positives, Recall indicates the proportion of detected vulnerabilities among all actual vulnerabilities, F1-score provides a balanced measure through the harmonic mean of Precision and Recall, and Accuracy reflects the overall correctness of predictions. For vulnerability explanation quality, we employed three key metrics: correctness, completeness, and conciseness, with detailed evaluation criteria specified in Section~\ref{subsubsec:evaluation_criteria}.

\subsection{Implementation Details}

We perform Continual Pre-training, Supervised Fine-Tuning and Direct Preference Optimization using LlamaFactory~\cite{zheng2024llamafactory} and DeepSpeed~\cite{rasley2020deepspeed} with \texttt{fp16} enabled.
We calculate loss with cross-entropy and optimize parameters using AdamW~\cite{adamw} with $\beta$=$(0.9, 0.99)$ and $\epsilon$=$1$e-$8$. For all our models, we employ full parameter tuning. 

During Continual Pre-training, we set the batch size to $64$ per device, gradient accumulation steps to $16$, epochs to $2$, learning rate to $1$e-$5$ with cosine decay, warmup steps to $0$, cutoff length to $2048$, and save steps to $500$. 

During MOE-Tuning, we set the batch size to $8$ per device, gradient accumulation steps to $8$, epochs to $3$, learning rate to $1$e-$5$ with cosine decay, warmup steps to $0$, cutoff length to $2048$, total experts to $8$ and activated experts to $2$(8-Top2).
All models were trained on a server equipped with two NVIDIA GeForce RTX H800 GPUs,  each with 80GB memory. 

For evaluating our MOS, we use greedy decoding with do\_sample set to \texttt{false} and set temperature to  \texttt{0}. If not specified, the hyperparameters for the number of experts and the number of activations in our experimental section were set to: total experts to 8 and activated experts to 2(8-Top2).
Peculiar \cite{wu2021peculiar} originally only detects reentrancy vulnerabilities, while PSCVFinder \cite{yu2023pscvfinder} detects reentrancy and timestamp dependency vulnerabilities. We extended both tools to detect two additional vulnerability types using the same code.

GPTLENS \cite{hu2023large} was reproduced following the original paper, with GPT-4 as the base model. The implementation maintained two Auditors with each generating up to 3 vulnerabilities, and a single Critic for consistent evaluation. The original prompt templates were followed strictly, with JSON format for output standardization. Temperature parameters were set according to the paper: 0.7 for Auditors to ensure diversity and 0 for the Critic to maintain consistency.

For iAudit \cite{ma2024combining} implementation, we focused on the Detector and Reasoner components responsible for vulnerability detection. Following the paper, we used LLaMAFactory to fine-tune the model with LoRA on our dataset. We adjusted the prompt templates for our specific vulnerability types while maintaining the original LabelName output format, with results manually compiled afterward.

For GPTScan \cite{sun2024gptscan}, we strictly followed the original paper's configuration, using GPT-3.5-turbo as the base model. As specified in the paper, the temperature parameter was adjusted from default 1 to 0 to reduce randomness, while maintaining empty sessions for each query. The static analysis component, following the original implementation, utilized ANTLR for code parsing and AST generation, along with crytic-compiler for control flow and data dependency analysis.

To mitigate the impact of randomness in large language model outputs on our experimental results, we adopted the "mimic-in-the-background" prompting strategy proposed by Sun et al. in GPTScan \cite{sun2024gptscan}. Specifically, for each input, we set the temperature parameter to 0 to enhance determinism and instructed the model to simulate five independent answers in the background, selecting the most frequently appearing response as the final output to reduce the model output randomness. This strategy was uniformly applied to our proposed MoS and all baseline models, including the LLaMA series, Qwen series, GPT series, and Claude series.

\subsection{Experimental Results}

\begin{table}[htbp]
\setlength{\tabcolsep}{5pt} 
    \caption{The performance of our method compared with baselines in terms of Accuracy, Precision, Recall and F1-score (Part 1). Note: LLM-based techniques use Instruct versions of Llama3.1 models.}
    \label{TAB1a}
    \centering
    \begin{tabular}{@{}c@{\hspace{0pt}}|cccc@{\hspace{0pt}}|cccc@{}}
        \toprule
        \raisebox{-1\height}{\centering Methods} & \multicolumn{4}{c|}{Reentrancy} & \multicolumn{4}{c}{Timestamp Dependency}\\ 
        & A(\%) & P(\%) & R(\%) & F1(\%) & A(\%) & P(\%) & R(\%) & F1(\%)\\ 
        \midrule
        Securify & 54.46 & 29.25 & 53.45 & 37.80 & -- & -- & -- & --\\ 
        Smartian & 57.14 & 26.83 & 37.93 & 31.43 & 21.55 & 43.75 & 20.35 & 27.78\\ 
        Osiris & 33.48 & 26.90 & 91.38 & 41.56 & 47.41 & 70.83 & 49.42 & 58.21\\ 
        Conkas & 65.62 & 36.62 & 44.83 & 40.31 & 28.02 & 52.94 & 26.16 & 35.02\\ 
        Oyente & 69.64 & 42.65 & 50.00 & 46.03 & 52.59 & 80.39 & 47.67 & 59.85\\ 
        Slither & 56.25 & 30.00 & 51.72 & 37.97 &53.88 & 76.00 & 55.23 & 63.96\\ 
        \midrule
        GCN & 73.21 & 74.47 & 73.18 & 73.82 & 75.91 & 74.93 & 77.55 & 76.22\\ 
        TMP & 76.45 & 76.04 & 75.30 & 75.67 & 78.84 & 78.68 & 76.09 & 77.36\\ 
        AME & 81.06 & 79.62 & 78.45 & 79.03 & 82.25 & 81.42 & 80.26 & 80.84\\
        SMS & 83.85 & 79.46 & 77.48 & 78.46 & 89.77 & 89.15 & 91.09  & 90.11\\ 
        DMT & 89.42 & 83.62 & 81.06 & 82.32 & 94.58 & 93.60 & 96.39 & 94.97\\ 
        \midrule
        Peculiar & 63.84 & 35.44 & 48.28 & 40.88 & 69.08 & 77.31 & 71.12 & 74.18\\ 
        PSCVFinder & 63.39 & 35.37 & 50.00 & 41.43 & 40.10 & 52.00 & 50.39 & 51.18\\ 
        \midrule
        LLaMA3.1-8B & 34.38 & 26.20 & 84.48 & 40.00 & 53.88 & 76.86 & 54.07 & 63.48\\ 
        LLaMA3.1-70B & 25.89 & 25.89 & 100.00 &41.13 & 74.14 & 74.14 & 100.00 & 85.15\\ 
        GPT-3.5-Turbo & 43.30 & 27.45 & 72.41 & 39.81 & 71.55 & 87.32 & 72.09 & 78.98\\ 
        GPT-4o & 45.54 & 28.95 & 75.86 & 41.90 & 76.29 & 89.80 & 76.74 & 82.76\\ 
        GPTScan & 50.00 & 29.85 & 68.97 & 41.67  & 78.45 & 90.67 & 79.07 & 84.47 \\ 
        GPTLens & 51.79 & 31.06 & 70.69 & 43.16 & 79.74 & 91.39 & 80.23 & 85.45\\
        iAudit & 79.91 & 59.42 & 70.69 & 64.57 & 81.30 & 87.65 & 86.63 & 87.13\\ 
        \textbf{MOS} & \textbf{95.41} & \textbf{97.96} & \textbf{84.21} & \textbf{90.57} & \textbf{95.26}$^*$ & \textbf{97.08} & \textbf{96.51} & \textbf{96.79}\\
        \bottomrule
    \end{tabular}
\end{table}

\begin{table}[!ht] 
\setlength{\tabcolsep}{5pt} 
    \caption{The performance of our method compared with baselines in terms of Accuracy, Precision, Recall and F1-score (Part 2). Note: LLM-based techniques use Instruct versions of Llama3.1 models.}
    \label{TAB1b}
    \centering
    \begin{tabular}{@{}c@{\hspace{0pt}}|cccc@{\hspace{0pt}}|cccc@{}}
        \toprule
        \raisebox{-1\height}{\centering Methods} & \multicolumn{4}{c|}{Overflow/Underflow} & \multicolumn{4}{c}{Delegatecall}\\ 
        & A(\%) & P(\%) & R(\%) & F1(\%) & A(\%) & P(\%) & R(\%) & F1(\%)\\ 
        \midrule
        Smartian & 77.42 & 54.55 & 75.00 & 63.16 & 76.92 & 55.41 & 82.00 & 66.13\\ 
        Osiris & 70.16 & 45.28 & 75.00 & 56.47 & -- & -- & -- & --\\ 
        Conkas & 19.76 & 13.51 & 39.06 & 20.08 & 80.22 & 58.75 & 94.00 & 72.31\\ 
        Oyente & 75.40 & 52.24 & 54.69 & 53.44 & 63.74 & 37.88 & 50.00 & 43.10\\ 
        Slither & 61.29 & 32.22 & 45.31 & 37.66 & 54.95 & 37.10 & 92.00 & 52.88\\ 
        \midrule
        GCN & 67.53 & 69.52 & 70.93 & 70.22 & 65.76 & 69.01 & 69.74 & 69.37\\ 
        TMP & 70.85 & 70.26 & 69.47 & 69.86 & 69.11 & 68.18 & 70.37 & 69.26\\ 
        AME & 73.24 & 71.36 & 71.59 & 71.47 & 72.85 & 70.25 & 69.40 & 69.82\\
        SMS & 79.36 & 78.14 & 72.98 & 75.47 & 78.82 & 76.97 & 73.69 & 75.29\\ 
        DMT & 85.64 & 85.44 &  74.32 & 79.49 & 82.76 & 84.61 & 77.93 & 81.13\\ 
        \midrule
        Peculiar & 75.91 & 64.29 & 60.00 & 62.07 & 88.46 & 77.36 & 82.00 & 79.61\\ 
        PSCVFinder & 51.09 & 33.58 & 50.00 & 40.18 & 89.56 & 79.25 & 84.00 & 81.57\\ 
        \midrule
        LLaMA3.1-8B & 54.84 & 31.82 & 65.62 & 42.86 & 63.74 & 38.89 & 56.00 & 45.90\\ 
        LLaMA3.1-70B & 78.63 & 56.32 & 76.56 & 64.90 & 65.93 & 44.55 & 98.00 & 61.25\\ 
        GPT-3.5-Turbo & 81.45 & 62.16 & 71.88 & 66.67 & 75.82 & 54.55 & 72.00 & 62.07 \\ 
        GPT-4o & 84.68 & 66.67 & 81.25 & 73.24 & 81.32 & 62.50 & 80.00 & 70.18\\ 
        GPTScan & 85.48 & 69.44 & 78.12 & 73.53 & 79.12 & 59.38 & 76.00 & 66.67 \\ 
        GPTLens & 85.48 & 67.95 & 82.81 & 74.65 & 82.97 & 65.08 & 82.00 & 72.57\\
        iAudit & 81.85 & 62.03 & 76.56 & 68.53 & 89.01 & 75.93 & 82.00 & 78.84\\ 
        \textbf{MOS} & \textbf{91.53} & \textbf{89.09} & \textbf{76.56} & \textbf{82.35} & \textbf{93.96} & \textbf{85.45} & \textbf{94.00} & \textbf{89.52}\\
        \bottomrule
    \end{tabular}
\end{table}

\begin{figure}[htbp]
    \centering
        \includegraphics[width=0.6\textwidth]{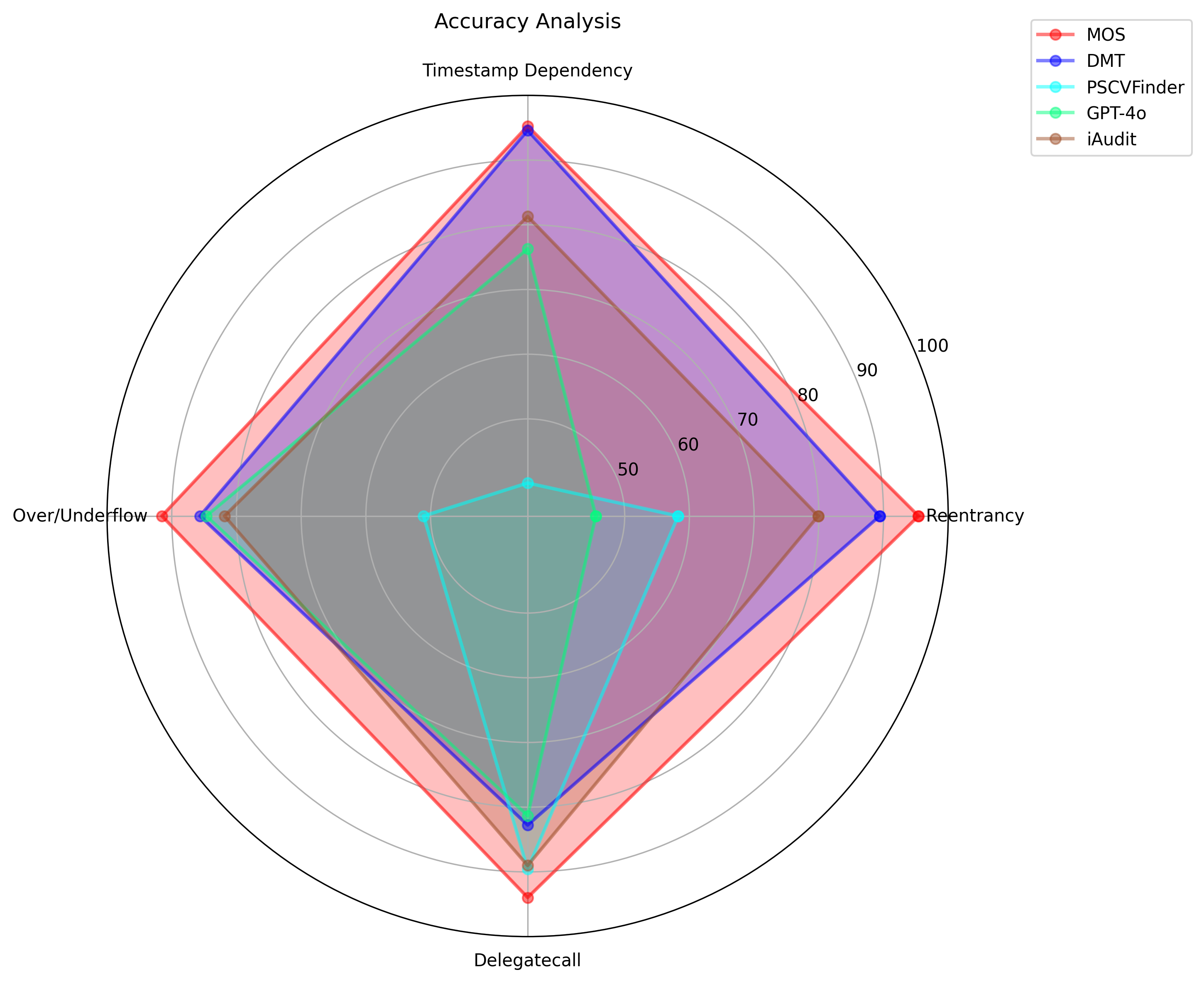}
        \caption{Performance of four optimal baselines in the different categories and MOS}
        \label{balance}
\end{figure}

\subsubsection{RQ1:Effectiveness of MOS}

In this RQ, we aim to investigate the effectiveness of MOS compared to baseline methods in smart contract vulnerability detection. In this experiment, our approach is trained using corresponding datasets at different stages. Based on existing methods for smart contract vulnerability detection, we compared our approach with different types of baseline methods across four different types of smart contract vulnerabilities. The results, as shown in the Table \ref{TAB1a} and Table \ref{TAB1b}, indicate that our method outperforms the state-of-the-art (SOTA) methods in detecting all four types of vulnerabilities.

\textbf{Effectiveness performance:}
In comparison to the best baseline method for reentrancy vulnerability detection, our approach achieved accuracy and F1 score of 95.41\% and 90.57\%, respectively, improving by 6.70\% and 10.02\%. For Timestamp Dependency vulnerability detection, our method achieved high scores of 95.26\% and 96.79\%, respectively.
However, in our tests, the output is not simply labels but a label-and-explanation format. Due to the inherent limitations of large models and some shortcomings of greedy search, there are instances of repeated outputs during testing. We have taken measures to alleviate this issue, such as using beam search during inference and introducing random seeds. Nevertheless, in the case of Timestamp Dependency vulnerability detection, some instances remain unresolved. For these duplicate results, we apply a majority voting mechanism to determine the final answer.

For the detection of Integer Overflow/Underflow vulnerability, we achieved an accuracy of 91.53\%, which is 6.88\% higher than the best baseline, DMT (85.64\%). We also achieved a F1 score of 82.35\%, which is 3.60 \% higher than the best baseline DMT. Lastly, for Delegatecall vulnerability detection, our approach attained accuracy and F1 score of 93.96\% and 89.52\%, respectively, surpassing the best baseline, PSCVFinder, by 4.91\% and 9.75\%.

Notably, MOS also consistently outperformed other popular methods such as Smartian, Oyente, Securify, and Slither across all vulnerability types, and showed significant improvements over more recent LLM-based approaches like GPTScan, GPTLens and iAudit in most metrics.

\textbf{Balanced performance:}
We selected the best-performing methods among the various baselines, and the results are shown in Fig. \ref{balance}. We observed that large models and methods based on pre-trained models exhibit a significant preference in vulnerability detection. There is a substantial gap in accuracy between the vulnerabilities that are detected well and those that are not. For instance, LLaMA3.1-70B achieves an accuracy of 78.63\% in Integer Overflow/Underflow vulnerability detection but only 25.89\% in Reentrancy vulnerability detection, a difference of 52.74\%. PSCVFinder performs best in Delegatecall vulnerability detection, achieving an accuracy of 89.56\%, but only 40.10\% in Timestamp Depesndency detection.

For methods based on LLMs, which incorporate mechanisms such as fine-tuning or Agents, for instance, GPTScan, GPTLens, and iAudit, have shown improvements both in performance and balance compared to the zero-shot capabilities of LLMs. However, issues of preference bias persist. For example, iAudit achieves an accuracy of 89.01\% in detecting delegatecall vulnerabilities but performs poorly in detecting Reentrancy vulnerabilities, with an accuracy of only 79.91\%.

In contrast, MOS not only outperforms other baselines in terms of average accuracy across the four vulnerabilities but also exhibits a significantly lower standard deviation of only 2.07, which is half that of the best baseline DMT (4.42).

\vspace{5pt} 
\noindent\begin{tikzpicture}
  \node[draw=black, thick, fill=gray!20, rounded corners, inner sep=10pt, text width=0.92\linewidth] {
    \textbf{Answer to RQ1:} Our method consistently outperformed 16 state-of-the-art baseline methods across all four types of vulnerabilities (reentrancy, timestamp dependency, integer overflow/underflow, and delegatecall).
    Moreover, our method shows a better balance on the detection of the four vulnerabilities, with better average performance and standard deviation than the other baselines.
  };
\end{tikzpicture}
\vspace{5pt} 

\begin{figure}[htbp]
    \centering
    \begin{subfigure}{0.23\textwidth}
        \includegraphics[width=\linewidth]{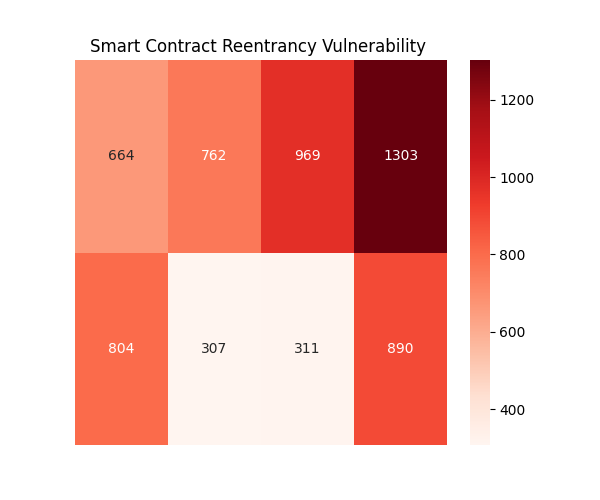}
        \caption{Reentrancy (1)}
    \end{subfigure}
    \begin{subfigure}{0.23\textwidth}
        \includegraphics[width=\linewidth]{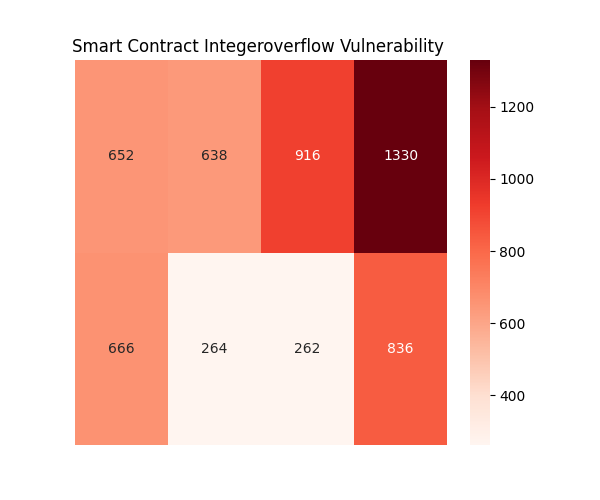}
        \caption{Integeroverflow (1)}
    \end{subfigure}
    \begin{subfigure}{0.23\textwidth}
        \includegraphics[width=\linewidth]{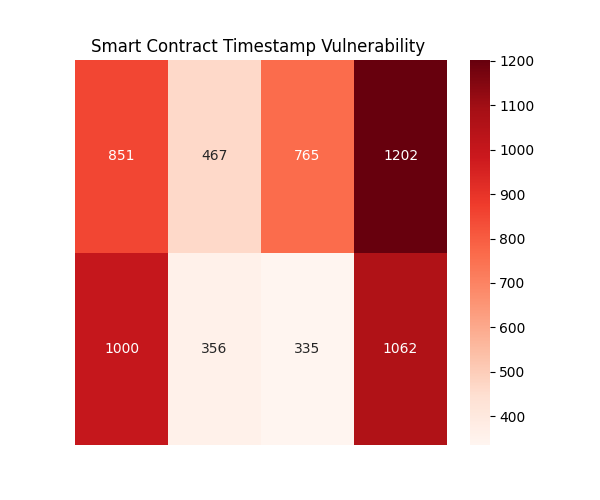}
        \caption{Timestamp (1)}
    \end{subfigure}
    \begin{subfigure}{0.23\textwidth}
        \includegraphics[width=\linewidth]{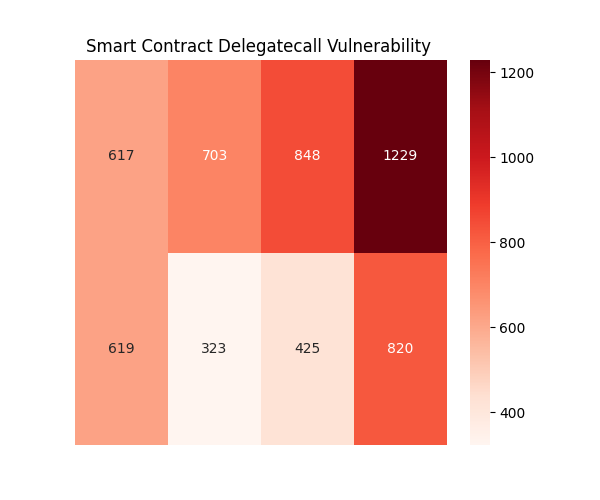}
        \caption{Delegatecall(1)}
    \end{subfigure}

    \begin{subfigure}{0.23\textwidth}
        \includegraphics[width=\linewidth]{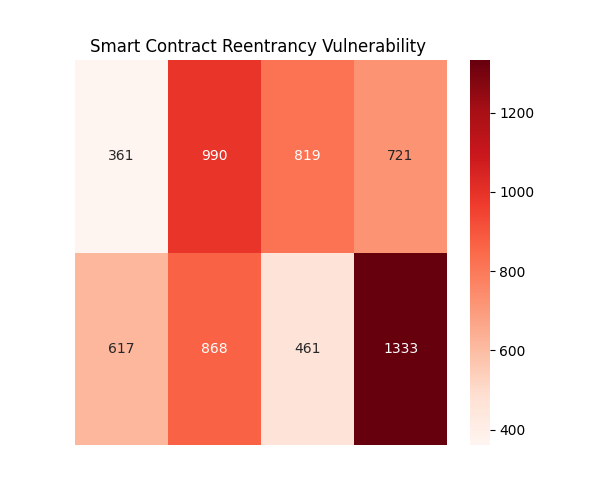}
        \caption{Reentrancy (14)}
    \end{subfigure}
    \begin{subfigure}{0.23\textwidth}
        \includegraphics[width=\linewidth]{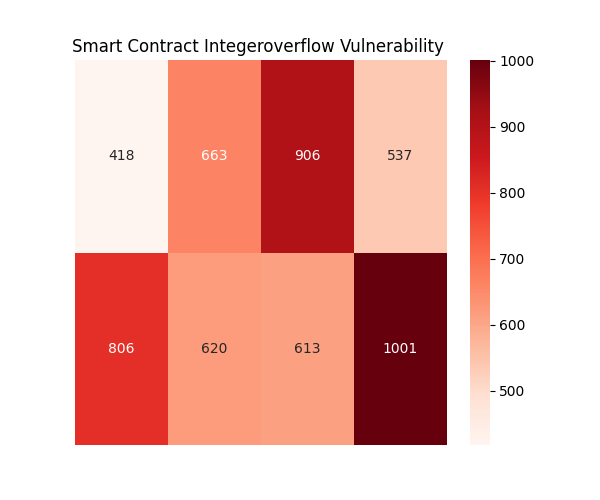}
        \caption{Integeroverflow (14)}
    \end{subfigure}
    \begin{subfigure}{0.23\textwidth}
        \includegraphics[width=\linewidth]{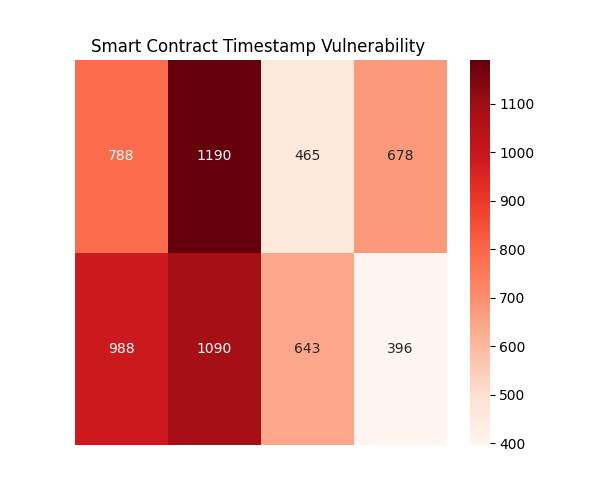}
        \caption{Timestamp (14)}
    \end{subfigure}
    \begin{subfigure}{0.23\textwidth}
        \includegraphics[width=\linewidth]{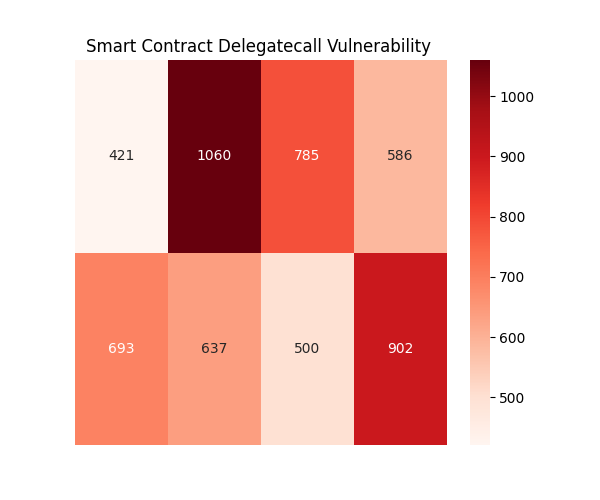}
        \caption{Delegatecall (14)}
    \end{subfigure}

    \begin{subfigure}{0.23\textwidth}
        \includegraphics[width=\linewidth]{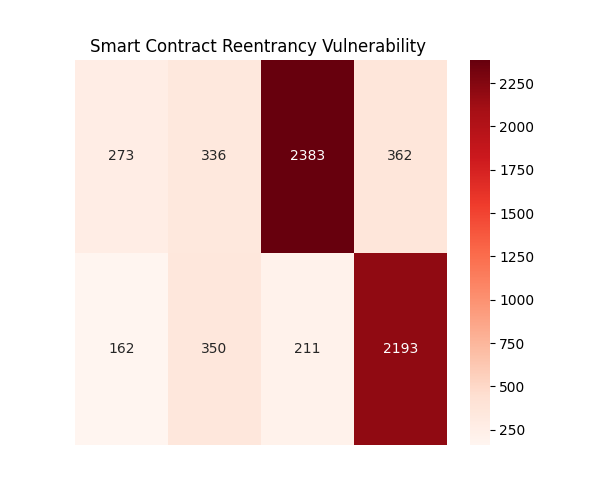}
        \caption{Reentrancy (28)}
        \label{i}
    \end{subfigure}
    \begin{subfigure}{0.23\textwidth}
        \includegraphics[width=\linewidth]{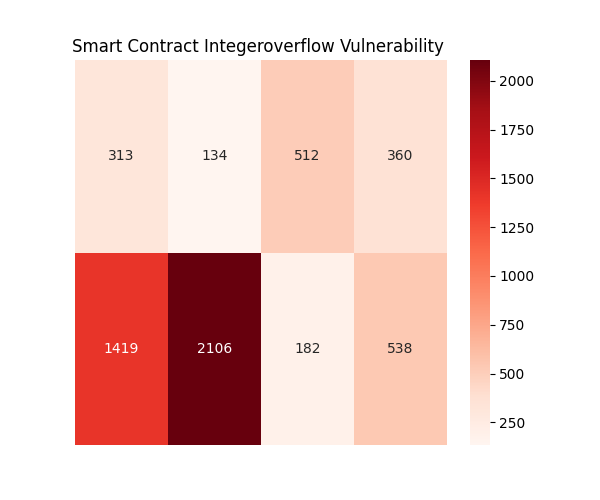}
        \caption{Integeroverflow (28)}
    \end{subfigure}
    \begin{subfigure}{0.23\textwidth}
        \includegraphics[width=\linewidth]{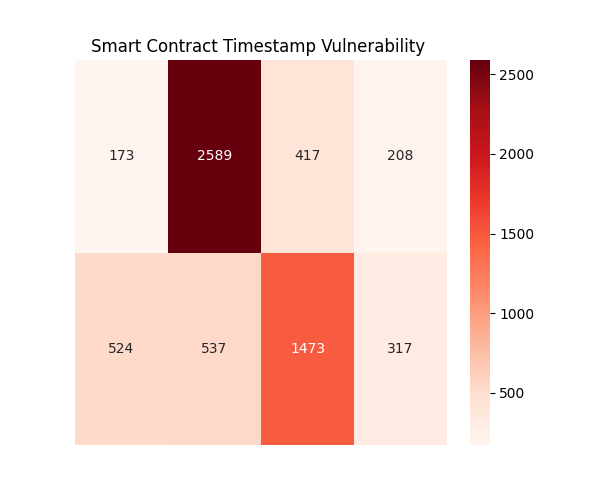}
        \caption{Timestamp (28)}
    \end{subfigure}
    \begin{subfigure}{0.23\textwidth}
        \includegraphics[width=\linewidth]{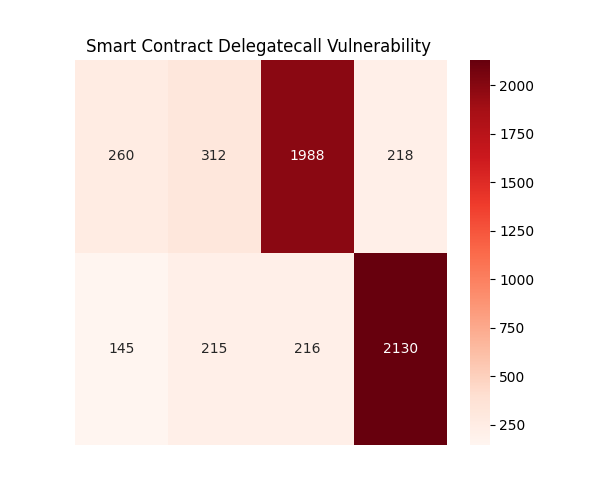}
        \caption{Delegatecall (28)}
        \label{l}
    \end{subfigure}

    \caption{Expert Specialization for different  vulnerabilities across different layers.For example, \textbf{Reentrancy(28)} represents the distribution at layer 28 for Reentrancy. }
    \label{expert special}
\end{figure}

\subsubsection{RQ2: Expert Specialization}

In this section, we investigate the specialized expert networks differentiated by the vulnerability-aware routing mechanism.

\textbf{Setting:}
We evaluate the specialized expert network using datasets comprising four distinct vulnerability types. Inspired by Zhu et al.\cite{zhu2024llama}, for each vulnerability type, we sample between 5.5k-6k tokens to facilitate detailed observation of expert differentiation patterns. Similarly to Zhu et al.\cite{zhu2024llama}, to examine the  specialization across different network depths, we sample the first layer, middle layer (layer 14), and final layer (layer 28) of the model. This experimental setup is designed to directly observe and assess the specialization tendencies of the Mixture of Experts network by analyzing expert allocation patterns in single-vulnerability scenarios.

\textbf{Result:}
As shown in Fig. \ref{expert special}, our analysis reveals a hierarchical pattern of expert specialization in vulnerability detection across network depths. The experimental results demonstrate significant patterns in expert specialization:
At the architectural level, we observe a progressive specialization phenomenon in which expert specialization exhibits a correlation with network depth. 
In the initial layer, we observe minimal expert differentiation across the four vulnerability types. Specifically, experts 1, 4, 5, and 8 consistently demonstrate higher selection frequencies across all vulnerability types. This homogeneous distribution pattern suggests that these experts at shallow layers may primarily focus on capturing general semantics and contract features in smart contracts. As we progress through deeper layers, the expert specialization becomes increasingly pronounced, culminating in highly differentiated patterns at layer 28. 
This deepest layer exhibits clear expert specialization for specific vulnerability types:
For reentrancy vulnerabilities, experts 3 and 8 demonstrate significantly higher activation frequencies. In the case of integer overflow vulnerabilities, experts 5 and 6 show predominant activation patterns.
Shallow layers predominantly focus on capturing universal features and semantics of smart contracts. Deeper layers develop highly specialized detection capabilities for specific vulnerability characteristics.
This observed specialization pattern provides evidence for the effectiveness of our vulnerability-aware routing mechanism in fostering expert differentiation and specialization throughout the network architecture.

However,the differentiation in expert selection exhibits non-mutually exclusive characteristics.
First, regarding the operational mechanism of selection, the system activates the two highest-scoring experts: primary and secondary specialists. This design enables individual experts to demonstrate primary expertise in reentrancy vulnerability features while maintaining secondary competence in delegatecall vulnerabilities.
As shown in Figs. \ref{i} and \ref{l}, for reentrancy vulnerability, expert 3 serves as the primary expert while expert 8 acts as the secondary expert. In contrast, for delegatecall vulnerability detection, expert 3 transitions to a secondary role in the vulnerability detection and handling process.
Consequently, the routing allocation process may assign different vulnerability types to the same expert, albeit with varying priority levels.
Second, concerning vulnerability characteristic similarities, reentrancy and delegatecall vulnerabilities demonstrate high correlation in expert selection, particularly pronounced in the final layer. This phenomenon can be attributed to two core factors.
On the one hand, as analyzed in the selection mechanism above, the secondary expertise effect enabled expert 3 to develop auxiliary competency for delegatecall vulnerability detection.
On the other hand,the fundamental similarity in risk patterns between these vulnerabilities: Both involve external contract calls; and their core risks stem from function call sequence errors, particularly in state change timing.
Lastly, we also observed that certain experts were rarely selected throughout the process. This observation may inform future MoE models by suggesting potential directions for expert compression or pruning.





\vspace{5pt} 
\noindent\begin{tikzpicture}
  \node[draw=black, thick, fill=gray!20, rounded corners, inner sep=10pt, text width=0.92\linewidth] {
  \textbf{Answer to RQ2:}  The experimental results reveal that MOS, through its vulnerability-aware routing mechanism and after training, differentiated the Mixture of Experts network into specialized expert networks for different vulnerabilities. Different experts exhibited corresponding inclinations towards different vulnerability types. During the specialization process, the Mixture of Experts network demonstrated a progressive specialization pattern. Additionally, due to the inherent characteristics of vulnerabilities, similar expert preferences may emerge for vulnerabilities sharing common features.};
\end{tikzpicture}
\vspace{5pt} 

\begin{figure}[htbp]
    
    \centering
    \begin{subfigure}{0.4\textwidth}
        \includegraphics[width=\linewidth]{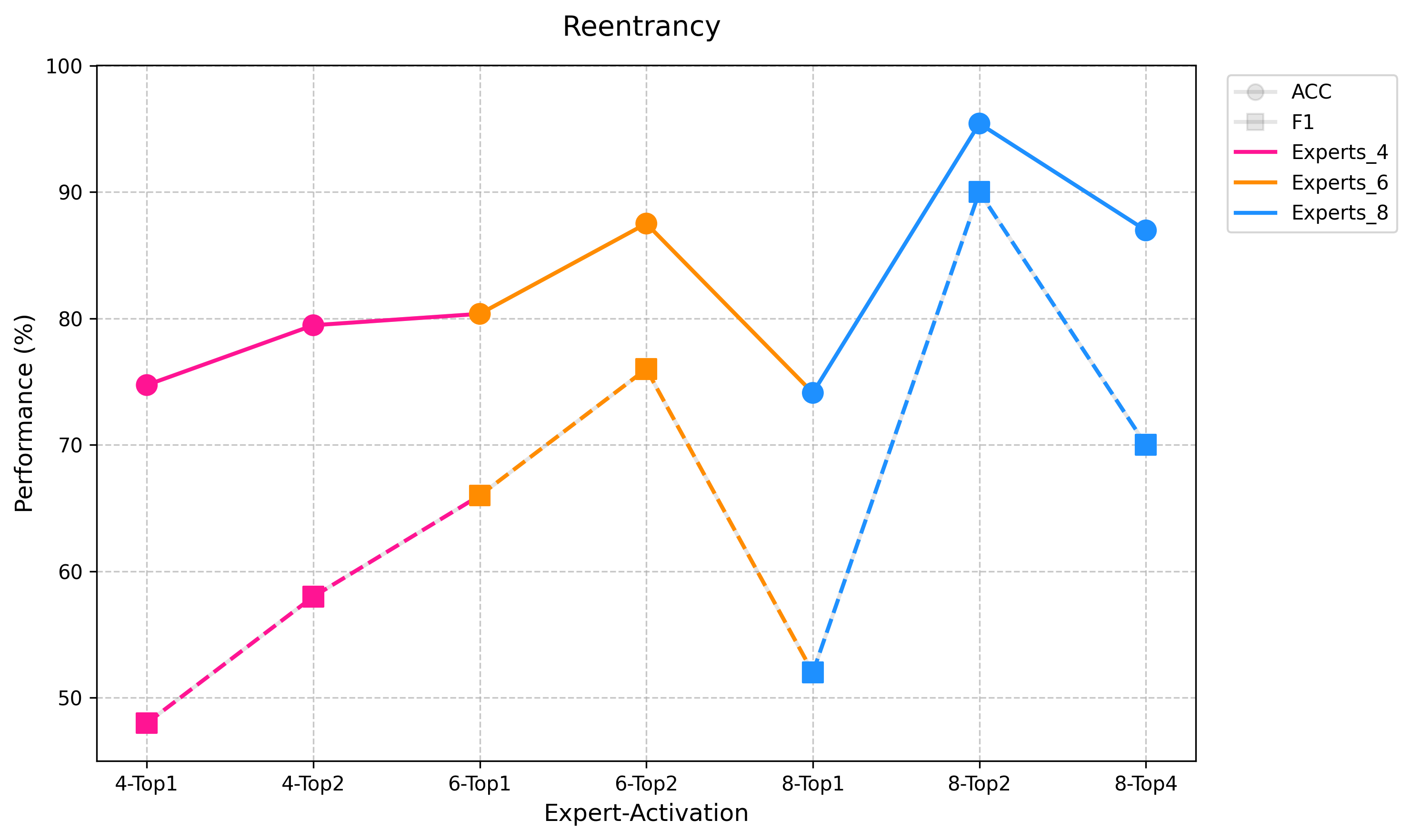}
        \caption{Reentrancy }
    \end{subfigure}
    \begin{subfigure}{0.4\textwidth}
        \includegraphics[width=\linewidth]{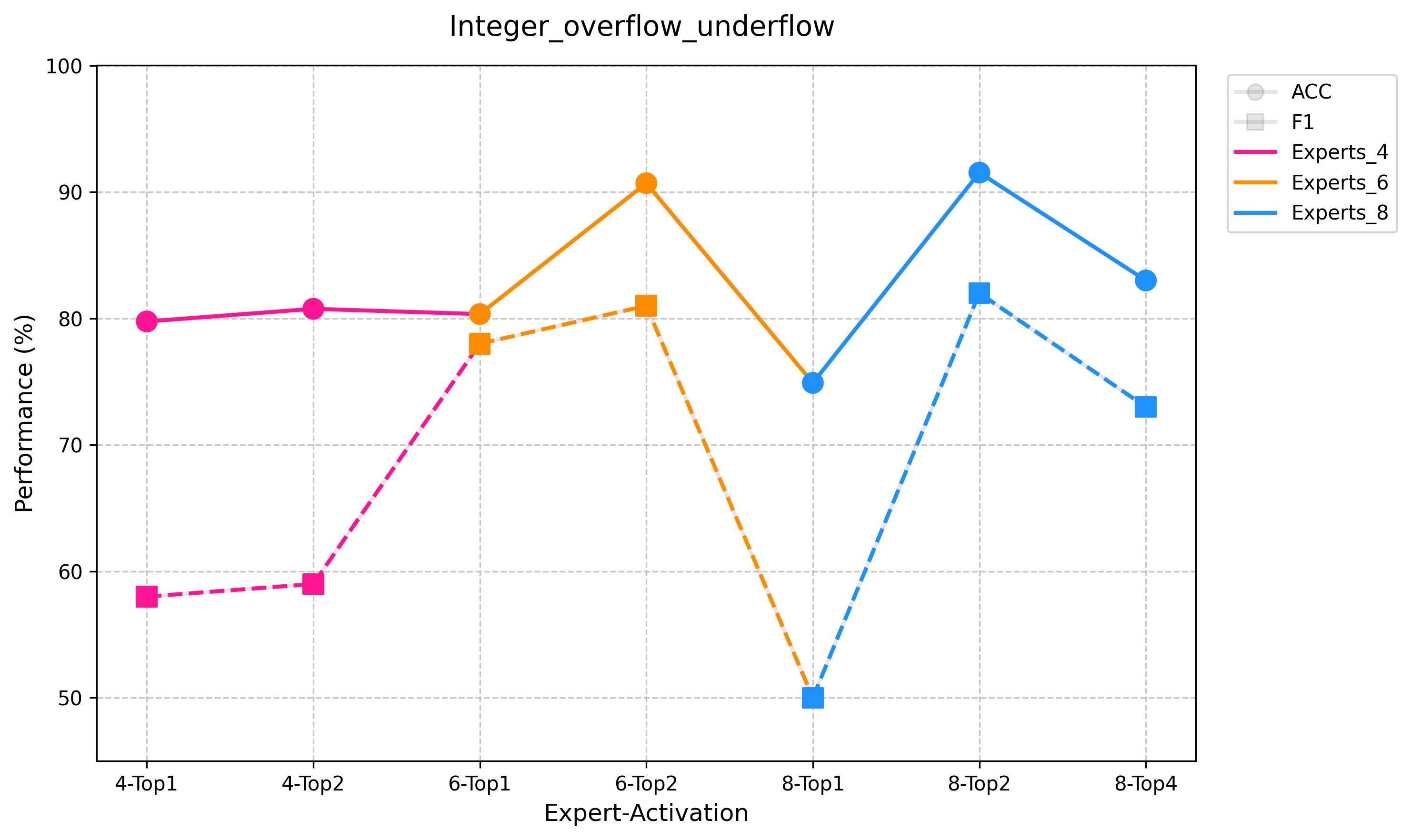}
        \caption{Integeroverflow }
    \end{subfigure}

    \begin{subfigure}{0.4\textwidth}
        \includegraphics[width=\linewidth]{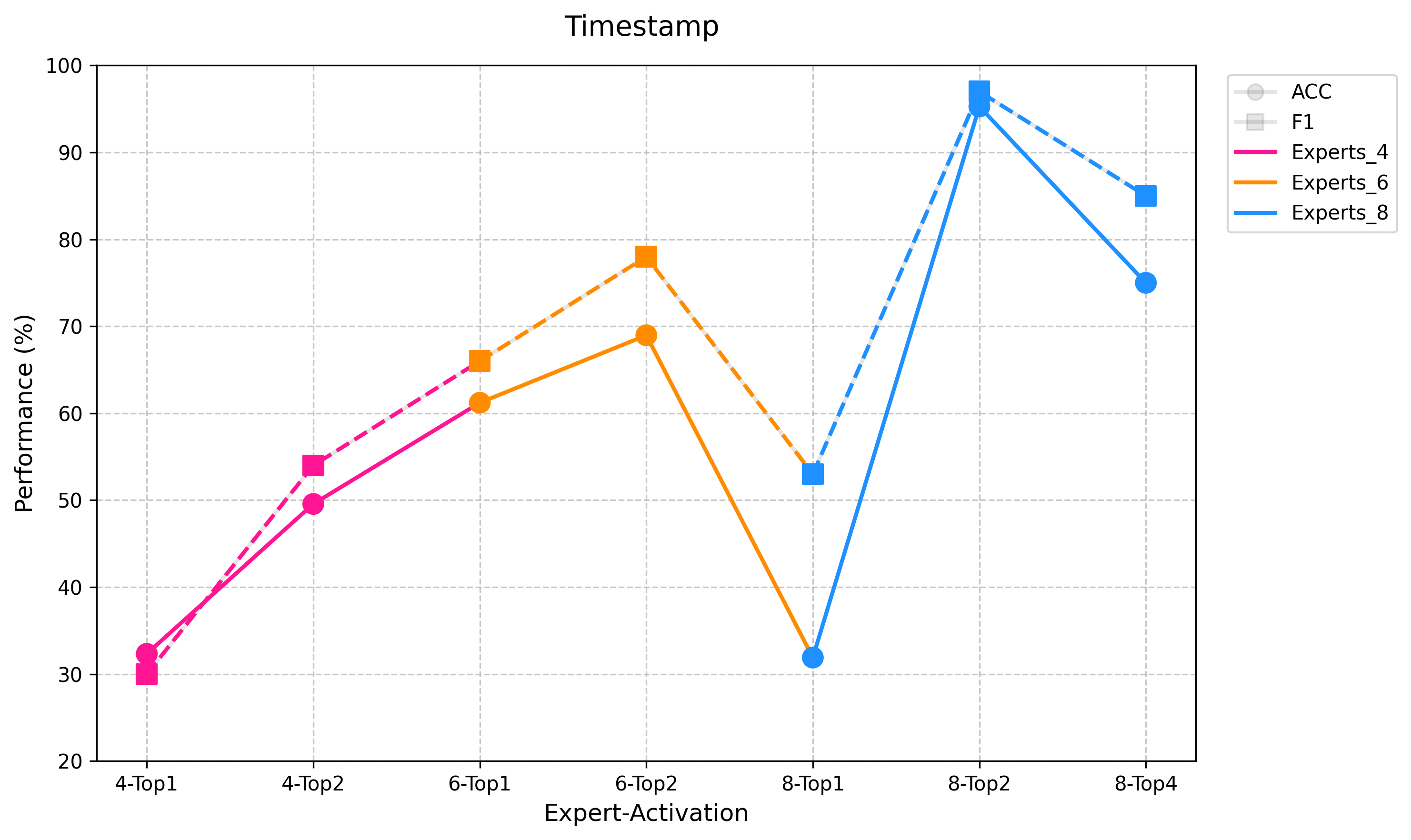}
        \caption{Timestamp }
    \end{subfigure}
    \begin{subfigure}{0.4\textwidth}
        \includegraphics[width=\linewidth]{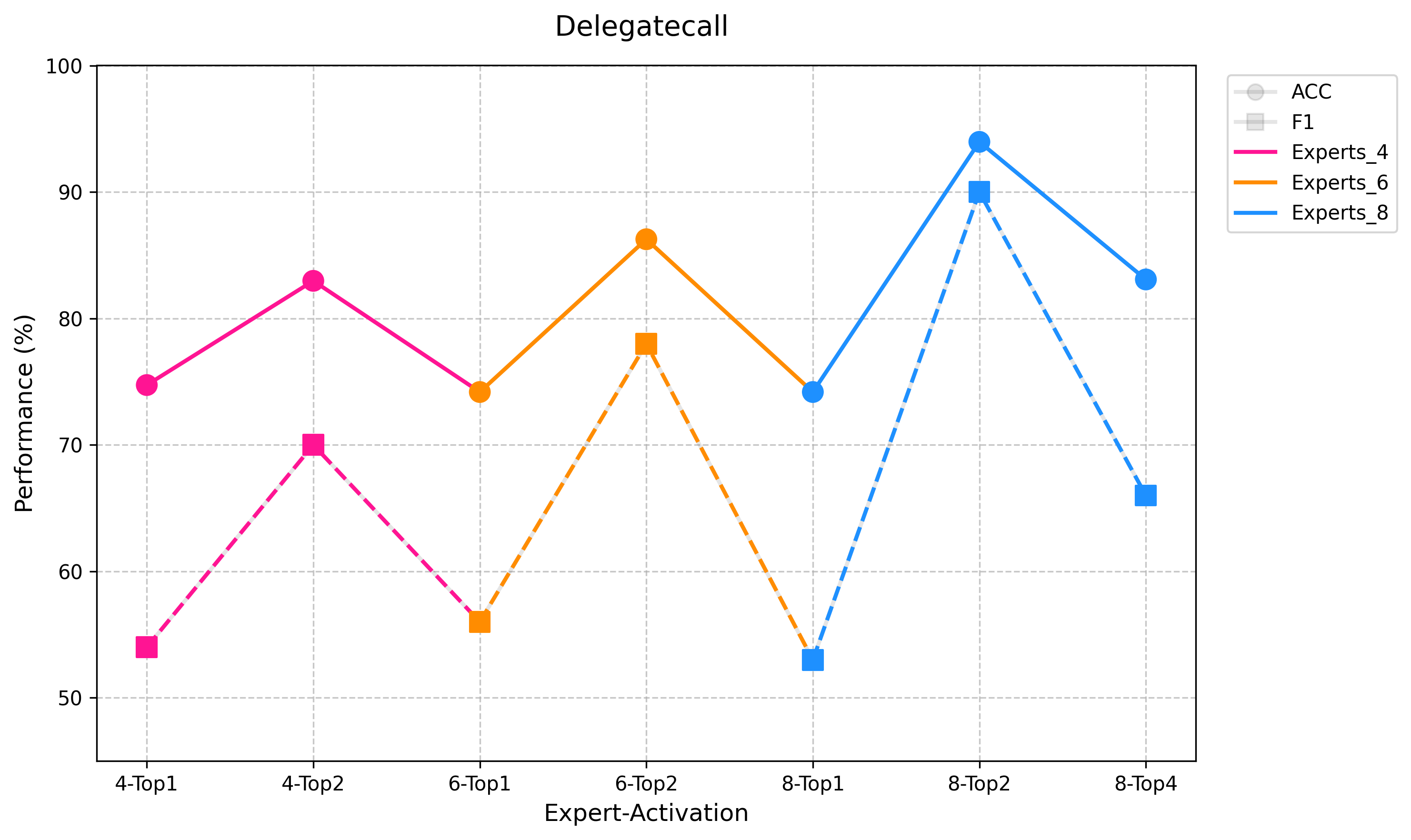}
        \caption{Delegatecall}
    \end{subfigure}
    \caption{Performance  of different hyperparameters of total experts and activations.The x-axis represents various combinations of experts and \textbf{6-Top2} indicates a total of 6 experts with the 2 highest-scoring experts activated.}
    \label{combaination}
\end{figure}

\begin{table}[htb]
\centering
\caption{Parameters and Time for Different Numbers of Experts and Activation. We trained three types of variant models, which 8E represent total 8 experts. \textbf{AEs} (Activated Experts): Number of concurrently active experts;\textbf{TPs} (Total Parameters): Complete parameter count in model architecture;\textbf{APs} (Activated Parameters): Parameters activated during inference;\textbf{Acc} and \textbf{F1}:average detection accuracy and F1 score for four vulnerability types}
\begin{tabular}{l c c c c c c}
\toprule
\textbf{Model} & \textbf{AEs} & \textbf{TPs} & \textbf{APs} & \textbf{Time} & \textbf{Acc(\%)} & \textbf{F1(\%)} \\
\midrule
MOS-8E-3B & 1 & $\sim$18B & $\sim$3B & $\sim$6.5h & 63.77 & 52.20 \\
\rowcolor{gray!20}MOS-8E-5B & 2 & $\sim$18B & $\sim$5B & $\sim$8h & \textbf{94.04} & \textbf{89.81} \\
MOS-8E-9.5B & 4 & $\sim$18B & $\sim$9.5B & $\sim$11h & 82.01 & 73.46 \\
\hline
MOS-4E-3B & 1 & $\sim$9.5B & $\sim$3B & $\sim$4h & 65.38 & 47.50 \\
MOS-4E-5B & 2 & $\sim$9.5B & $\sim$5B & $\sim$4.5h & 73.19 & 60.27 \\
\hline
MOS-6E-3B & 1 & $\sim$14B & $\sim$3B & $\sim$5h & 74.02 & 63.53 \\
MOS-6E-5B & 2 & $\sim$14B & $\sim$5B & $\sim$6h & 83.35 & 78.25 \\
\bottomrule
\end{tabular}
\label{tab:time}
\end{table}

\subsubsection{RQ3:Expert Combination}


To systematically evaluate the impact of the hyperparameters on MOS performance, we conducted experimental analysis exploring various combinations of total experts and activation patterns, where the x-axis represents various combinations of experts and \textbf{6-Top2} indicates a total of 6 experts with the 2 highest-scoring experts activated during inference. As shown in Fig. \ref{combaination}, the results demonstrate the performance variations across four distinct vulnerability detection tasks under different expert-activation configurations.


Our analysis reveals a positive correlation between the number of total experts and the performance of the model when multiple experts are activated. Among all investigated configurations, the architecture with 8 experts and Top-2 routing mechanism consistently achieved superior performance across all of vulnerability categories. Specifically, for reentrancy vulnerability detection, this optimal configuration demonstrated significant improvements with accuracy and F1 score increases of 9.1\% and 18.4\% respectively, compared to the second-best configuration. In the context of Timestamp Dependency vulnerability detection, the performance gains were even more pronounced, with the 8-experts Top-2 configuration showing a substantial 26.6\% improvement in accuracy and a 14.1\% enhancement in F1 score. Similarly, for Delegatecall vulnerability detection, the optimal configuration outperformed the next best alternative by 8.1\% in accuracy and 15.3\% in F1 score. For integer overflow vulnerabilities, the optimal configuration achieved improvements of 0.84\% and 2.01\% in accuracy and F1 score respectively.

\textbf{Activated Experts:}
However, a notable observation is that in the 8-expert configuration, models with 4 activated experts demonstrate inferior performance compared to those with 2 activated experts.
To elucidate this phenomenon, we conducted a detailed analysis of the routing mechanism, specifically examining the token distribution patterns and softmax probability distributions. Our experimental results reveal that in the 4-activated-expert configuration, the cumulative probability mass of the top two selected experts consistently approaches 90\%, indicating that the additional activated experts beyond the top two contribute marginally to the final prediction. The probability distribution analysis shows that the remaining experts receive negligible attention weights, resulting in minimal impact on the model's decision-making process.
Furthermore, we identified that increasing the number of activated experts adversely affects the model's ability to maintain semantic coherence. We conducted sampling analysis of vulnerability detection results from 8-Top2 and 8-Top4, randomly selecting 10 samples for each vulnerability type for further examination. Our analysis revealed that under 8-Top2, the vulnerability detection outputs were normal and coherent statements. However, 8-Top4 exhibited various anomalies, including garbled characters (5 instances), null characters (9 instances), and discontinuous statements (13 instances). The situation was particularly severe in detecting reentrancy and Delegatecall vulnerabilities, where anomalies accounted for 70\% and 80\% of the cases, respectively.





\textbf{Total Experts:}
Moreover, we found that the 8-expert configuration with single-expert activation exhibits the poorest performance metrics across all experimental combinations, underperforming even the 4-expert single-activation configuration, despite possessing double the parameter capacity.
Although the 8-expert configuration maintains twice the total parameter count compared to its 4-expert counterpart, both configurations activate an identical number of parameters during inference due to the single-expert constraint. This observation suggests that the capacity of the parameters alone does not determine the effectiveness of the model in sparse activation scenarios.
Through the analysis of the model behavior, we identified that single-expert activation effectively transforms the MoE architecture into a functionally dense model. The routing mechanism becomes functionally redundant in single-expert scenarios and continues to introduce token-level decision boundaries. This not only adversely affects model performance but also leads to anomalies in the semantic continuity of the model's output.



\textbf{Time for Experts:}
Additionally, we conducted a statistical analysis of the activation parameters and training time for models with different expert combinations, as shown in the Table \ref{tab:time}. 
It is evident that, for the same total number of experts, the number of activated experts significantly affects both training and inference times. This is because during both training and inference, varying numbers of activated experts correspond to different activation parameters.
At the same time, for models with the same number of activation parameters, there are considerable differences in training time depending on the total number of experts. This can be attributed to the fact that during training, different experts undergo a process of specialization or differentiation, which is a global process and thus is influenced by the total number of experts.

\vspace{5pt} 
\noindent\begin{tikzpicture}
  \node[draw=black, thick, fill=gray!20, rounded corners, inner sep=10pt, text width=0.92\linewidth] {
\textbf{Answer to RQ3:} We investigated the impact of hyperparameters on MOS, and the results demonstrated that well-chosen expert combinations can positively influence vulnerability detection performance. Inappropriate combination patterns may even result in model degradation, diminishing expert advantages and negatively affecting model performance.
  };
\end{tikzpicture}
\vspace{5pt} 


\begin{table}[htbp]
\centering
\caption{Ablation Study of MOS.}
\begin{tabular}{l c c c c c c}
\toprule
\textbf{Types} & \textbf{Metric} & \textbf{Base} & \textbf{w/o moe-tuning} & \textbf{w/o cpt} & \textbf{w/o both} \\
\midrule
RE & Acc(\%) & 95.41 & 75.00  & 74.11  & 70.13 \\
 & F1(\%) & 90.57 & 52.29 & 50.20  & 46.30 \\
\hline
TD & Acc(\%) & 95.26 & 65.52 & 42.64 & 68.53 \\
 & F1(\%) & 96.79 & 72.97 & 58.08 & 75.59 \\
\hline
IO & Acc(\%) & 91.53 & 80.16  & 75.30 & 82.19 \\
 & F1(\%) & 82.35 & 61.02 & 50.81 & 64.52 \\
\hline
DE & Acc(\%) & 93.96 & 85.71 & 74.73 & 74.73\\
 & F1(\%) & 89.52  & 76.48 & 55.24 & 55.24 \\
\bottomrule
\end{tabular}
\label{tab:Ablation}
\end{table}

\subsubsection{RQ4:Ablation Study}
Our ablation study elucidates the crucial roles of MOE-Tuning and continual pre-training (cpt) in the MOS architecture. The base model represents  full performance of MOS, "w/o moe-tuning" indicates the model without MOE-Tuning, "w/o cpt" represents the LLM without continual pre-training, and "w/o both" shows the performance without both components.

As shown in the Table \ref{tab:Ablation}
For reentrancy vulnerabilities, the model demonstrates a balanced dual-dependency characteristic. The full model achieves an accuracy of 95.41\% and an F1 score of 90.57\%, with the removal of any single stage leading to a significant performance drop. When both stages are removed, the accuracy decreases to 70.13\%, and the F1 score drops to 46.30\%. We analyze that for reentrancy vulnerability detection, the model not only relies on strong prior learning but also requires the routing preference capabilities provided by MoE-tuning. This aligns with the fundamental characteristics of reentrancy vulnerability detection, which necessitates the model’s ability to accurately distinguish and understand the subtle differences in the "check-effect-interaction" pattern.

Similar to reentrancy vulnerability detection, the model also exhibits a dual-stage dependency for timestamp vulnerability detection. However, the MoE model's reliance on prior knowledge is more pronounced in this case. When the CPT stage is removed, the model’s accuracy drops by over 50\% and F1 score drops to 58.08\%  for this particular vulnerability. We analyze that this is because the expert layer needs to understand the dual complexity of timestamp vulnerabilities in order to accurately distinguish the security risk levels under different time-dependency scenarios. This presents a challenge for the MoE model without prior knowledge. During this process, if fine-tuning is applied to the expert layer without prior knowledge, it results in insensitivity in the routing process for expert allocation, thereby leading to a decline in model performance.

Integer overflow and delegatecall vulnerability detection exhibit similar dependency characteristics. Compared to the significant performance drop caused by removing the CPT stage, the impact of removing MOE-Tuning is relatively smaller. Integer overflow detection requires a deep understanding of Solidity's type system and arithmetic rules, making prior knowledge essential for the model. Research has shown that proper initialization is crucial for MoE models. Delegate call vulnerabilities involve complex storage layouts and context inheritance mechanisms, which rely on the expert layer being well-trained to achieve expert specialization and accurate routing. Therefore, removing the CPT stage causes a significant decline in the performance of the MoE expert model in detecting these two types of vulnerabilities.

\vspace{5pt} 
\noindent\begin{tikzpicture}
  \node[draw=black, thick, fill=gray!20, rounded corners, inner sep=10pt, text width=0.92\linewidth] {
    \textbf{Answer to RQ4:} The experimental results reveal that the two-stage design (continual pre-training and MOE-Tuning) makes a positive contribution to the detection performance of all four types of vulnerabilities. 
    The prior knowledge provided by continual pre-training, combined with the expert specialization and routing training from MOE-Tuning, enables the model to accurately detect and understand different vulnerabilities, achieving state-of-the-art (SOTA) performance across all vulnerability detections
  };
\end{tikzpicture}
\vspace{5pt} 

\subsubsection{RQ5:Explanation Evaluation}

\begin{figure}[htbp]
    \centering
    \begin{subfigure}{0.32\textwidth}
        \includegraphics[width=\linewidth]{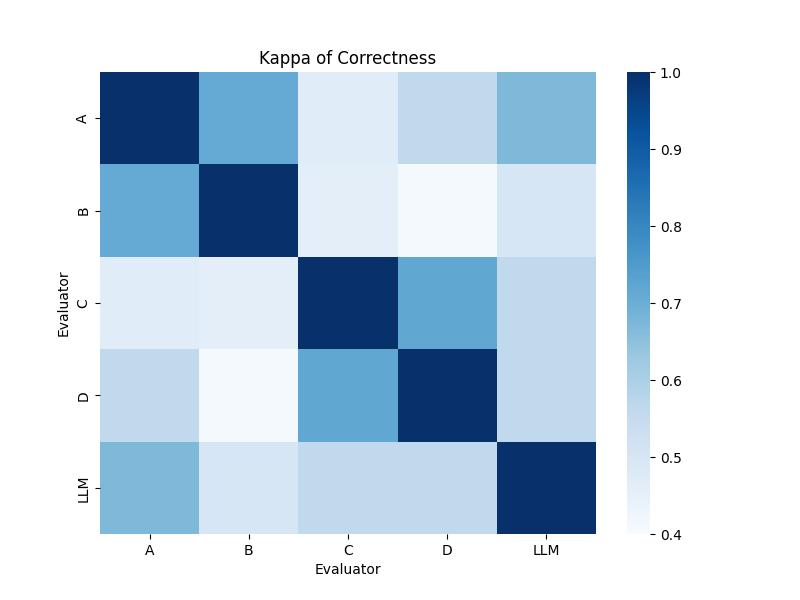}
        \caption{Correctness }
    \end{subfigure}
    \begin{subfigure}{0.32\textwidth}
        \includegraphics[width=\linewidth]{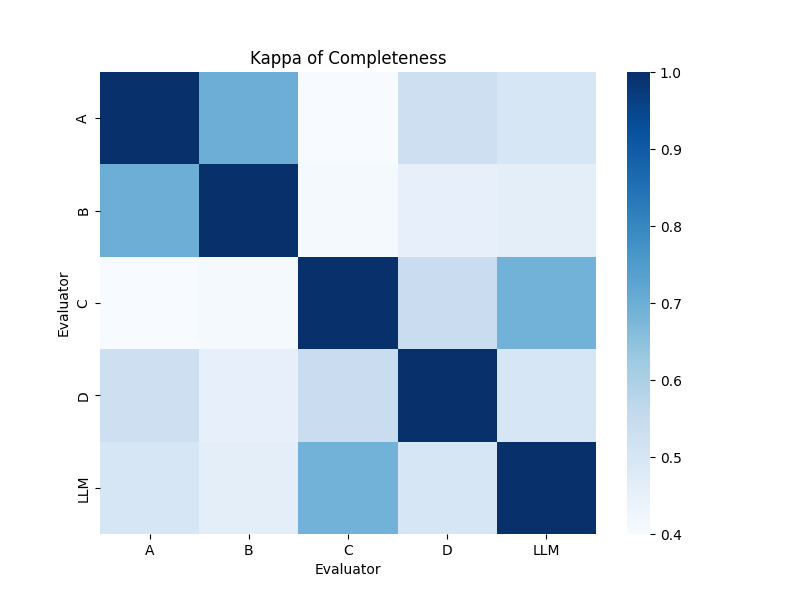}
        \caption{Completeness }
    \end{subfigure}
    \begin{subfigure}{0.32\textwidth}
        \includegraphics[width=\linewidth]{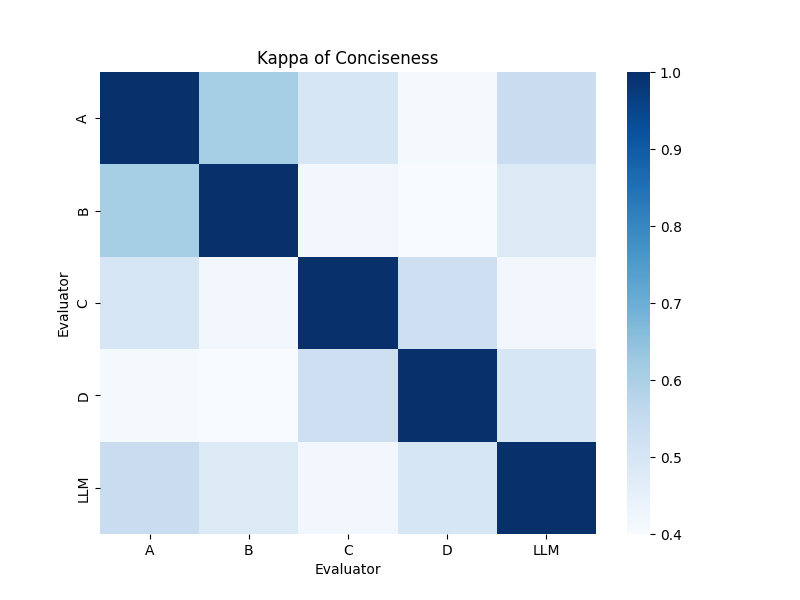}
        \caption{Conciseness }
    \end{subfigure}

    \caption{Kappa coefficient of explanation evaluation}
    \label{kappa}
    
\end{figure}

Since existing automated evaluation metrics, such as BLEU and ROUGE, are solely based on rule-based text matching and lack in-depth analysis of the underlying meaning of explanations, they fail to adequately reflect the effectiveness of generated explanations. Therefore, we adopt a combined approach of human evaluation and large language model (LLM) assessment to provide a more authentic evaluation.

\textbf{Setting:}
The evaluation process begins with conducting a Kappa consistency test on the evaluators (evaluation experts and large language models, LLMs) to ensure that disagreements do not occur frequently and achieve more consistent and reliable results.
The value of the Kappa coefficient ranges from -1 to 1, where a value of 0.75 or higher indicates strong consistency, and a value of 0.4 or higher indicates medium consistency\cite{hubert1977kappa}. 
Subsequently, the generated explanations are assessed based on three dimensions: correctness, completeness, and conciseness, using both LLM evaluation and manual evaluation on a 4-point Likert scale\cite{joshi2015likert}. The final scores are averaged and rounded down to the nearest whole number. 
Additionally, samples with a score difference exceeding one point or those showing simultaneous positive and negative evaluations are defined as disagreements. For such cases, additional experts are invited to discuss with the initial evaluators to reach a consensus on the final score.

\textbf{Result:}
As shown in Fig. \ref{kappa}, we conducted a Kappa consistency test on four human experts and the LLM across the three dimensions of accuracy, completeness, and conciseness. The results indicate strong consistency among the selected experts, as well as between the experts and the LLM. Specifically, the Kappa coefficients for all three dimensions exceed 0.4, with some reaching as high as 0.8, demonstrating good consistency and credibility in the scoring process with minimal disagreement. 

Furthermore, we observed that consistency in accuracy is higher than in completeness and conciseness. Upon analysis, this discrepancy can be attributed to the more objective nature of accuracy compared to completeness and conciseness, which rely more heavily on the subjective judgment of the evaluators.

The evaluation results shown in Table \ref{tab:ratings} validate the effectiveness of MOS's MOE-Tuning approach and vulnerability-aware routing mechanism. For Reentrancy vulnerabilities, our method achieved 79.46\% positive ratings (scores 3 and 4, totaling 178 cases) versus 20.54\% negative ratings (scores 1 and 2, totaling 46 cases), compared to baseline's 50.89\% positive (114 cases) versus 49.11\% negative (110 cases), with significant improvements in completeness (80.80\% vs 62.05\%) and conciseness (96.88\% vs 74.11\%) for positive ratings (scores 3 and 4), demonstrating our expert networks' superiority in capturing complex state-dependent characteristics. The remarkable improvements in Timestamp Dependency cases (positive ratings of 88.36\% vs 59.05\% in correctness) showcase our vulnerability-aware routing mechanism's excellence in identifying and analyzing temporal logic vulnerabilities. For Integer Overflow/Underflow, the substantial improvements in positive ratings for correctness (89.92\% vs 66.53\%) and completeness (88.71\% vs 64.92\%) demonstrate how MOE-Tuning enables expert networks to develop expertise in complex arithmetic vulnerability patterns. The evaluation results for Delegatecall (70.88\% vs 61.54\% in correctness, 76.37\% vs 64.84\% in completeness) also demonstrate the model's effectiveness in identifying and analyzing this type of vulnerability.

These improvements align perfectly with MOS's core design principles: the mixture-of-experts architecture enables different expert networks to focus on specific vulnerability patterns, while the vulnerability-aware routing mechanism ensures smart contract is analyzed by the most relevant experts. Across all these vulnerability types requiring deep domain knowledge, from state management in Reentrancy to temporal logic in Timestamp Dependency, from arithmetic validation in Integer operations to access control in Delegatecall, our method significantly outperforms the baseline approach in terms of correctness, completeness, and conciseness.

\vspace{5pt} 
\noindent\begin{tikzpicture}
  \node[draw=black, thick, fill=gray!20, rounded corners, inner sep=10pt, text width=0.92\linewidth] {
    \textbf{Answer to RQ5:} Through a combined approach of LLM evaluation and human evaluation with verified consistency (Kappa > 0.4), MOS demonstrates consistently superior performance in generating vulnerability explanations. The evaluation results across different vulnerability types show significant improvements in correctness, completeness, and conciseness compared to the baseline.
  };
\end{tikzpicture}
\vspace{5pt} 

\begin{table}[htbp]
\centering
\caption{Ratings of Correctness, Completeness, and Conciseness}
\label{tab:ratings}
\setlength{\tabcolsep}{5pt}
\begin{tabular}{|l|cccc|cccc|cccc|}
\hline
\multirow{2}{*}{Vulnerability} & \multicolumn{4}{c|}{Correctness} & \multicolumn{4}{c|}{Completeness} & \multicolumn{4}{c|}{Conciseness} \\
& 1 & 2 & 3 & 4 & 1 & 2 & 3 & 4 & 1 & 2 & 3 & 4 \\
\hline\hline
\multicolumn{13}{|c|}{\cellcolor{gray!20}\textbf{Reentrancy}} \\
\hline
Baseline & 2 & 108 & 20 & 94 & 19 & 66 & 83 & 56 & 6 & 52 & 30 & 136 \\
Ours & 4 & 42 & 52 & 126 & 4 & 39 & 101 & 80 & 0 & 7 & 72 & 145 \\
\hline\hline
\multicolumn{13}{|c|}{\cellcolor{gray!20}\textbf{Timestamp Dependency}} \\
\hline
Baseline & 11 & 84 & 87 & 50 & 16 & 53 & 132 & 31 & 12 & 47 & 120 & 53 \\
Ours & 2 & 25 & 23 & 182 & 2 & 15 & 67 & 148 & 1 & 3 & 123 & 105 \\
\hline\hline
\multicolumn{13}{|c|}{\cellcolor{gray!20}\textbf{Integer Overflow/Underflow}} \\
\hline
Baseline & 21 & 62 & 106 & 59 & 8 & 79 & 146 & 15 & 6 & 49 & 160 & 33 \\
Ours & 7 & 18 & 42 & 181 & 6 & 22 & 67 & 153 & 1 & 11 & 70 & 166 \\
\hline\hline
\multicolumn{13}{|c|}{\cellcolor{gray!20}\textbf{Delegatecall}} \\
\hline
Baseline & 12 & 58 & 62 & 50 & 9 & 55 & 99 & 19 & 9 & 40 & 99 & 34 \\
Ours & 17 & 36 & 32 & 97 & 17 & 26 & 54 & 85 & 4 & 21 & 77 & 80 \\
\hline\hline
\multicolumn{13}{|c|}{\cellcolor{gray!20}\textbf{Total}} \\
\hline
Baseline & 46 & 312 & 275 & 253 & 52 & 253 & 460 & 121 & 33 & 188 & 409 & 256 \\
Ours & 30 & 121 & 149 & 586 & 29 & 102 & 289 & 466 & 6 & 42 & 342 & 496 \\
\hline
\end{tabular}
\end{table}

\subsubsection{RQ6: Empirical Assessment of Detection Limitations}

To comprehensively evaluate MOS's practical capabilities and constraints, we conducted an in-depth examination of detection discrepancies across four critical vulnerability categories (Reentrancy, Timestamp Dependency, Integer Over/Underflow, and Delegatecall). For each category, we analyzed 10 representative misclassified cases from our test suite, with findings documented in Table~\ref{tab:error_analysis}.

\begin{table}[!t]
\caption{Error Analysis of MOS's Smart Contract Vulnerability Detection}
\label{tab:error_analysis}
\centering
\begin{tabular}{p{3.4cm}|p{2.2cm}|p{5.5cm}|c}
\hline
\textbf{Vulnerability Type} & \textbf{Error Type} & \textbf{Error Pattern} & \textbf{Numbers} \\
\hline
\multirow{2}{*}{Reentrancy} & False Positive & Non-critical State Change (2) / Secure Pattern Misidentification (2) & 4 \\
& False Negative & Execution Order Misjudgment & 6 \\
\hline
\multirow{2}{*}{Timestamp Dependency} & False Positive & Standard Time Check Misclassification & 5 \\
& False Negative & Critical Time Precision Oversight & 5 \\
\hline
\multirow{2}{*}{Integer Over/Underflow} & False Positive & Safe Arithmetic Misidentification & 7 \\
& False Negative & Hidden Overflow Oversight & 3 \\
\hline
\multirow{2}{*}{Delegatecall} & False Positive & Excessive Protection Misjudgment & 5 \\
& False Negative & Dangerous Delegatecall Oversight & 5 \\
\hline
\end{tabular}
\end{table}

Our systematic investigation across four critical vulnerability types revealed distinct patterns of detection inconsistencies. For Reentrancy, MOS demonstrated accuracy challenges, with 4 cases of unnecessary alerts evenly distributed between non-critical state changes (2 cases) and secure pattern implementations (2 cases), while 6 missed vulnerabilities involved execution order misjudgments embedded within complex control flows. The system's performance in Timestamp Dependency detection showed conservative tendencies, generating 5 overcautious flags on standard temporal mechanisms (particularly in DeFi applications' time locks and cooldown periods) and missing 5 critical cases in sophisticated multi-phase protocols where timestamp manipulation could significantly impact token economics. Integer Over/Underflow detection exhibited an imbalanced pattern with 7 cases of overprotective warnings on safe arithmetic implementations and 3 missed risks involving hidden overflow scenarios, notably struggling with contracts implementing thorough SafeMath operations or complex cross-function arithmetic interactions. For Delegatecall analysis, MOS showed a balanced distribution of detection errors, with 5 unnecessary alerts on well-protected implementations with comprehensive access controls and 5 potentially dangerous missed cases.

\vspace{5pt} 
\noindent\begin{tikzpicture}
  \node[draw=black, thick, fill=gray!20, rounded corners, inner sep=10pt, text width=0.92\linewidth] {
    \textbf{Answer to RQ6:} The empirical analysis reveals certain limitations in Mos's vulnerability detection capabilities. Systematic analysis indicates that the tool tends to generate false positives when handling secure patterns and false negatives in complex scenarios, suggesting room for improvement in accurate vulnerability detection.
  };
\end{tikzpicture}
\vspace{5pt} 

\section{Limitations}
\label{limitations}

Despite MOS's excellent experimental performance, the first major limitation concerns its computational requirements. As detailed in the implementation specifications, the training process requires substantial computing resources, with MOE-Tuning requiring at least two NVIDIA GeForce RTX H800 GPUs (80GB VRAM each). Although only about 5B parameters are activated during model inference, the total number of model parameters reaches 18B. While MOE models demonstrate excellent performance in terms of computational efficiency and can significantly reduce computational load (through sparse activation of experts), they still have considerable memory requirements (as all experts need to be loaded). We utilized one NVIDIA GeForce RTX H800 GPU for our implementation.

Another limitation centers on the challenges of prompt engineering and its optimization. Our approach utilized consistent prompt templates across both training and inference stages, but this may not represent the most effective strategy. There could be room for improvement through alternative prompt formulations that we haven't explored. While we achieved reasonable results with our current prompting method, different approaches, such as modified instruction structures or incorporating few-shot examples, might lead to superior model performance. The challenge lies in the complexity of prompt design itself, particularly when it comes to selecting appropriate examples for in-context learning, as these choices can dramatically influence how large language models behave. In future work, we will explore and develop more sophisticated prompting methodologies to further enhance the model's performance, as this represents a crucial area for continued research and improvement.



\section{Related Work}
\label{related_work}

\subsection{Smart Contract Vulnerability Detection}

Research in smart contract vulnerability detection has evolved significantly. Early approaches relied on program analysis techniques: some used symbolic execution to explore execution paths and identify vulnerabilities \cite{luu2016making}, while others employed pattern matching against known vulnerability signatures \cite{mueller2017mythril, tikhomirov2018smartcheck} or applied formal verification with custom compliance patterns \cite{tsankov2018securify}. Wang et al. \cite{wang2024efficiently} proposed SliSE, a two-stage approach combining program slicing and symbolic execution to detect reentrancy vulnerabilities in complex smart contracts. The first stage performs program slicing based on Inter-contract Program Dependency Graph to identify suspicious vulnerabilities, while the second stage employs symbolic execution to verify these suspicious paths.

The field then progressed to machine learning, particularly deep learning models, offering improved accuracy and generalization. Notable examples include bidirectional LSTMs with attention mechanisms for specific vulnerability detection \cite{qian2020towards}, and deep learning for measuring similarity to known vulnerable contracts \cite{gao2019smartembed}. Graph neural networks (GNNs) gained popularity due to their ability to capture contract structural information \cite{zhuang2020smart, liu2021smart}.

The latest frontier involves Large Language Models (LLMs). Studies by Chen et al. \cite{chen2023chatgpt} and David et al. \cite{david2023you} evaluated LLMs on real-world datasets, revealing potential and challenges like high false positive rates. Hu et al. \cite{hu2023large} explored LLMs' reasoning capabilities in this domain, while Sun et al. \cite{sun2024gptscan} introduced GPTScan, combining GPT with program analysis. Chen et al. \cite{chen2024identifying} conducted a comprehensive study on identifying security issues in smart contract code snippets from Stack Overflow. They developed SOChecker, a tool that combines LLM-based code completion with traditional program analysis methods. Sun et al. \cite{sun2024llm4vuln} proposed LLM4Vuln, a unified evaluation framework that decouples LLMs' vulnerability reasoning capabilities from other aspects such as knowledge retrieval and tool support. The framework introduces components including knowledge enhancement, context supplementation, and specialized prompt schemes to improve smart contract vulnerability detection effectiveness. Wu et al. \cite{wu2024advscanner} propose AdvSCanner, an approach combining Large Language Models (LLM) and static analysis to generate adversarial smart contracts for reentrancy vulnerabilities. The system extracts attack flows through static analysis to guide LLM generation, incorporating a self-reflection component that collects feedback for refinement.

Despite these advancements, current approaches face limitations. Many rely on limited datasets, and LLM-based methods struggle with domain knowledge. Additionally, providing explanations for detected vulnerabilities remains challenging, especially for deep learning-based methods.

\subsection{Mixture of Experts }
Mixture of Experts (MoE)  \cite{jacobs1991adaptive} is a hybrid model consisting of multiple sub-models, known as experts, which are integrated together. The key concept of MoE is the use of a router to determine the token set that each expert handles, thereby reducing interference between different types of samples\cite{lin2024moe}.

Traditionally, dense models feed all parameters to each input token. In this way, the growing model capacity brings increased computational cost. To alleviate this issue,sparse models attempt to activate a subset of parameters for each input and these activated parameters are referred as experts\cite{zhu2024llama}.
In Shazeer et al.\cite{shazeer2017outrageously} research, MoE was first proven effective in modern deep learning. This work added an MoE layer which was stacked between LSTM, resulting in state-of-the art results in language modeling and machine translation benchmarks. Subsequently, the MoE layer is introduced to the transformer architecture as a substitute for the FFN layers.
GShard \cite{lepikhin2020gshard} applied MoE to transformers and significantly improved machine translation across 100 languages. Switch Transformers \cite{fedus2022switch} further scaled language models to trillion-parameter sizes through a simple yet effective MoE layer design.
Traditional MoE models are prone to load imbalance, where only a few experts are frequently used while others are rarely activated. To optimize training, BASE layers \cite{lewis2021base}, HASH layers \cite{roller2021hash}, and expert selection strategies  explore methods for constructing MoE models that fully utilize the model's capabilities. Recently, in terms of model architecture, Xue et al. \cite{xue2024openmoe} investigated the use of an improved UL2 training objective to train a decoder-only MoE. Mixtral\cite{jiang2024mixtral} is another decoder-style MoE model that selects two experts from eight, using a token-selection routing mechanism.

However, these methods primarily focus on studying the computational performance and cost of MoE models, with few studies applying MoE as a method for specific tasks in the software engineering field. 
Wu et al.\cite{wu2024ismell} applied the MoE model to code smell detection by leveraging the strengths of experts to select static analysis tools. However, they did not utilize the experts to specifically handle different types of Code Smells. Fundamentally, the approach still relied on the tools for detecting code smells. This tool-dependent approach has limited generalization ability when encountering novel code smell types that existing tools cannot detect.
In contrast to previous research, our study proposes the application of MoE to the specific task of smart contract vulnerability detection and introduces a novel design aimed at improving the model's performance on this task. By treating the MoE experts as specialized detectors for different types of vulnerabilities, we aim to enhance the detection of various vulnerabilities through expert specialization.

\section{Conclusion}
\label{conclusion}
In this paper, we propose MOS, a smart contract vulnerability detection framework based on Mixture-of-Experts Tuning (MOE-Tuning). It performs detection and explanation generation for different vulnerabilities through specialized expert networks and vulnerability-aware routing. The routing mechanism activates the most relevant expert networks by analyzing code features and their compatibility with experts, addressing the limitations of predefined patterns in traditional program analysis approaches. Expert networks process corresponding vulnerabilities based on routing activation. First, we conduct continual pretraining on a large-scale smart contract dataset to enhance the model's domain understanding capability and provide solid initialization for specialized expert networks. Subsequently, we introduce the routing mechanism to achieve expert network specialization through MOE-Tuning. We have conducted extensive experiments to validate the effectiveness of MOS. Through a combination of human evaluation and LLM evaluation, we demonstrate the superiority of MOS in three dimensions: correctness, completeness, and conciseness. In future research, we will focus on expanding the types of vulnerabilities covered.



\bibliographystyle{ACM-Reference-Format}
\bibliography{software}


\begin{thebibliography}{77}


\ifx \showCODEN    \undefined \def \showCODEN     #1{\unskip}     \fi
\ifx \showDOI      \undefined \def \showDOI       #1{#1}\fi
\ifx \showISBNx    \undefined \def \showISBNx     #1{\unskip}     \fi
\ifx \showISBNxiii \undefined \def \showISBNxiii  #1{\unskip}     \fi
\ifx \showISSN     \undefined \def \showISSN      #1{\unskip}     \fi
\ifx \showLCCN     \undefined \def \showLCCN      #1{\unskip}     \fi
\ifx \shownote     \undefined \def \shownote      #1{#1}          \fi
\ifx \showarticletitle \undefined \def \showarticletitle #1{#1}   \fi
\ifx \showURL      \undefined \def \showURL       {\relax}        \fi
\providecommand\bibfield[2]{#2}
\providecommand\bibinfo[2]{#2}
\providecommand\natexlab[1]{#1}
\providecommand\showeprint[2][]{arXiv:#2}

\bibitem[Alharby and Van~Moorsel(2017)]%
        {alharby2017blockchain}
\bibfield{author}{\bibinfo{person}{Maher Alharby} {and} \bibinfo{person}{Aad Van~Moorsel}.} \bibinfo{year}{2017}\natexlab{}.
\newblock \showarticletitle{Blockchain-based smart contracts: A systematic mapping study}.
\newblock \bibinfo{journal}{\emph{arXiv preprint arXiv:1710.06372}} (\bibinfo{year}{2017}).
\newblock


\bibitem[Alrashedy and Binjahlan(2023)]%
        {alrashedy2023language}
\bibfield{author}{\bibinfo{person}{Kamel Alrashedy} {and} \bibinfo{person}{Ahmed Binjahlan}.} \bibinfo{year}{2023}\natexlab{}.
\newblock \showarticletitle{Language Models are Better Bug Detector Through Code-Pair Classification}.
\newblock \bibinfo{journal}{\emph{arXiv preprint arXiv:2311.07957}} (\bibinfo{year}{2023}).
\newblock


\bibitem[Bose et~al\mbox{.}(2022)]%
        {bose2022sailfish}
\bibfield{author}{\bibinfo{person}{Priyanka Bose}, \bibinfo{person}{Dipanjan Das}, \bibinfo{person}{Yanju Chen}, \bibinfo{person}{Yu Feng}, \bibinfo{person}{Christopher Kruegel}, {and} \bibinfo{person}{Giovanni Vigna}.} \bibinfo{year}{2022}\natexlab{}.
\newblock \showarticletitle{Sailfish: Vetting smart contract state-inconsistency bugs in seconds}. In \bibinfo{booktitle}{\emph{2022 IEEE Symposium on Security and Privacy (SP)}}. IEEE, \bibinfo{pages}{161--178}.
\newblock


\bibitem[Cai et~al\mbox{.}(2023)]%
        {cai2023combine}
\bibfield{author}{\bibinfo{person}{Jie Cai}, \bibinfo{person}{Bin Li}, \bibinfo{person}{Jiale Zhang}, \bibinfo{person}{Xiaobing Sun}, {and} \bibinfo{person}{Bing Chen}.} \bibinfo{year}{2023}\natexlab{}.
\newblock \showarticletitle{Combine sliced joint graph with graph neural networks for smart contract vulnerability detection}.
\newblock \bibinfo{journal}{\emph{Journal of Systems and Software}}  \bibinfo{volume}{195} (\bibinfo{year}{2023}), \bibinfo{pages}{111550}.
\newblock


\bibitem[Chen et~al\mbox{.}(2023)]%
        {chen2023chatgpt}
\bibfield{author}{\bibinfo{person}{Chong Chen}, \bibinfo{person}{Jianzhong Su}, \bibinfo{person}{Jiachi Chen}, \bibinfo{person}{Yanlin Wang}, \bibinfo{person}{Tingting Bi}, \bibinfo{person}{Yanli Wang}, \bibinfo{person}{Xingwei Lin}, \bibinfo{person}{Ting Chen}, {and} \bibinfo{person}{Zibin Zheng}.} \bibinfo{year}{2023}\natexlab{}.
\newblock \showarticletitle{When chatgpt meets smart contract vulnerability detection: How far are we?}
\newblock \bibinfo{journal}{\emph{arXiv preprint arXiv:2309.05520}} (\bibinfo{year}{2023}).
\newblock


\bibitem[Chen et~al\mbox{.}(2020)]%
        {chen2020survey}
\bibfield{author}{\bibinfo{person}{Huashan Chen}, \bibinfo{person}{Marcus Pendleton}, \bibinfo{person}{Laurent Njilla}, {and} \bibinfo{person}{Shouhuai Xu}.} \bibinfo{year}{2020}\natexlab{}.
\newblock \showarticletitle{A survey on ethereum systems security: Vulnerabilities, attacks, and defenses}.
\newblock \bibinfo{journal}{\emph{ACM Computing Surveys (CSUR)}} \bibinfo{volume}{53}, \bibinfo{number}{3} (\bibinfo{year}{2020}), \bibinfo{pages}{1--43}.
\newblock


\bibitem[Chen et~al\mbox{.}(2024)]%
        {chen2024identifying}
\bibfield{author}{\bibinfo{person}{Jiachi Chen}, \bibinfo{person}{Chong Chen}, \bibinfo{person}{Jiang Hu}, \bibinfo{person}{John Grundy}, \bibinfo{person}{Yanlin Wang}, \bibinfo{person}{Ting Chen}, {and} \bibinfo{person}{Zibin Zheng}.} \bibinfo{year}{2024}\natexlab{}.
\newblock \showarticletitle{Identifying Smart Contract Security Issues in Code Snippets from Stack Overflow}. In \bibinfo{booktitle}{\emph{Proceedings of the 33rd ACM SIGSOFT International Symposium on Software Testing and Analysis}}. \bibinfo{pages}{1198--1210}.
\newblock


\bibitem[Choi et~al\mbox{.}(2021)]%
        {choi2021smartian}
\bibfield{author}{\bibinfo{person}{Jaeseung Choi}, \bibinfo{person}{Doyeon Kim}, \bibinfo{person}{Soomin Kim}, \bibinfo{person}{Gustavo Grieco}, \bibinfo{person}{Alex Groce}, {and} \bibinfo{person}{Sang~Kil Cha}.} \bibinfo{year}{2021}\natexlab{}.
\newblock \showarticletitle{Smartian: Enhancing smart contract fuzzing with static and dynamic data-flow analyses}. In \bibinfo{booktitle}{\emph{2021 36th IEEE/ACM International Conference on Automated Software Engineering (ASE)}}. IEEE, \bibinfo{pages}{227--239}.
\newblock


\bibitem[Cifarelli et~al\mbox{.}(2023)]%
        {cifarelli2023safurai}
\bibfield{author}{\bibinfo{person}{Davide Cifarelli}, \bibinfo{person}{Leonardo Boiardi}, \bibinfo{person}{Alessandro Puppo}, {and} \bibinfo{person}{Leon Jovanovic}.} \bibinfo{year}{2023}\natexlab{}.
\newblock \showarticletitle{Safurai-Csharp: Harnessing Synthetic Data to improve language-specific Code LLM}.
\newblock \bibinfo{journal}{\emph{arXiv preprint arXiv:2311.03243}} (\bibinfo{year}{2023}).
\newblock


\bibitem[David et~al\mbox{.}(2023)]%
        {david2023you}
\bibfield{author}{\bibinfo{person}{Isaac David}, \bibinfo{person}{Liyi Zhou}, \bibinfo{person}{Kaihua Qin}, \bibinfo{person}{Dawn Song}, \bibinfo{person}{Lorenzo Cavallaro}, {and} \bibinfo{person}{Arthur Gervais}.} \bibinfo{year}{2023}\natexlab{}.
\newblock \showarticletitle{Do you still need a manual smart contract audit?}
\newblock \bibinfo{journal}{\emph{arXiv preprint arXiv:2306.12338}} (\bibinfo{year}{2023}).
\newblock


\bibitem[Destefanis et~al\mbox{.}(2018)]%
        {destefanis2018smart}
\bibfield{author}{\bibinfo{person}{Giuseppe Destefanis}, \bibinfo{person}{Michele Marchesi}, \bibinfo{person}{Marco Ortu}, \bibinfo{person}{Roberto Tonelli}, \bibinfo{person}{Andrea Bracciali}, {and} \bibinfo{person}{Robert Hierons}.} \bibinfo{year}{2018}\natexlab{}.
\newblock \showarticletitle{Smart contracts vulnerabilities: a call for blockchain software engineering?}. In \bibinfo{booktitle}{\emph{2018 International Workshop on Blockchain Oriented Software Engineering (IWBOSE)}}. IEEE, \bibinfo{pages}{19--25}.
\newblock


\bibitem[Dhillon et~al\mbox{.}(2017)]%
        {dhillon2017dao}
\bibfield{author}{\bibinfo{person}{Vikram Dhillon}, \bibinfo{person}{David Metcalf}, \bibinfo{person}{Max Hooper}, \bibinfo{person}{Vikram Dhillon}, \bibinfo{person}{David Metcalf}, {and} \bibinfo{person}{Max Hooper}.} \bibinfo{year}{2017}\natexlab{}.
\newblock \showarticletitle{The DAO hacked}.
\newblock \bibinfo{journal}{\emph{blockchain enabled applications: Understand the blockchain Ecosystem and How to Make it work for you}} (\bibinfo{year}{2017}), \bibinfo{pages}{67--78}.
\newblock


\bibitem[Dubey et~al\mbox{.}(2024)]%
        {dubey2024llama}
\bibfield{author}{\bibinfo{person}{Abhimanyu Dubey}, \bibinfo{person}{Abhinav Jauhri}, \bibinfo{person}{Abhinav Pandey}, \bibinfo{person}{Abhishek Kadian}, \bibinfo{person}{Ahmad Al-Dahle}, \bibinfo{person}{Aiesha Letman}, \bibinfo{person}{Akhil Mathur}, \bibinfo{person}{Alan Schelten}, \bibinfo{person}{Amy Yang}, \bibinfo{person}{Angela Fan}, {et~al\mbox{.}}} \bibinfo{year}{2024}\natexlab{}.
\newblock \showarticletitle{The llama 3 herd of models}.
\newblock \bibinfo{journal}{\emph{arXiv preprint arXiv:2407.21783}} (\bibinfo{year}{2024}).
\newblock


\bibitem[Durieux et~al\mbox{.}(2020)]%
        {durieux2020empirical}
\bibfield{author}{\bibinfo{person}{Thomas Durieux}, \bibinfo{person}{Jo{\~a}o~F Ferreira}, \bibinfo{person}{Rui Abreu}, {and} \bibinfo{person}{Pedro Cruz}.} \bibinfo{year}{2020}\natexlab{}.
\newblock \showarticletitle{Empirical review of automated analysis tools on 47,587 ethereum smart contracts}. In \bibinfo{booktitle}{\emph{Proceedings of the ACM/IEEE 42nd International conference on software engineering}}. \bibinfo{pages}{530--541}.
\newblock


\bibitem[Fedus et~al\mbox{.}(2022)]%
        {fedus2022switch}
\bibfield{author}{\bibinfo{person}{William Fedus}, \bibinfo{person}{Barret Zoph}, {and} \bibinfo{person}{Noam Shazeer}.} \bibinfo{year}{2022}\natexlab{}.
\newblock \showarticletitle{Switch transformers: Scaling to trillion parameter models with simple and efficient sparsity}.
\newblock \bibinfo{journal}{\emph{Journal of Machine Learning Research}} \bibinfo{volume}{23}, \bibinfo{number}{120} (\bibinfo{year}{2022}), \bibinfo{pages}{1--39}.
\newblock


\bibitem[Feist et~al\mbox{.}(2019)]%
        {feist2019slither}
\bibfield{author}{\bibinfo{person}{Josselin Feist}, \bibinfo{person}{Gustavo Grieco}, {and} \bibinfo{person}{Alex Groce}.} \bibinfo{year}{2019}\natexlab{}.
\newblock \showarticletitle{Slither: a static analysis framework for smart contracts}. In \bibinfo{booktitle}{\emph{2019 IEEE/ACM 2nd International Workshop on Emerging Trends in Software Engineering for Blockchain (WETSEB)}}. IEEE, \bibinfo{pages}{8--15}.
\newblock


\bibitem[Feng et~al\mbox{.}(2020)]%
        {feng2020codebert}
\bibfield{author}{\bibinfo{person}{Zhangyin Feng}, \bibinfo{person}{Daya Guo}, \bibinfo{person}{Duyu Tang}, \bibinfo{person}{Nan Duan}, \bibinfo{person}{Xiaocheng Feng}, \bibinfo{person}{Ming Gong}, \bibinfo{person}{Linjun Shou}, \bibinfo{person}{Bing Qin}, \bibinfo{person}{Ting Liu}, \bibinfo{person}{Daxin Jiang}, {et~al\mbox{.}}} \bibinfo{year}{2020}\natexlab{}.
\newblock \showarticletitle{CodeBERT: A Pre-Trained Model for Programming and Natural Languages}. In \bibinfo{booktitle}{\emph{Findings of the Association for Computational Linguistics: EMNLP 2020}}. \bibinfo{pages}{1536--1547}.
\newblock


\bibitem[Ferreira et~al\mbox{.}(2020)]%
        {ferreira2020smartbugs}
\bibfield{author}{\bibinfo{person}{Jo{\~a}o~F Ferreira}, \bibinfo{person}{Pedro Cruz}, \bibinfo{person}{Thomas Durieux}, {and} \bibinfo{person}{Rui Abreu}.} \bibinfo{year}{2020}\natexlab{}.
\newblock \showarticletitle{Smartbugs: A framework to analyze solidity smart contracts}. In \bibinfo{booktitle}{\emph{Proceedings of the 35th IEEE/ACM International Conference on Automated Software Engineering}}. \bibinfo{pages}{1349--1352}.
\newblock


\bibitem[Gao et~al\mbox{.}(2019b)]%
        {gao2019easyflow}
\bibfield{author}{\bibinfo{person}{Jianbo Gao}, \bibinfo{person}{Han Liu}, \bibinfo{person}{Chao Liu}, \bibinfo{person}{Qingshan Li}, \bibinfo{person}{Zhi Guan}, {and} \bibinfo{person}{Zhong Chen}.} \bibinfo{year}{2019}\natexlab{b}.
\newblock \showarticletitle{Easyflow: Keep ethereum away from overflow}. In \bibinfo{booktitle}{\emph{2019 IEEE/ACM 41st International Conference on Software Engineering: Companion Proceedings (ICSE-Companion)}}. IEEE, \bibinfo{pages}{23--26}.
\newblock


\bibitem[Gao et~al\mbox{.}(2019a)]%
        {gao2019smartembed}
\bibfield{author}{\bibinfo{person}{Zhipeng Gao}, \bibinfo{person}{Vinoj Jayasundara}, \bibinfo{person}{Lingxiao Jiang}, \bibinfo{person}{Xin Xia}, \bibinfo{person}{David Lo}, {and} \bibinfo{person}{John Grundy}.} \bibinfo{year}{2019}\natexlab{a}.
\newblock \showarticletitle{Smartembed: A tool for clone and bug detection in smart contracts through structural code embedding}. In \bibinfo{booktitle}{\emph{2019 IEEE International Conference on Software Maintenance and Evolution (ICSME)}}. IEEE, \bibinfo{pages}{394--397}.
\newblock


\bibitem[Guo et~al\mbox{.}(2020)]%
        {guo2020graphcodebert}
\bibfield{author}{\bibinfo{person}{Daya Guo}, \bibinfo{person}{Shuo Ren}, \bibinfo{person}{Shuai Lu}, \bibinfo{person}{Zhangyin Feng}, \bibinfo{person}{Duyu Tang}, \bibinfo{person}{LIU Shujie}, \bibinfo{person}{Long Zhou}, \bibinfo{person}{Nan Duan}, \bibinfo{person}{Alexey Svyatkovskiy}, \bibinfo{person}{Shengyu Fu}, {et~al\mbox{.}}} \bibinfo{year}{2020}\natexlab{}.
\newblock \showarticletitle{GraphCodeBERT: Pre-training Code Representations with Data Flow}. In \bibinfo{booktitle}{\emph{International Conference on Learning Representations}}.
\newblock


\bibitem[Heged{\H{u}}s(2018)]%
        {hegedHus2018towards}
\bibfield{author}{\bibinfo{person}{P{\'e}ter Heged{\H{u}}s}.} \bibinfo{year}{2018}\natexlab{}.
\newblock \showarticletitle{Towards analyzing the complexity landscape of solidity based ethereum smart contracts}. In \bibinfo{booktitle}{\emph{Proceedings of the 1st International Workshop on Emerging Trends in Software Engineering for Blockchain}}. \bibinfo{pages}{35--39}.
\newblock


\bibitem[Hewa et~al\mbox{.}(2021)]%
        {hewa2021survey}
\bibfield{author}{\bibinfo{person}{Tharaka Hewa}, \bibinfo{person}{Mika Ylianttila}, {and} \bibinfo{person}{Madhusanka Liyanage}.} \bibinfo{year}{2021}\natexlab{}.
\newblock \showarticletitle{Survey on blockchain based smart contracts: Applications, opportunities and challenges}.
\newblock \bibinfo{journal}{\emph{Journal of Network and Computer Applications}}  \bibinfo{volume}{177} (\bibinfo{year}{2021}), \bibinfo{pages}{102857}.
\newblock


\bibitem[Hu et~al\mbox{.}(2023)]%
        {hu2023large}
\bibfield{author}{\bibinfo{person}{Sihao Hu}, \bibinfo{person}{Tiansheng Huang}, \bibinfo{person}{Fatih {\.I}lhan}, \bibinfo{person}{Selim~Furkan Tekin}, {and} \bibinfo{person}{Ling Liu}.} \bibinfo{year}{2023}\natexlab{}.
\newblock \showarticletitle{Large language model-powered smart contract vulnerability detection: New perspectives}.
\newblock \bibinfo{journal}{\emph{arXiv preprint arXiv:2310.01152}} (\bibinfo{year}{2023}).
\newblock


\bibitem[Hubert(1977)]%
        {hubert1977kappa}
\bibfield{author}{\bibinfo{person}{Lawrence Hubert}.} \bibinfo{year}{1977}\natexlab{}.
\newblock \showarticletitle{Kappa revisited.}
\newblock \bibinfo{journal}{\emph{Psychological Bulletin}} \bibinfo{volume}{84}, \bibinfo{number}{2} (\bibinfo{year}{1977}), \bibinfo{pages}{289}.
\newblock


\bibitem[Jacobs et~al\mbox{.}(1991)]%
        {jacobs1991adaptive}
\bibfield{author}{\bibinfo{person}{Robert~A Jacobs}, \bibinfo{person}{Michael~I Jordan}, \bibinfo{person}{Steven~J Nowlan}, {and} \bibinfo{person}{Geoffrey~E Hinton}.} \bibinfo{year}{1991}\natexlab{}.
\newblock \showarticletitle{Adaptive mixtures of local experts}.
\newblock \bibinfo{journal}{\emph{Neural computation}} \bibinfo{volume}{3}, \bibinfo{number}{1} (\bibinfo{year}{1991}), \bibinfo{pages}{79--87}.
\newblock


\bibitem[Jiang et~al\mbox{.}(2024)]%
        {jiang2024mixtral}
\bibfield{author}{\bibinfo{person}{Albert~Q Jiang}, \bibinfo{person}{Alexandre Sablayrolles}, \bibinfo{person}{Antoine Roux}, \bibinfo{person}{Arthur Mensch}, \bibinfo{person}{Blanche Savary}, \bibinfo{person}{Chris Bamford}, \bibinfo{person}{Devendra~Singh Chaplot}, \bibinfo{person}{Diego de~las Casas}, \bibinfo{person}{Emma~Bou Hanna}, \bibinfo{person}{Florian Bressand}, {et~al\mbox{.}}} \bibinfo{year}{2024}\natexlab{}.
\newblock \showarticletitle{Mixtral of experts}.
\newblock \bibinfo{journal}{\emph{arXiv preprint arXiv:2401.04088}} (\bibinfo{year}{2024}).
\newblock


\bibitem[Joshi et~al\mbox{.}(2015)]%
        {joshi2015likert}
\bibfield{author}{\bibinfo{person}{Ankur Joshi}, \bibinfo{person}{Saket Kale}, \bibinfo{person}{Satish Chandel}, {and} \bibinfo{person}{D~Kumar Pal}.} \bibinfo{year}{2015}\natexlab{}.
\newblock \showarticletitle{Likert scale: Explored and explained}.
\newblock \bibinfo{journal}{\emph{British journal of applied science \& technology}} \bibinfo{volume}{7}, \bibinfo{number}{4} (\bibinfo{year}{2015}), \bibinfo{pages}{396--403}.
\newblock


\bibitem[Kipf and Welling(2016)]%
        {kipf2016semi}
\bibfield{author}{\bibinfo{person}{Thomas~N Kipf} {and} \bibinfo{person}{Max Welling}.} \bibinfo{year}{2016}\natexlab{}.
\newblock \showarticletitle{Semi-supervised classification with graph convolutional networks}.
\newblock \bibinfo{journal}{\emph{arXiv preprint arXiv:1609.02907}} (\bibinfo{year}{2016}).
\newblock


\bibitem[Lepikhin et~al\mbox{.}(2020)]%
        {lepikhin2020gshard}
\bibfield{author}{\bibinfo{person}{Dmitry Lepikhin}, \bibinfo{person}{HyoukJoong Lee}, \bibinfo{person}{Yuanzhong Xu}, \bibinfo{person}{Dehao Chen}, \bibinfo{person}{Orhan Firat}, \bibinfo{person}{Yanping Huang}, \bibinfo{person}{Maxim Krikun}, \bibinfo{person}{Noam Shazeer}, {and} \bibinfo{person}{Zhifeng Chen}.} \bibinfo{year}{2020}\natexlab{}.
\newblock \showarticletitle{Gshard: Scaling giant models with conditional computation and automatic sharding}.
\newblock \bibinfo{journal}{\emph{arXiv preprint arXiv:2006.16668}} (\bibinfo{year}{2020}).
\newblock


\bibitem[Lewis et~al\mbox{.}(2021)]%
        {lewis2021base}
\bibfield{author}{\bibinfo{person}{Mike Lewis}, \bibinfo{person}{Shruti Bhosale}, \bibinfo{person}{Tim Dettmers}, \bibinfo{person}{Naman Goyal}, {and} \bibinfo{person}{Luke Zettlemoyer}.} \bibinfo{year}{2021}\natexlab{}.
\newblock \showarticletitle{Base layers: Simplifying training of large, sparse models}. In \bibinfo{booktitle}{\emph{International Conference on Machine Learning}}. PMLR, \bibinfo{pages}{6265--6274}.
\newblock


\bibitem[Lewis et~al\mbox{.}(2020)]%
        {lewis2020retrieval}
\bibfield{author}{\bibinfo{person}{Patrick Lewis}, \bibinfo{person}{Ethan Perez}, \bibinfo{person}{Aleksandra Piktus}, \bibinfo{person}{Fabio Petroni}, \bibinfo{person}{Vladimir Karpukhin}, \bibinfo{person}{Naman Goyal}, \bibinfo{person}{Heinrich K{\"u}ttler}, \bibinfo{person}{Mike Lewis}, \bibinfo{person}{Wen-tau Yih}, \bibinfo{person}{Tim Rockt{\"a}schel}, {et~al\mbox{.}}} \bibinfo{year}{2020}\natexlab{}.
\newblock \showarticletitle{Retrieval-augmented generation for knowledge-intensive nlp tasks}.
\newblock \bibinfo{journal}{\emph{Advances in Neural Information Processing Systems}}  \bibinfo{volume}{33} (\bibinfo{year}{2020}), \bibinfo{pages}{9459--9474}.
\newblock


\bibitem[Lin et~al\mbox{.}(2024)]%
        {lin2024moe}
\bibfield{author}{\bibinfo{person}{Bin Lin}, \bibinfo{person}{Zhenyu Tang}, \bibinfo{person}{Yang Ye}, \bibinfo{person}{Jiaxi Cui}, \bibinfo{person}{Bin Zhu}, \bibinfo{person}{Peng Jin}, \bibinfo{person}{Junwu Zhang}, \bibinfo{person}{Munan Ning}, {and} \bibinfo{person}{Li Yuan}.} \bibinfo{year}{2024}\natexlab{}.
\newblock \showarticletitle{Moe-llava: Mixture of experts for large vision-language models}.
\newblock \bibinfo{journal}{\emph{arXiv preprint arXiv:2401.15947}} (\bibinfo{year}{2024}).
\newblock


\bibitem[Liu et~al\mbox{.}(2021a)]%
        {liu2021smart}
\bibfield{author}{\bibinfo{person}{Zhenguang Liu}, \bibinfo{person}{Peng Qian}, \bibinfo{person}{Xiang Wang}, \bibinfo{person}{Lei Zhu}, \bibinfo{person}{Qinming He}, {and} \bibinfo{person}{Shouling Ji}.} \bibinfo{year}{2021}\natexlab{a}.
\newblock \showarticletitle{Smart contract vulnerability detection: from pure neural network to interpretable graph feature and expert pattern fusion}.
\newblock \bibinfo{journal}{\emph{arXiv preprint arXiv:2106.09282}} (\bibinfo{year}{2021}).
\newblock


\bibitem[Liu et~al\mbox{.}(2021b)]%
        {liu2021combining}
\bibfield{author}{\bibinfo{person}{Zhenguang Liu}, \bibinfo{person}{Peng Qian}, \bibinfo{person}{Xiaoyang Wang}, \bibinfo{person}{Yuan Zhuang}, \bibinfo{person}{Lin Qiu}, {and} \bibinfo{person}{Xun Wang}.} \bibinfo{year}{2021}\natexlab{b}.
\newblock \showarticletitle{Combining graph neural networks with expert knowledge for smart contract vulnerability detection}.
\newblock \bibinfo{journal}{\emph{IEEE Transactions on Knowledge and Data Engineering}} (\bibinfo{year}{2021}).
\newblock


\bibitem[Liu et~al\mbox{.}(2023)]%
        {liu2023rethinking}
\bibfield{author}{\bibinfo{person}{Zhenguang Liu}, \bibinfo{person}{Peng Qian}, \bibinfo{person}{Jiaxu Yang}, \bibinfo{person}{Lingfeng Liu}, \bibinfo{person}{Xiaojun Xu}, \bibinfo{person}{Qinming He}, {and} \bibinfo{person}{Xiaosong Zhang}.} \bibinfo{year}{2023}\natexlab{}.
\newblock \showarticletitle{Rethinking smart contract fuzzing: Fuzzing with invocation ordering and important branch revisiting}.
\newblock \bibinfo{journal}{\emph{IEEE Transactions on Information Forensics and Security}}  \bibinfo{volume}{18} (\bibinfo{year}{2023}), \bibinfo{pages}{1237--1251}.
\newblock


\bibitem[Loshchilov and Hutter(2017)]%
        {adamw}
\bibfield{author}{\bibinfo{person}{Ilya Loshchilov} {and} \bibinfo{person}{Frank Hutter}.} \bibinfo{year}{2017}\natexlab{}.
\newblock \showarticletitle{Decoupled weight decay regularization}.
\newblock \bibinfo{journal}{\emph{arXiv preprint arXiv:1711.05101}} (\bibinfo{year}{2017}).
\newblock


\bibitem[Luo et~al\mbox{.}(2024)]%
        {luo2024scvhunter}
\bibfield{author}{\bibinfo{person}{Feng Luo}, \bibinfo{person}{Ruijie Luo}, \bibinfo{person}{Ting Chen}, \bibinfo{person}{Ao Qiao}, \bibinfo{person}{Zheyuan He}, \bibinfo{person}{Shuwei Song}, \bibinfo{person}{Yu Jiang}, {and} \bibinfo{person}{Sixing Li}.} \bibinfo{year}{2024}\natexlab{}.
\newblock \showarticletitle{SCVHunter: Smart Contract Vulnerability Detection Based on Heterogeneous Graph Attention Network}. In \bibinfo{booktitle}{\emph{2024 IEEE/ACM 46th International Conference on Software Engineering (ICSE)}}. IEEE Computer Society, \bibinfo{pages}{954--954}.
\newblock


\bibitem[Luu et~al\mbox{.}(2016)]%
        {luu2016making}
\bibfield{author}{\bibinfo{person}{Loi Luu}, \bibinfo{person}{Duc-Hiep Chu}, \bibinfo{person}{Hrishi Olickel}, \bibinfo{person}{Prateek Saxena}, {and} \bibinfo{person}{Aquinas Hobor}.} \bibinfo{year}{2016}\natexlab{}.
\newblock \showarticletitle{Making smart contracts smarter}. In \bibinfo{booktitle}{\emph{Proceedings of the 2016 ACM SIGSAC conference on computer and communications security}}. \bibinfo{pages}{254--269}.
\newblock


\bibitem[Ma et~al\mbox{.}(2024)]%
        {ma2024combining}
\bibfield{author}{\bibinfo{person}{Wei Ma}, \bibinfo{person}{Daoyuan Wu}, \bibinfo{person}{Yuqiang Sun}, \bibinfo{person}{Tianwen Wang}, \bibinfo{person}{Shangqing Liu}, \bibinfo{person}{Jian Zhang}, \bibinfo{person}{Yue Xue}, {and} \bibinfo{person}{Yang Liu}.} \bibinfo{year}{2024}\natexlab{}.
\newblock \showarticletitle{Combining Fine-Tuning and LLM-based Agents for Intuitive Smart Contract Auditing with Justifications}.
\newblock \bibinfo{journal}{\emph{arXiv preprint arXiv:2403.16073}} (\bibinfo{year}{2024}).
\newblock


\bibitem[Mehar et~al\mbox{.}(2019)]%
        {mehar2019understanding}
\bibfield{author}{\bibinfo{person}{Muhammad~Izhar Mehar}, \bibinfo{person}{Charles~Louis Shier}, \bibinfo{person}{Alana Giambattista}, \bibinfo{person}{Elgar Gong}, \bibinfo{person}{Gabrielle Fletcher}, \bibinfo{person}{Ryan Sanayhie}, \bibinfo{person}{Henry~M Kim}, {and} \bibinfo{person}{Marek Laskowski}.} \bibinfo{year}{2019}\natexlab{}.
\newblock \showarticletitle{Understanding a revolutionary and flawed grand experiment in blockchain: the DAO attack}.
\newblock \bibinfo{journal}{\emph{Journal of Cases on Information Technology (JCIT)}} \bibinfo{volume}{21}, \bibinfo{number}{1} (\bibinfo{year}{2019}), \bibinfo{pages}{19--32}.
\newblock


\bibitem[Mueller(2017)]%
        {mueller2017mythril}
\bibfield{author}{\bibinfo{person}{B Mueller}.} \bibinfo{year}{2017}\natexlab{}.
\newblock \bibinfo{title}{Mythril-Reversing and bug hunting framework for the Ethereum blockchain}.
\newblock


\bibitem[OpenAI(2023)]%
        {openai2023gpt35}
\bibfield{author}{\bibinfo{person}{OpenAI}.} \bibinfo{year}{2023}\natexlab{}.
\newblock \bibinfo{title}{GPT-3.5}.
\newblock \bibinfo{howpublished}{\url{https://platform.openai.com/docs/models/gpt-3-5}}.
\newblock


\bibitem[OpenAI(2024)]%
        {openai2024gpt4o}
\bibfield{author}{\bibinfo{person}{OpenAI}.} \bibinfo{year}{2024}\natexlab{}.
\newblock \bibinfo{title}{GPT-4o}.
\newblock \bibinfo{howpublished}{\url{https://openai.com/index/hello-gpt-4o}}.
\newblock


\bibitem[{OWASP Foundation}(2023)]%
        {owasp2023}
\bibfield{author}{\bibinfo{person}{{OWASP Foundation}}.} \bibinfo{year}{2023}\natexlab{}.
\newblock \bibinfo{title}{{OWASP Smart Contract Top 10}}.
\newblock \bibinfo{howpublished}{\url{https://owasp.org/www-project-smart-contract-top-10/}}.
\newblock
\newblock
\shownote{Accessed: 2024-12-04}.


\bibitem[Praitheeshan et~al\mbox{.}(2019)]%
        {praitheeshan2019security}
\bibfield{author}{\bibinfo{person}{Purathani Praitheeshan}, \bibinfo{person}{Lei Pan}, \bibinfo{person}{Jiangshan Yu}, \bibinfo{person}{Joseph Liu}, {and} \bibinfo{person}{Robin Doss}.} \bibinfo{year}{2019}\natexlab{}.
\newblock \showarticletitle{Security analysis methods on ethereum smart contract vulnerabilities: a survey}.
\newblock \bibinfo{journal}{\emph{arXiv preprint arXiv:1908.08605}} (\bibinfo{year}{2019}).
\newblock


\bibitem[Qian et~al\mbox{.}(2020)]%
        {qian2020towards}
\bibfield{author}{\bibinfo{person}{Peng Qian}, \bibinfo{person}{Zhenguang Liu}, \bibinfo{person}{Qinming He}, \bibinfo{person}{Roger Zimmermann}, {and} \bibinfo{person}{Xun Wang}.} \bibinfo{year}{2020}\natexlab{}.
\newblock \showarticletitle{Towards automated reentrancy detection for smart contracts based on sequential models}.
\newblock \bibinfo{journal}{\emph{IEEE Access}}  \bibinfo{volume}{8} (\bibinfo{year}{2020}), \bibinfo{pages}{19685--19695}.
\newblock


\bibitem[Qian et~al\mbox{.}(2023)]%
        {qian2023cross}
\bibfield{author}{\bibinfo{person}{Peng Qian}, \bibinfo{person}{Zhenguang Liu}, \bibinfo{person}{Yifang Yin}, {and} \bibinfo{person}{Qinming He}.} \bibinfo{year}{2023}\natexlab{}.
\newblock \showarticletitle{Cross-modality mutual learning for enhancing smart contract vulnerability detection on bytecode}. In \bibinfo{booktitle}{\emph{Proceedings of the ACM Web Conference 2023}}. \bibinfo{pages}{2220--2229}.
\newblock


\bibitem[Rasley et~al\mbox{.}(2020)]%
        {rasley2020deepspeed}
\bibfield{author}{\bibinfo{person}{Jeff Rasley}, \bibinfo{person}{Samyam Rajbhandari}, \bibinfo{person}{Olatunji Ruwase}, {and} \bibinfo{person}{Yuxiong He}.} \bibinfo{year}{2020}\natexlab{}.
\newblock \showarticletitle{Deepspeed: System optimizations enable training deep learning models with over 100 billion parameters}. In \bibinfo{booktitle}{\emph{Proceedings of the 26th ACM SIGKDD International Conference on Knowledge Discovery \& Data Mining}}. \bibinfo{pages}{3505--3506}.
\newblock


\bibitem[Roller et~al\mbox{.}(2021)]%
        {roller2021hash}
\bibfield{author}{\bibinfo{person}{Stephen Roller}, \bibinfo{person}{Sainbayar Sukhbaatar}, \bibinfo{person}{Jason Weston}, {et~al\mbox{.}}} \bibinfo{year}{2021}\natexlab{}.
\newblock \showarticletitle{Hash layers for large sparse models}.
\newblock \bibinfo{journal}{\emph{Advances in Neural Information Processing Systems}}  \bibinfo{volume}{34} (\bibinfo{year}{2021}), \bibinfo{pages}{17555--17566}.
\newblock


\bibitem[Roumeliotis et~al\mbox{.}(2023)]%
        {roumeliotis2023llama}
\bibfield{author}{\bibinfo{person}{Konstantinos~I Roumeliotis}, \bibinfo{person}{Nikolaos~D Tselikas}, {and} \bibinfo{person}{Dimitrios~K Nasiopoulos}.} \bibinfo{year}{2023}\natexlab{}.
\newblock \showarticletitle{Llama 2: Early Adopters' Utilization of Meta's New Open-Source Pretrained Model}.
\newblock  (\bibinfo{year}{2023}).
\newblock


\bibitem[Shazeer et~al\mbox{.}(2017)]%
        {shazeer2017outrageously}
\bibfield{author}{\bibinfo{person}{Noam Shazeer}, \bibinfo{person}{Azalia Mirhoseini}, \bibinfo{person}{Krzysztof Maziarz}, \bibinfo{person}{Andy Davis}, \bibinfo{person}{Quoc Le}, \bibinfo{person}{Geoffrey Hinton}, {and} \bibinfo{person}{Jeff Dean}.} \bibinfo{year}{2017}\natexlab{}.
\newblock \showarticletitle{Outrageously large neural networks: The sparsely-gated mixture-of-experts layer}.
\newblock \bibinfo{journal}{\emph{arXiv preprint arXiv:1701.06538}} (\bibinfo{year}{2017}).
\newblock


\bibitem[Storhaug et~al\mbox{.}(2023)]%
        {storhaug2023efficient}
\bibfield{author}{\bibinfo{person}{Andr{\'e} Storhaug}, \bibinfo{person}{Jingyue Li}, {and} \bibinfo{person}{Tianyuan Hu}.} \bibinfo{year}{2023}\natexlab{}.
\newblock \showarticletitle{Efficient avoidance of vulnerabilities in auto-completed smart contract code using vulnerability-constrained decoding}. In \bibinfo{booktitle}{\emph{2023 IEEE 34th International Symposium on Software Reliability Engineering (ISSRE)}}. IEEE, \bibinfo{pages}{683--693}.
\newblock


\bibitem[Sun et~al\mbox{.}(2024a)]%
        {sun2024llm4vuln}
\bibfield{author}{\bibinfo{person}{Yuqiang Sun}, \bibinfo{person}{Daoyuan Wu}, \bibinfo{person}{Yue Xue}, \bibinfo{person}{Han Liu}, \bibinfo{person}{Wei Ma}, \bibinfo{person}{Lyuye Zhang}, \bibinfo{person}{Yang Liu}, {and} \bibinfo{person}{Yingjiu Li}.} \bibinfo{year}{2024}\natexlab{a}.
\newblock \showarticletitle{Llm4vuln: A unified evaluation framework for decoupling and enhancing llms' vulnerability reasoning}.
\newblock \bibinfo{journal}{\emph{arXiv preprint arXiv:2401.16185}} (\bibinfo{year}{2024}).
\newblock


\bibitem[Sun et~al\mbox{.}(2024b)]%
        {sun2024gptscan}
\bibfield{author}{\bibinfo{person}{Yuqiang Sun}, \bibinfo{person}{Daoyuan Wu}, \bibinfo{person}{Yue Xue}, \bibinfo{person}{Han Liu}, \bibinfo{person}{Haijun Wang}, \bibinfo{person}{Zhengzi Xu}, \bibinfo{person}{Xiaofei Xie}, {and} \bibinfo{person}{Yang Liu}.} \bibinfo{year}{2024}\natexlab{b}.
\newblock \showarticletitle{GPTScan: Detecting Logic Vulnerabilities in Smart Contracts by Combining GPT with Program Analysis}.
\newblock \bibinfo{journal}{\emph{Proc. IEEE/ACM ICSE}} (\bibinfo{year}{2024}).
\newblock


\bibitem[Sun et~al\mbox{.}(2023)]%
        {sun2023real}
\bibfield{author}{\bibinfo{person}{Yuqiang Sun}, \bibinfo{person}{Zhengzi Xu}, \bibinfo{person}{Chengwei Liu}, \bibinfo{person}{Yiran Zhang}, {and} \bibinfo{person}{Yang Liu}.} \bibinfo{year}{2023}\natexlab{}.
\newblock \showarticletitle{Who is the Real Hero? Measuring Developer Contribution via Multi-dimensional Data Integration}. In \bibinfo{booktitle}{\emph{2023 38th IEEE/ACM International Conference on Automated Software Engineering (ASE)}}. IEEE, \bibinfo{pages}{825--836}.
\newblock


\bibitem[Swan(2015)]%
        {swan2015blockchain}
\bibfield{author}{\bibinfo{person}{Melanie Swan}.} \bibinfo{year}{2015}\natexlab{}.
\newblock \bibinfo{booktitle}{\emph{Blockchain: Blueprint for a new economy}}.
\newblock \bibinfo{publisher}{" O'Reilly Media, Inc."}.
\newblock


\bibitem[Szabo(1996)]%
        {szabo1996smart}
\bibfield{author}{\bibinfo{person}{Nick Szabo}.} \bibinfo{year}{1996}\natexlab{}.
\newblock \showarticletitle{Smart contracts: building blocks for digital markets}.
\newblock \bibinfo{journal}{\emph{EXTROPY: The Journal of Transhumanist Thought,(16)}} \bibinfo{volume}{18}, \bibinfo{number}{2} (\bibinfo{year}{1996}), \bibinfo{pages}{28}.
\newblock


\bibitem[Tikhomirov et~al\mbox{.}(2018)]%
        {tikhomirov2018smartcheck}
\bibfield{author}{\bibinfo{person}{Sergei Tikhomirov}, \bibinfo{person}{Ekaterina Voskresenskaya}, \bibinfo{person}{Ivan Ivanitskiy}, \bibinfo{person}{Ramil Takhaviev}, \bibinfo{person}{Evgeny Marchenko}, {and} \bibinfo{person}{Yaroslav Alexandrov}.} \bibinfo{year}{2018}\natexlab{}.
\newblock \showarticletitle{Smartcheck: Static analysis of ethereum smart contracts}. In \bibinfo{booktitle}{\emph{Proceedings of the 1st International Workshop on Emerging Trends in Software Engineering for Blockchain}}. \bibinfo{pages}{9--16}.
\newblock


\bibitem[Torres et~al\mbox{.}(2018)]%
        {torres2018osiris}
\bibfield{author}{\bibinfo{person}{Christof~Ferreira Torres}, \bibinfo{person}{Julian Sch{\"u}tte}, {and} \bibinfo{person}{Radu State}.} \bibinfo{year}{2018}\natexlab{}.
\newblock \showarticletitle{Osiris: Hunting for integer bugs in ethereum smart contracts}. In \bibinfo{booktitle}{\emph{Proceedings of the 34th Annual Computer Security Applications Conference}}. \bibinfo{pages}{664--676}.
\newblock


\bibitem[Torres et~al\mbox{.}(2019)]%
        {torres2019art}
\bibfield{author}{\bibinfo{person}{Christof~Ferreira Torres}, \bibinfo{person}{Mathis Steichen}, {et~al\mbox{.}}} \bibinfo{year}{2019}\natexlab{}.
\newblock \showarticletitle{The art of the scam: Demystifying honeypots in ethereum smart contracts}. In \bibinfo{booktitle}{\emph{28th USENIX Security Symposium (USENIX Security 19)}}. \bibinfo{pages}{1591--1607}.
\newblock


\bibitem[Tsankov et~al\mbox{.}(2018)]%
        {tsankov2018securify}
\bibfield{author}{\bibinfo{person}{Petar Tsankov}, \bibinfo{person}{Andrei Dan}, \bibinfo{person}{Dana Drachsler-Cohen}, \bibinfo{person}{Arthur Gervais}, \bibinfo{person}{Florian Buenzli}, {and} \bibinfo{person}{Martin Vechev}.} \bibinfo{year}{2018}\natexlab{}.
\newblock \showarticletitle{Securify: Practical security analysis of smart contracts}. In \bibinfo{booktitle}{\emph{Proceedings of the 2018 ACM SIGSAC Conference on Computer and Communications Security}}. \bibinfo{pages}{67--82}.
\newblock


\bibitem[Wang et~al\mbox{.}(2023)]%
        {wang2023generating}
\bibfield{author}{\bibinfo{person}{Chong Wang}, \bibinfo{person}{Yiling Lou}, \bibinfo{person}{Junwei Liu}, {and} \bibinfo{person}{Xin Peng}.} \bibinfo{year}{2023}\natexlab{}.
\newblock \showarticletitle{Generating variable explanations via zero-shot prompt learning}. In \bibinfo{booktitle}{\emph{2023 38th IEEE/ACM International Conference on Automated Software Engineering (ASE)}}. IEEE, \bibinfo{pages}{748--760}.
\newblock


\bibitem[Wang et~al\mbox{.}(2021)]%
        {wang2021codet5}
\bibfield{author}{\bibinfo{person}{Yue Wang}, \bibinfo{person}{Weishi Wang}, \bibinfo{person}{Shafiq Joty}, {and} \bibinfo{person}{Steven~CH Hoi}.} \bibinfo{year}{2021}\natexlab{}.
\newblock \showarticletitle{CodeT5: Identifier-aware Unified Pre-trained Encoder-Decoder Models for Code Understanding and Generation}. In \bibinfo{booktitle}{\emph{Proceedings of the 2021 Conference on Empirical Methods in Natural Language Processing}}. \bibinfo{pages}{8696--8708}.
\newblock


\bibitem[Wang et~al\mbox{.}(2024)]%
        {wang2024efficiently}
\bibfield{author}{\bibinfo{person}{Zexu Wang}, \bibinfo{person}{Jiachi Chen}, \bibinfo{person}{Yanlin Wang}, \bibinfo{person}{Yu Zhang}, \bibinfo{person}{Weizhe Zhang}, {and} \bibinfo{person}{Zibin Zheng}.} \bibinfo{year}{2024}\natexlab{}.
\newblock \showarticletitle{Efficiently detecting reentrancy vulnerabilities in complex smart contracts}.
\newblock \bibinfo{journal}{\emph{Proceedings of the ACM on Software Engineering}} \bibinfo{volume}{1}, \bibinfo{number}{FSE} (\bibinfo{year}{2024}), \bibinfo{pages}{161--181}.
\newblock


\bibitem[Wood et~al\mbox{.}(2014)]%
        {wood2014ethereum}
\bibfield{author}{\bibinfo{person}{Gavin Wood} {et~al\mbox{.}}} \bibinfo{year}{2014}\natexlab{}.
\newblock \showarticletitle{Ethereum: A secure decentralised generalised transaction ledger}.
\newblock \bibinfo{journal}{\emph{Ethereum project yellow paper}} \bibinfo{volume}{151}, \bibinfo{number}{2014} (\bibinfo{year}{2014}), \bibinfo{pages}{1--32}.
\newblock


\bibitem[Wu et~al\mbox{.}(2024a)]%
        {wu2024ismell}
\bibfield{author}{\bibinfo{person}{Di Wu}, \bibinfo{person}{Fangwen Mu}, \bibinfo{person}{Lin Shi}, \bibinfo{person}{Zhaoqiang Guo}, \bibinfo{person}{Kui Liu}, \bibinfo{person}{Weiguang Zhuang}, \bibinfo{person}{Yuqi Zhong}, {and} \bibinfo{person}{Li Zhang}.} \bibinfo{year}{2024}\natexlab{a}.
\newblock \showarticletitle{iSMELL: Assembling LLMs with Expert Toolsets for Code Smell Detection and Refactoring}. In \bibinfo{booktitle}{\emph{Proceedings of the 39th IEEE/ACM International Conference on Automated Software Engineering}}. \bibinfo{pages}{1345--1357}.
\newblock


\bibitem[Wu et~al\mbox{.}(2021)]%
        {wu2021peculiar}
\bibfield{author}{\bibinfo{person}{Hongjun Wu}, \bibinfo{person}{Zhuo Zhang}, \bibinfo{person}{Shangwen Wang}, \bibinfo{person}{Yan Lei}, \bibinfo{person}{Bo Lin}, \bibinfo{person}{Yihao Qin}, \bibinfo{person}{Haoyu Zhang}, {and} \bibinfo{person}{Xiaoguang Mao}.} \bibinfo{year}{2021}\natexlab{}.
\newblock \showarticletitle{Peculiar: Smart contract vulnerability detection based on crucial data flow graph and pre-training techniques}. In \bibinfo{booktitle}{\emph{2021 IEEE 32nd International Symposium on Software Reliability Engineering (ISSRE)}}. IEEE, \bibinfo{pages}{378--389}.
\newblock


\bibitem[Wu et~al\mbox{.}(2024b)]%
        {wu2024advscanner}
\bibfield{author}{\bibinfo{person}{Yin Wu}, \bibinfo{person}{Xiaofei Xie}, \bibinfo{person}{Chenyang Peng}, \bibinfo{person}{Dijun Liu}, \bibinfo{person}{Hao Wu}, \bibinfo{person}{Ming Fan}, \bibinfo{person}{Ting Liu}, {and} \bibinfo{person}{Haijun Wang}.} \bibinfo{year}{2024}\natexlab{b}.
\newblock \showarticletitle{AdvSCanner: Generating Adversarial Smart Contracts to Exploit Reentrancy Vulnerabilities Using LLM and Static Analysis}. In \bibinfo{booktitle}{\emph{Proceedings of the 39th IEEE/ACM International Conference on Automated Software Engineering}}. \bibinfo{pages}{1019--1031}.
\newblock


\bibitem[Xue et~al\mbox{.}(2024)]%
        {xue2024openmoe}
\bibfield{author}{\bibinfo{person}{Fuzhao Xue}, \bibinfo{person}{Zian Zheng}, \bibinfo{person}{Yao Fu}, \bibinfo{person}{Jinjie Ni}, \bibinfo{person}{Zangwei Zheng}, \bibinfo{person}{Wangchunshu Zhou}, {and} \bibinfo{person}{Yang You}.} \bibinfo{year}{2024}\natexlab{}.
\newblock \showarticletitle{Openmoe: An early effort on open mixture-of-experts language models}.
\newblock \bibinfo{journal}{\emph{arXiv preprint arXiv:2402.01739}} (\bibinfo{year}{2024}).
\newblock


\bibitem[Yu et~al\mbox{.}(2023a)]%
        {yu2023pscvfinder}
\bibfield{author}{\bibinfo{person}{Lei Yu}, \bibinfo{person}{Junyi Lu}, \bibinfo{person}{Xianglong Liu}, \bibinfo{person}{Li Yang}, \bibinfo{person}{Fengjun Zhang}, {and} \bibinfo{person}{Jiajia Ma}.} \bibinfo{year}{2023}\natexlab{a}.
\newblock \showarticletitle{PSCVFinder: A Prompt-Tuning Based Framework for Smart Contract Vulnerability Detection}. In \bibinfo{booktitle}{\emph{2023 IEEE 34th International Symposium on Software Reliability Engineering (ISSRE)}}. IEEE, \bibinfo{pages}{556--567}.
\newblock


\bibitem[Yu et~al\mbox{.}(2023b)]%
        {yu2023money}
\bibfield{author}{\bibinfo{person}{Lei Yu}, \bibinfo{person}{Fengjun Zhang}, \bibinfo{person}{Jiajia Ma}, \bibinfo{person}{Li Yang}, \bibinfo{person}{Yuanzhe Yang}, {and} \bibinfo{person}{Wei Jia}.} \bibinfo{year}{2023}\natexlab{b}.
\newblock \showarticletitle{Who Are the Money Launderers? Money Laundering Detection on Blockchain via Mutual Learning-Based Graph Neural Network}. In \bibinfo{booktitle}{\emph{2023 International Joint Conference on Neural Networks (IJCNN)}}. IEEE, \bibinfo{pages}{1--8}.
\newblock


\bibitem[Zhang et~al\mbox{.}(2022)]%
        {zhang2022reentrancy}
\bibfield{author}{\bibinfo{person}{Zhuo Zhang}, \bibinfo{person}{Yan Lei}, \bibinfo{person}{Meng Yan}, \bibinfo{person}{Yue Yu}, \bibinfo{person}{Jiachi Chen}, \bibinfo{person}{Shangwen Wang}, {and} \bibinfo{person}{Xiaoguang Mao}.} \bibinfo{year}{2022}\natexlab{}.
\newblock \showarticletitle{Reentrancy Vulnerability Detection and Localization: A Deep Learning Based Two-phase Approach}. In \bibinfo{booktitle}{\emph{37th IEEE/ACM International Conference on Automated Software Engineering}}. \bibinfo{pages}{1--13}.
\newblock


\bibitem[Zheng et~al\mbox{.}(2024)]%
        {zheng2024llamafactory}
\bibfield{author}{\bibinfo{person}{Yaowei Zheng}, \bibinfo{person}{Richong Zhang}, \bibinfo{person}{Junhao Zhang}, \bibinfo{person}{Yanhan Ye}, {and} \bibinfo{person}{Zheyan Luo}.} \bibinfo{year}{2024}\natexlab{}.
\newblock \showarticletitle{Llamafactory: Unified efficient fine-tuning of 100+ language models}.
\newblock \bibinfo{journal}{\emph{arXiv preprint arXiv:2403.13372}} (\bibinfo{year}{2024}).
\newblock


\bibitem[Zhu et~al\mbox{.}(2024)]%
        {zhu2024llama}
\bibfield{author}{\bibinfo{person}{Tong Zhu}, \bibinfo{person}{Xiaoye Qu}, \bibinfo{person}{Daize Dong}, \bibinfo{person}{Jiacheng Ruan}, \bibinfo{person}{Jingqi Tong}, \bibinfo{person}{Conghui He}, {and} \bibinfo{person}{Yu Cheng}.} \bibinfo{year}{2024}\natexlab{}.
\newblock \showarticletitle{Llama-moe: Building mixture-of-experts from llama with continual pre-training}. In \bibinfo{booktitle}{\emph{Proceedings of the 2024 Conference on Empirical Methods in Natural Language Processing}}. \bibinfo{pages}{15913--15923}.
\newblock


\bibitem[Zhuang et~al\mbox{.}(2020)]%
        {zhuang2020smart}
\bibfield{author}{\bibinfo{person}{Yuan Zhuang}, \bibinfo{person}{Zhenguang Liu}, \bibinfo{person}{Peng Qian}, \bibinfo{person}{Qi Liu}, \bibinfo{person}{Xiang Wang}, {and} \bibinfo{person}{Qinming He}.} \bibinfo{year}{2020}\natexlab{}.
\newblock \showarticletitle{Smart Contract Vulnerability Detection using Graph Neural Network.}. In \bibinfo{booktitle}{\emph{IJCAI}}. \bibinfo{pages}{3283--3290}.
\newblock


\bibitem[Zou et~al\mbox{.}(2019)]%
        {zou2019smart}
\bibfield{author}{\bibinfo{person}{Weiqin Zou}, \bibinfo{person}{David Lo}, \bibinfo{person}{Pavneet~Singh Kochhar}, \bibinfo{person}{Xuan-Bach~Dinh Le}, \bibinfo{person}{Xin Xia}, \bibinfo{person}{Yang Feng}, \bibinfo{person}{Zhenyu Chen}, {and} \bibinfo{person}{Baowen Xu}.} \bibinfo{year}{2019}\natexlab{}.
\newblock \showarticletitle{Smart contract development: Challenges and opportunities}.
\newblock \bibinfo{journal}{\emph{IEEE Transactions on Software Engineering}} \bibinfo{volume}{47}, \bibinfo{number}{10} (\bibinfo{year}{2019}), \bibinfo{pages}{2084--2106}.
\newblock


\end{thebibliography}

\end{document}